# Transforming the Hybrid Cloud for Emerging AI Workloads

White Paper

Prepared by:
IBM Research and University of Illinois Urbana-Champaign
IBM-Illinois Discovery Accelerator Institute (IIDAI)

Deming Chen*, Alaa Youssef+, Ruchi Pendse+, André Schleife*, Bryan K. Clark*, Hendrik Hamann+, Jingrui He*, Teodoro Laino+, Lav Varshney*, Yuxiong Wang*, Avirup Sil+, Reyhaneh Jabbarvand*, Tianyin Xu*, Volodymyr Kindratenko*, Carlos Costa+, Sarita Adve*, Charith Mendis*, Minjia Zhang*, Santiago Núñez-Corrales*, Raghu Ganti+, Mudhakar Srivatsa+, Nam Sung Kim*, Josep Torrellas*, Jian Huang*, Seetharami Seelam+, Klara Nahrstedt*, Tarek Abdelzaher*, Tamar Eilam+, Huimin Zhao*, Matteo Manica+, Ravishankar Iyer*, Martin Hirzel+, Vikram Adve*, Darko Marinov*, Hubertus Franke+, Hanghang Tong*, Elizabeth Ainsworth*, Han Zhao*, Deepak Vasisht*, Minh Do*, Sahil Suneja+, Fabio Oliveira+, Giovanni Pacifici+, Ruchir Puri+, Priya Nagpurkar+

* : University of Illinois Urbana-Champaign. + : IBM Research

Corresponding Authors: Deming Chen (dchen@illinois.edu); Alaa Youssef (asyousse@us.ibm.com)

## ABSTRACT

This white paper, developed through close collaboration between IBM Research and University of Illinois Urbana-Champaign researchers within the IBM-Illinois Discovery Accelerator Institute (IIDAI), envisions transforming hybrid cloud systems to meet the growing complexity of AI workloads through innovative, full-stack co-design approaches, emphasizing usability, manageability, affordability, adaptability, efficiency, and scalability. By integrating cutting-edge technologies such as generative and agentic AI, cross-layer automation and optimization, unified control plane, and composable and adaptive system architecture, the proposed framework addresses critical challenges in energy efficiency, performance, and cost-effectiveness. Incorporating quantum computing as it matures will enable quantum-accelerated simulations for materials science, climate modeling, and other high-impact domains. Collaborative efforts between academia and industry are central to this vision, driving advancements in foundation models for material design and climate solutions, scalable multimodal data processing, and enhanced physics-based AI emulators for applications like weather forecasting and carbon sequestration. Research priorities include advancing AI agentic systems, LLM as an Abstraction (LLMaaA), AI model optimization and unified abstractions across heterogeneous infrastructure, end-to-end edge-cloud transformation, efficient programming model, middleware and platform, secure infrastructure, application-adaptive cloud systems, and new quantum-classical collaborative workflows. These ideas and solutions encompass both theoretical and practical research questions, requiring coordinated input and support from the research community. This joint initiative aims to establish hybrid clouds as secure, efficient, and sustainable platforms, fostering breakthroughs in AI-driven applications and scientific discovery across academia, industry, and society.

## Table of Contents

# 1 Executive Summary

Our goal over the next 5-10 years is to identify new computing, storage, and communication elements, subsystems, and innovations across all stack levels to transform the hybrid cloud for emerging artificial intelligence (AI) workloads. In this white paper, we intend to answer the following two key questions: (1) What is possible? (2) What would be the outcome? We envision a reimagined hybrid cloud system featuring a fully integrated and optimized stack that supports a wide range of AI frameworks, runtimes, tools, and various hardware resources, including cache-coherent interconnects, SmartNiCs, AI accelerators, and quantum computers - just to name a few. This will ensure that AI remains a dynamic and transformative force in years to come. Our aspiration is to achieve a 100-1000x improvement in performance/watt when all the pieces come together.

Generative AI (Gen AI), foundation models (FMs), and large language models (LLMs) represent significant advancements in AI technology. These innovations have the potential to revolutionize various sectors by enabling more sophisticated and context-aware applications. Market research anticipates substantial growth in the AI sector over the next 5-10 years. The demand for AI-driven solutions is expected to skyrocket as industries recognize the potential of Gen AI, FMs, and LLMs to drive efficiency, innovation, and competitive advantage.

These AI models need terabytes or petabytes of data to train and often require more advanced memory architectures and storage solutions to handle the vast amounts of data and model parameters. Meanwhile, given the layers of such AI models are interdependent, and transformer architectures and attention mechanisms require close coordination between tokens in a sequence, such tightly coupled computations necessitate high-speed interconnections and synchronization across numerous hardware components. In addition, such AI models are rapidly advancing and evolving, experiencing exponential increase in model size and data, requiring heterogenous hardware environment with various AI accelerators, and demanding high throughput, high energy efficiency, and dependable performance and accuracy. All these differ from traditional high-performance computing (HPC) workloads in their complexity, computational demands, infrastructure needs, and breadth of applications.

Most importantly, the compute system to support such AI workloads must be affordable to ensure broad accessibility and adoption. Cost-effective solutions leveraging cloud-based technologies and scalable hardware systems are essential to support the widespread use of the advanced AI models. Additionally, the complexity of these systems necessitates innovative approaches to manage and simplify their design, deployment, and operation. Overall, managing complexity, ensuring cost-effectiveness, maintaining sustainability and efficiency, and being adaptive for rapidly changing workload characteristics and demands are key challenges that must be addressed to drive the momentum of AI advancements.

To address these key challenges, we call upon the research community to optimize and redesign application, middleware, platform, infrastructure, and hardware layers of cloud systems, while also implementing intelligent and scalable full-stack system management solutions. Incorporating new enabling technologies, such as agentic AI, unified control plane, cross-layer automation and integration, and composable and adaptive design approaches, will be critical to support these advancements.

To maximize the usability of AI technologies, it is important to develop new application interfaces that make it easier for users (and developers) to interact with and use AI models and frameworks. We propose LLM-as-an-Abstraction (LLMaaA), which relies on agentic AI frameworks to enable a natural language-based user interface for building, deploying, and managing complex applications. Efficient scaling and optimization of emerging AI models are essential for maximizing their impact on real-world applications. New reconfigurability and adaptability features should be explored across various system levels to dynamically optimize cloud resources for specific workload requirements and integrate emerging technologies efficiently. The expansion of cloud computing to the edge, and the development of smart cooperative edge-to-cloud computing models, further enhances the capabilities of AI systems, enabling real-time processing and decision making at the point of need. This evolution brings additional challenges related to data distribution, security, privacy, and latency.

We propose cross-layer automation and AI-driven integration to improve resource allocation, scheduling, and performance monitoring. This will enhance efficiency, flexibility, and scalability for complex AI workloads. Advancing programmable, AI-optimized networking and data management systems for large-scale models will be key to ensuring smooth, secure data flow across hybrid cloud environments. Leveraging emerging hardware technologies like CXL and GPUDirect through system co-design will enhance performance and energy efficiency. Future AI systems should also prioritize security, robustness, and energy management, with novel frameworks that ensure adaptability, sustainability, and resilience against evolving security threats in hybrid cloud environments.

As quantum computing hardware becomes increasingly available and reliable toward producing actual utility for applications, we foresee its increasing integration into Hybrid Clouds alongside AI. An emerging example for such hardware includes using the modern quantum computer as a powerful accelerator for simulations. To make this an everyday reality, we envision advancing quantum computing by demonstrating its utility through diverse scientific use cases: initially targeting quantum chemistry, materials science, and physics, and the goal is to expand the space of applications and to showcase how quantum computing can solve domain-specific problems by generalizing these solutions to other fields. This involves integrating quantum algorithms into classical computational workflows, exploring their performance and noise characteristics, developing new quantum algorithms, and testing error mitigation techniques.

Leveraging the new cloud system, we specifically target two important AI-driven scientific computing applications. First, for material discovery, future research will focus on using AI foundation models (FMs) for inverse material design, exploring vast chemical spaces, and developing intelligent agents to enhance problem-solving in material synthesis and optimization. Second, the long-term vision for climate and sustainability research focuses on developing scalable and efficient FMs that can process multimodal, high-dimensional data. This includes optimizing model size, composability, and pre-training strategies. Another key direction is improving physics-based AI emulators for applications like weather forecasting and carbon sequestration. Additionally, exploring the intersection of quantum computing with climate solutions presents new opportunities for scientific discovery in areas such as climate impact analysis and greenhouse gas emission quantification.

In summary, we are committed to fostering and developing a new generation of ideas and solutions that will revolutionize how we conceptualize, design, and operate future hybrid clouds for emerging AI frameworks and models while paving the road for seamless integration of quantum-bound workloads. Our focus is on ease of use, affordability, adaptability, and ubiquity at very large scales. These ideas and solutions encompass both theoretical and practical research questions, requiring coordinated input and support from the research community. Our role is to spearhead and nurture the environment that will lead to the realization of these transformative technologies.

## 2 Introduction

IBM and the Grainger College of Engineering at the University of Illinois Urbana-Champaign launched the IBM-Illinois Discovery Accelerator Institute in 2021 to combine the strengths of academia and industry to spur breakthroughs in the rapidly growing areas of hybrid cloud, AI, quantum computing, and AI-driven scientific computing applications for material discovery, climate and sustainability. Subsequently, the Institute announced a new call for proposals for multi-year projects in March of 2023.

In this white paper, for each of the focus areas, we outline challenges and opportunities, state-of-the-art solutions based on ongoing IBM-Illinois collaborations, and the long-term vision and the technical approach and research priorities of the Institute.

The growing adoption of AI-driven applications is leading to a significant surge in demand for high-performance computing resources. Recently, LLMs, with billions of parameters and trained on terabytes of data, have shown outstanding capabilities in handling a plethora of tasks. We expect this trend to continue with increase in number of parameters, amount of training data, and modalities of foundation models. The demand for more performant and optimized cloud platforms and infrastructure continues to rise, driven by the need for efficient training, fine-tuning, and inference of advanced AI models. Simultaneously, the maturation of quantum computing technologies necessitates thinking about their integration, challenges, and expected potential in the context of hybrid cloud systems. In this paper, we show how the resulting cost, complexity, and fragmentation of existing technologies and solutions cannot meet such demand, and we lay out our grand vision for a converged, adaptive, high-performant, affordable, consumable, and ubiquitously accessible hybrid cloud system.

Over the past sixty years, supercomputing has been instrumental to fundamental discoveries in scientific areas such as physics, astronomy, and medicine. Researchers and engineers have used supercomputing to study, simulate, and predict complex systems such as weather, financial systems, air travel, and more. Over the last fifteen years, supercomputing has enabled new fields, such as big data analytics and AI foundation models. Today, more than ever, the ability to drive new drugs or material discoveries, understand and manage complex systems like the earth's climate, or create ever more powerful intelligent automation depends on the ever-increased power of AI-enabled applications and systems.

However, the growing power of AI systems is being increasingly challenged by their rising complexity and cost. When developing future AI-enabled applications, teams must navigate the intricacies of large distributed systems, manage heterogeneous platforms and infrastructures, and now face the added complexity introduced by quantum computing. Balancing these demands while ensuring efficiency and affordability presents a significant challenge for future development. As a result, such applications require large teams and development cycles spanning months and years. What is even more problematic, the cost and amount of specialized computing resources required have grown significantly over time, making it no longer affordable for many university and scientific institutions. Furthermore, the energy required to train large AI models and run today's AI-enabled applications makes them unsustainable. The effort to build the AI platforms and applications of tomorrow will likely reach a regime of diminishing returns in multiple dimensions unless new ideas arise.

In this paper, we will highlight a new research program that the IBM-Illinois Discovery Accelerator Institute (IIDAI) has launched to tackle these challenges and change the current path of AI computing to make it more efficient, affordable, and accessible to enable the next generation of AI-enabled applications.

We are committed to fostering and developing a new generation of ideas that will revolutionize how we conceptualize, design, and operate the future hybrid cloud for emerging AI workloads. Our vision focuses on creating cloud systems that prioritize ease of use, affordability, adaptability, and ubiquity at massive scales. These ideas encompass both theoretical foundations and applied research questions, which require the collective input and support of the broader research community. Our role is to lead, catalyze, and nurture the conditions for these innovations to come to fruition.

### 2.1 Evolution of High-Performance Computing

For many years, the field of High-Performance Computing (HPC) has been dedicated to enabling scientific exploration and discovery through large-scale simulations and modeling in domains such as physics, biology, climate science, and many other fields. Traditional HPC applications were often developed as monolithic programs written in languages like FORTRAN and ran on single, large systems, such as those provided by Cray supercomputers.

Distributed computing clusters built with x86 machines emerged and delivered the needed performance for HPC simulations using frameworks for parallel processing like MPI. Such systems were crucial to fulfill the potential of supercomputers anticipated by the Atkins report in 2003 [1]. Initiatives such as TeraGrid and XSEDE then made these compute resources accessible to thousands of research scientists, reinvigorating science as a whole. This in turn enabled widespread access to modeling and simulation, a de facto third leg of science [2] alongside theory and experiment. Scientific computing applications continued to evolve and started to encompass new areas such as data analytics and machine learning (ML), which introduced corresponding heterogeneity in supporting platforms and infrastructures.

## 2.2 Today’s AI-infused Scientific Computing Applications, Platforms, and Infrastructures

Today, scientific computing applications span multiple computing paradigms, many platforms, and a variety of heterogeneous infrastructure requirements, as shown in Figure 1. Besides classical simulations, big data analytics, AI and machine learning are being used heavily. Each application component requires dedicated platform-level support, for instance, Spark for data analytics, PyTorch for AI training, and Ray for machine learning processing and model serving. Additionally, the rise of massively parallel distributed clusters and specialized hardware accelerators like GPUs, TPUs, and FPGAs has enabled highly efficient processing of specific tasks. These resources, essential for handling AI workloads, are now readily accessible in the cloud, making the combination of Cloud + AI a promising approach for advancing many HPC applications.

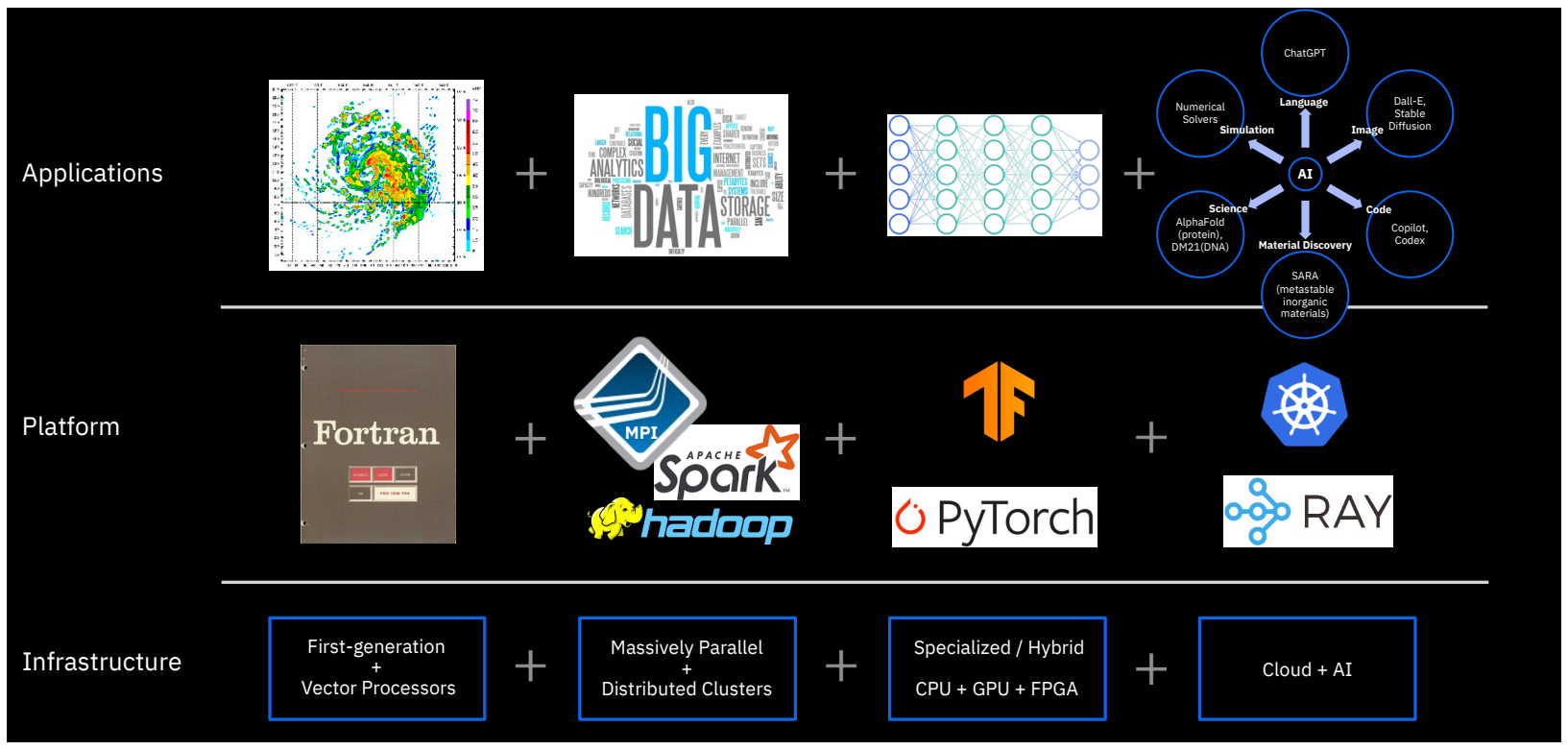

*Figure 1: Current state of AI-infused scientific applications, platforms, and infrastructure.*

AI-infused simulation is a fast-growing application paradigm in HPC. For example, in a cancer research application that simulates protein interactions at different scales of resolution with physics and molecular dynamics ensembles, the infusion of AI techniques in simulations accelerates search-space exploration by orders of magnitude. To accomplish such complex tasks, the combination of batch schedulers, such as Flux and Slurm, on massively parallel computers like IBM Summit [54], along with specialized accelerators and distributed cloud clusters for AI model training and inference, is essential.

## 2.3 AI-centric Hybrid Cloud Computing of the Future

Tomorrow, the future hybrid cloud will enable next-generation applications, which embrace fast-growing AI workloads, such as large language models and other types of Gen-AI workloads, big data analysis, AI-driven simulations, quantum resources and algorithms, to tackle new challenging problems in more efficient and integrated ways. Figure 2 illustrates this future trend.

More precisely, we envision systems that exhibit the following properties:

- Easy to consume by scientists, and will need to seamlessly leverage Classical, AI, and Quantum computing capabilities, from within the same hybrid cloud context. They will automatically map high-level task descriptions to the low-level optimized constructs.
- Affordability of such systems is a critical requirement to democratize access to resources. We envision a new breed of systems and specialized computing elements with 100x cost/performance improvements over today’s technology.
- Ubiquitous access to ensure demand fulfilment with abundant resource availability everywhere, is yet another requirement for these systems. Accessing these systems will be more similar to plugging into a standardized outlet than performing a request based on a detailed specification.

The table below highlights how evolving needs in HPC systems have been incrementally addressed through specific technical advancements. Despite the success of these responses, continued reliance on traditional approaches will fall short when scaling to more complex hybrid systems. We believe that new, innovative ideas are essential to tackle the future challenges of ease of use, adaptability, affordability, and ubiquity at greater problem and system scales.

| Dimension | Past | Current | Technical response |
|---|---|---|---|
| Data volume | Small number of parameters | Very large datasets | Multiple scales and localities of data storage |
| Data format | Structured, homogeneous | Unstructured, heterogeneous | Multimode information storage and retrieval mechanisms |
| Computational structure | Stencil-based, static, SIMD-friendly | Problem-specific, dynamic, varying distributed workloads | Hybrid and adaptive architectures (CPU, GPU, TPU, AIU, FPGA, QPU) |
| Problem type | Linear or linear-approximate | Non-linear | Advanced statistical and numerical methods |
| Problem dynamics | Deterministic | Stochastic | Stochastic numerical methods |
| Workflow composition | Homogeneous | Heterogeneous | Multiple scales of scheduling and division of tasks with better programmability |
| Counterfactual modeling | Parameter variation | Parameter variation + combinatorial explosion | Approximate algorithms, AI, quantum algorithms |
| Solution reachability | High | Very low | Programmability, portability, standard API |
| Extent of aims | Impact to scientific and applied matters | Impact to societal grand challenges | Integrated, specialized, and accessible cyberinfrastructure |
| Urgency of solutions | Low to medium | Medium to high | High-performance, high-throughput computing |

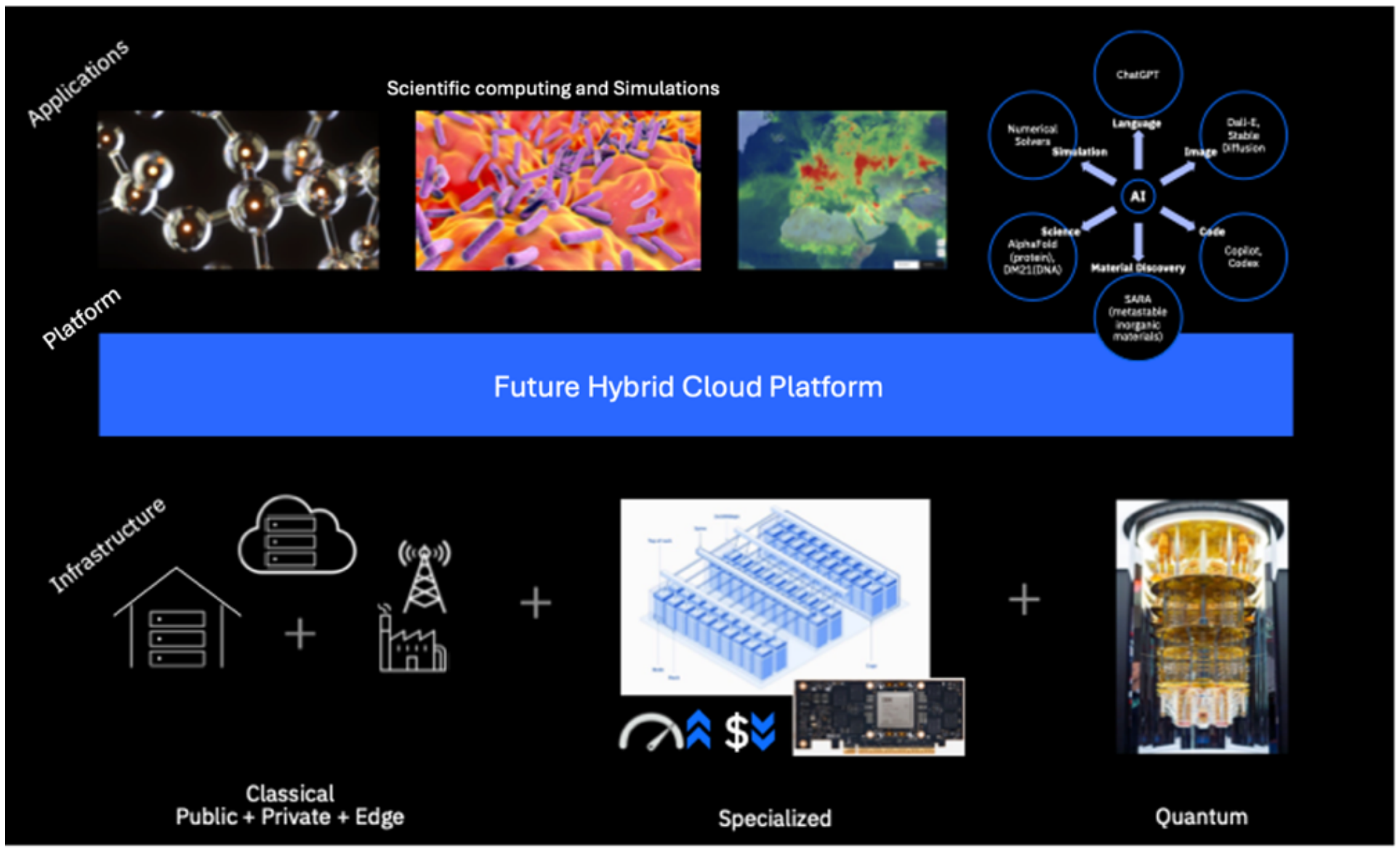

Figure 2: Hybrid Cloud Platform of the Future.

# 3 Challenges of AI-centric Hybrid Cloud Computing

AI-centric hybrid cloud systems are complex, difficult to model at operationally relevant levels of detail, costly, and unsustainable due to the inherent challenges of their massive scale. These issues, which will be explored in the following subsections, present significant barriers to future progress. In short, current knowledge, infrastructure, and tools are insufficient to propel future computing systems forward by orders of magnitude without encountering severely diminishing returns. Addressing these challenges requires fresh approaches to sustain scalability and efficiency at larger system scales.

## 3.1 Complexity and Difficult Consumability

In the words of the late Richard Hamming, "the purpose of computing is insight, not numbers" [3]. The co-evolution of scientific goals and computational infrastructure has been a defining factor in shaping research-driven software and hardware over the past 80 years. We've progressed from calculating non-elementary integrals with a few parameters to analyzing petabytes of video streams in near real-time, thanks to a positive feedback loop between the ever-growing challenges in research and the continuous advancements in computational capacity. In essence, as the complexity, volatility, and uncertainty of today's major research challenges continue to grow, the development and application of future computing systems are poised to generate significant, transformative insights.

Hardware heterogeneity in contemporary AI-centric hybrid cloud computing has been on the rise, causing increased complexity for the acquisition, integration, programming and administration of these systems. A mix of CPUs, GPUs, TPUs, FPGAs, and other specialized hardware accelerators that provide higher performance at higher energy efficiency entails multiple programming paradigms, new physical and logical constraints, and even higher hardware procurement uncertainty across market value chains. To make matters worse, the need to access increasingly scarce compute resources requires bursting from on-premise private clouds to multiple cloud providers, which do not offer a uniform management interface and introduce variance in the offered resource types, their cost, and cost models.

These distributed computing platforms come with their own challenges in scaling, distributed data movement, consistency and synchronization, scheduling and resource management, fault tolerance, and resiliency, just to name a few. Management and operations of each of these distributed computing platforms require deep talent and skills. A single team having to manage multiple of these platforms might find the task daunting.

HPC and AI systems today use a variety of programming models, frameworks, and libraries at different layers of abstraction. This wide range includes low-level GPU programming models like CUDA and NCCL, all the way to higher-level frameworks and abstractions like OpenMP, PyTorch, Spark and Ray. In addition, many workflow managers are in use today, such as Airflow, Makeflow, Pegasus, Argo, Parsl, etc. Moreover, a mixture of schedulers for bare-metal (e.g., LSF, Slurm), virtualized, and container systems (e.g., Kubernetes, OpenShift) are used. There is no single, unified, easy way to program or administer these systems. Such a diverse composition of interfaces, programming models, runtimes, and infrastructure systems leads to difficult consumability. A unified abstraction layer is therefore needed.

## 3.2 Complexity of Engineering Very Large-scale Systems

Increasing demands on AI and computing infrastructure have affected the resilience, operation, programmability, and design of AI-centric cloud systems. We anticipate that future systems will achieve unprecedented levels of integration, growing by orders of magnitude compared to today's systems. Reflecting on the behavior of these systems, we can draw the following key observations.

Very high hardware and data volumes necessitate thinking at the thermodynamic limit. Current error rates per bit reach $10^{-24}$ in the best-case scenario. In a system running at 1 exaflop ($10^{18}$), one would need $10^{6}$ seconds (~11.6 days) to observe an error. These error rates and number of bits are commensurate with Avogadro's number and Boltzmann's constant respectively, which justify thinking about computing as a phenomenon at thermodynamic limit through the tools of statistical physics. Moving from exaflops to zettaflops and beyond will justify even more such treatment into higher data units (e.g., bits to bytes).

Very large hybrid hardware platforms and complex software workflows follow known scaling laws driven by constraints. As the number and diversity of processing elements grow, the structure and communication patterns of these computing systems converge to hierarchical modularity. In particular, patterns of specialization and information integration have been shown to follow Rent's rule [4]. Rentian scaling arises in systems whose evolution is driven by the economics of constraints (e.g., energy consumption, complexity) and utility (e.g., efficiency, functionality).

Very large hybrid hardware platforms and complex software workflows also emerge compositionally. The properties of individual nodes and processing elements determine the ability of an entire system to compute at scale. The laws that govern the result depend to a large extent on the resources and modes of interaction these elements provide. For instance, space and time comprise the main resources for CPU, GPU, TPU, AIU, and FPGA architectures, while QPUs (Quantum Processing Units) add superposition, interference and entanglement. The ability to understand and predict system properties based on those of its components will be essential, yet increasingly complex as more diverse elements are introduced.

Very large hybrid hardware platforms increase software complexity. The number of possible bugs in a program appears to be directly proportional to the number of lines of code and the number of different kinds of compute elements involved. Beyond correctness, this growing complexity produces two classes of bottlenecks: one in which the probability of finding the skill set to write scientific software decreases as more technologies are added, and another one in which optimizing code execution becomes increasingly difficult and can fall outside of human ability in the future.

### 3.3 Limited Affordability and Adaptability

AI-centered computing is expensive. Moreover, the total expenses of running the system are not fixed, but depend on various factors that change over time and are influenced by market forces. These include the costs of operation, maintenance, and energy consumption, which may increase (typically) or decrease (seldom) depending on the demand and supply.

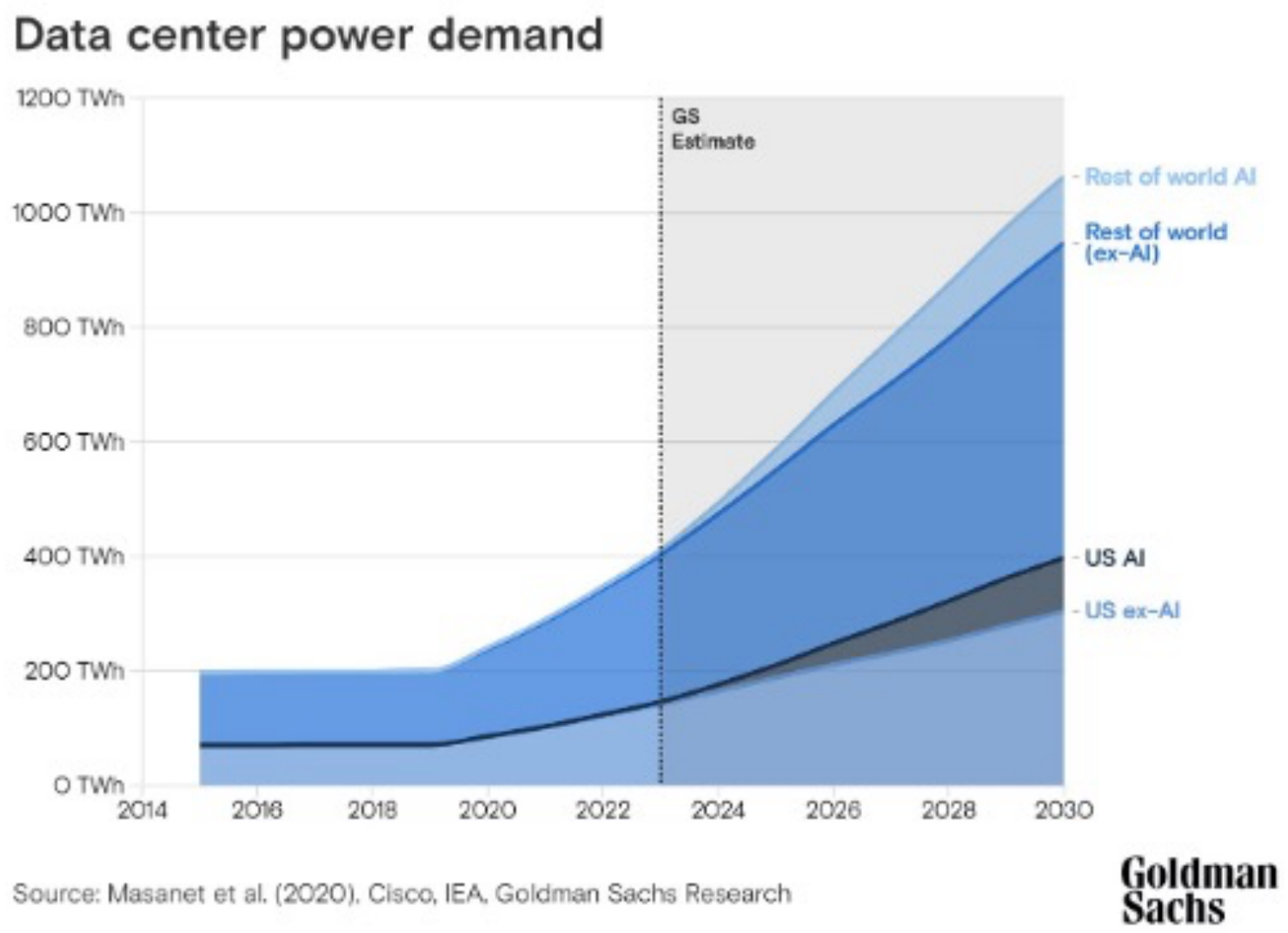

*Figure 3: Estimated increase in data center power demand driven by AI.*

We believe that an industrial-academic collaboration is the best approach to achieve the best advanced technical results at reasonable cost. We have seen many successful cases of this model, where the synergy between the two sectors led to breakthroughs that would not have been possible otherwise. It is encouraging to see that very large-scale academic collaboration has led to results matching focused industry innovation. For example, the BigScience Large Open-science Open-access Multilingual Language Model (BLOOM) matches the size of GPT-3. BLOOM resulted from collaboration between ~1000 researchers, 60 countries, and 250 institutions.

The training of large language models is unsustainable. The energy required for training is estimated to double every 3-4 months, and a single training of a large language model such as GPT-3 may emit the equivalent carbon of 10 cars throughout their lifetimes. What is worse is that pre-training is just the beginning of the journey. Once a model is deployed it needs to be retrained to maintain accuracy. Re-training can occur on a daily, weekly, or monthly basis, and billions of inference jobs contribute significantly to carbon emissions. It's also important to note that the manufacturing of specialized devices, such as GPUs, carries a substantial environmental impact, including a large carbon footprint, high water usage, and the release of toxic materials. These factors must be taken into account when considering the sustainability of AI-centric cloud systems.

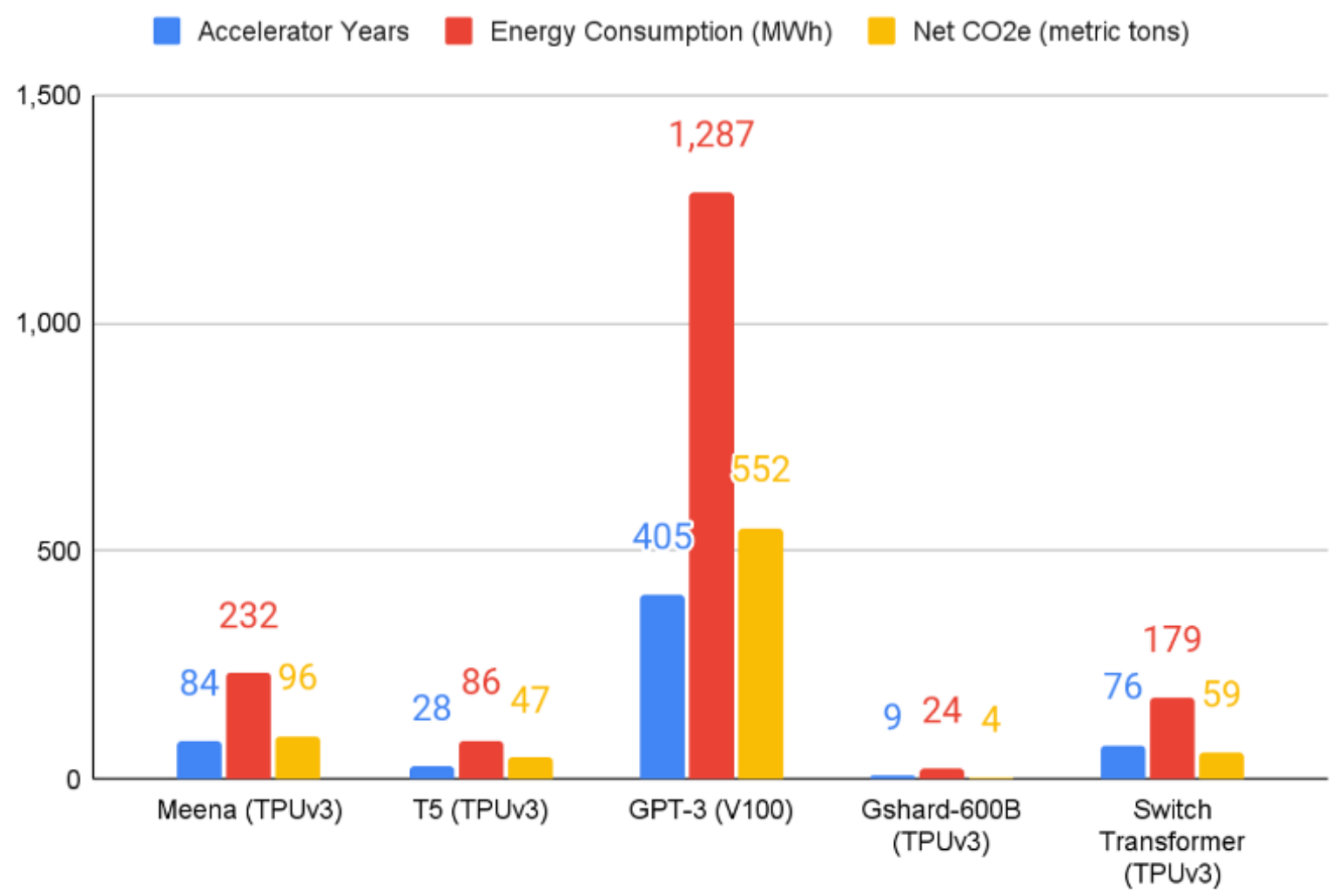

*Figure 4: AI model training in accelerator years, energy consumption, and CO2e emission.*

AI is expected to be the dominant factor in 160% increase in data center power demand, by 2030. This is illustrated in Figure 3. A ChatGPT query needs nearly 10 times as much electricity to process as a Google search [5]. Figure 4 shows that OpenAI trained the 175B GPT-3 by consuming 1.287 GWh, using 405 V100 GPU years, and emitting over 550 metric tons of CO2 equivalent gases [6].

Figure 5 shows that around 2015-2016 a new trend of large-scale models emerged. This new trend began with AlphaGo in late 2015 and continues up to the present day. The trend of increasing compute in these large-scale models appears to double every 4 to 9 months [7].

IceCube was an experiment conducted by an Antarctic observatory dedicated to detecting and analyzing neutrinos, with the goal of performing a month's worth of simulation in 1 hour. This required 80,000 GPUs, and the team set out to acquire them from AWS. They did not succeed, and ended up acquiring only 51,000 GPUs across AWS, Azure, and Google. This failed large-scale experiment illustrates the scarcity of the resources needed for these HPC projects, despite availability of funds to support such research work. In addition, even when these compute resources may be present, logistics across market value chains may pose conditions adverse to the prompt acquisition of relevant parts.

One advantage of a hybrid cloud system is its flexibility to offload workloads from a private cloud to a public cloud when additional resources are needed. However, offloading (or bursting) workloads to public cloud services presents notable challenges, particularly around cost management and price fluctuations, especially when seeking additional GPUs or specialized AI accelerators. The unpredictability of finding the right accelerator type, knowing where it is available, and assessing its cost can create barriers. This uncertainty impacts budgeting and cost management as prices for GPU or other accelerator resources can vary greatly based on availability, demand, and location. Such variability makes it difficult to plan and control expenses, adding complexity to managing hybrid cloud operations effectively. This issue emphasizes the need for more predictable resource allocation and transparent cost models in hybrid cloud services.

Increased specialization and reliance on special-purpose accelerators to cater to application-specific needs lead to more efficient systems. However, such systems lack the flexibility needed to adapt swiftly to different application and workload demands. The current HPC or cloud systems have some limited ability for reconfiguration and programmability. However, the existing capabilities focus on point solutions and do not offer a systematic and holistic treatment that coordinates different levels of such capabilities that can take place in the system hierarchy. Today's large-scale distributed computing platforms lack capabilities to reconfigure, reprogram, and adapt themselves to serve dynamically changing workloads while ensuring highly optimized performance and efficiency.

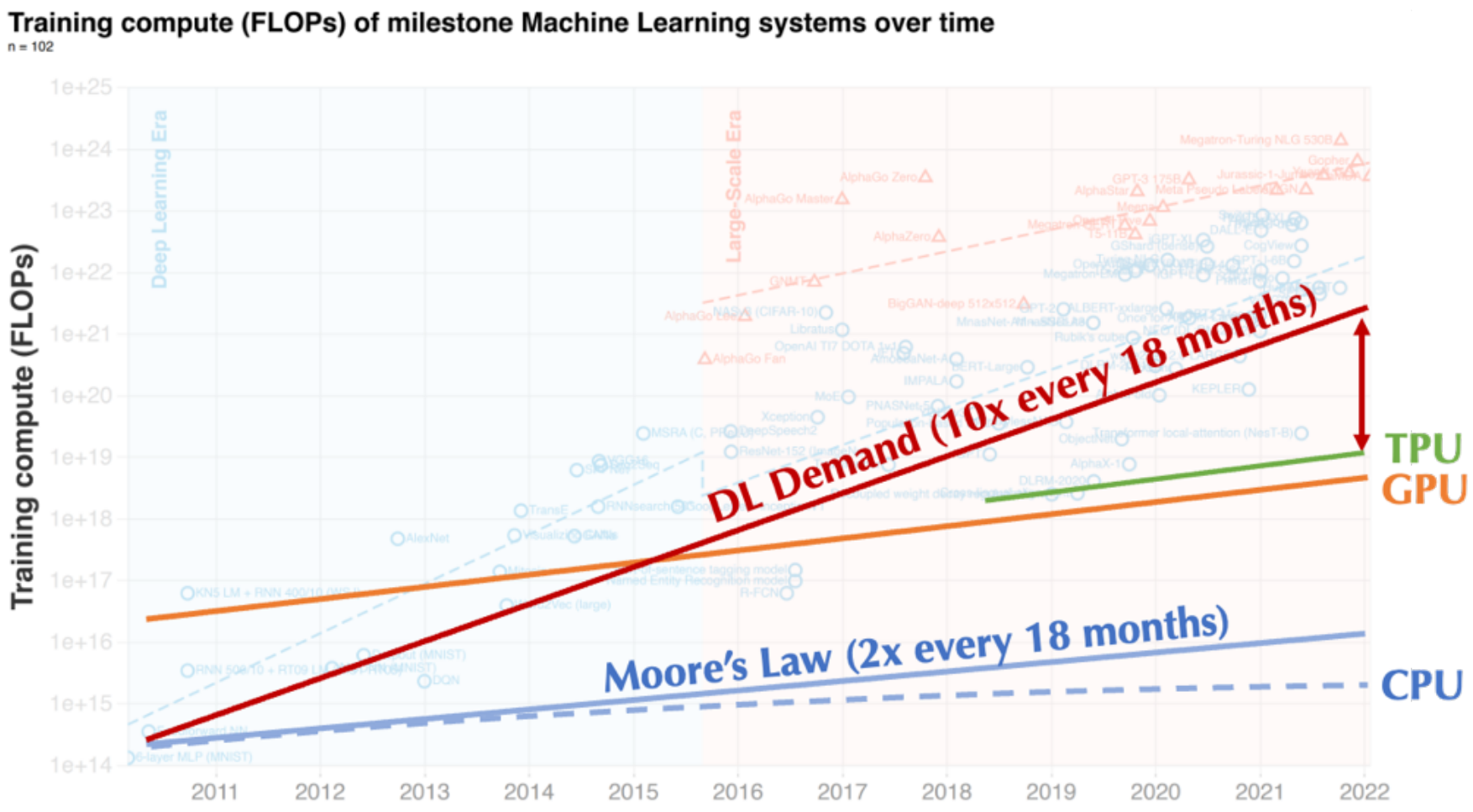

*Figure 5: Trends in training compute demand of ML systems, showing the emergence of a new trend of large-scale models, contrasted to CPU, GPU, and TPU compute trends.*

# 4 Transforming the Hybrid Cloud for the Future

## 4.1 A Broad Vision

The IBM-Illinois shared vision for IIDAI is to transform the hybrid cloud systematically, addressing the various challenges outlined in Section 3. Our goal is to identify new computing, storage, and communication elements, sub-systems, and innovations across the various layers of the whole cloud computing stack. We envision a reimagined hybrid cloud platform and infrastructure featuring a fully integrated and optimized stack that supports a wide range of AI frameworks, runtime middleware, tools, and hardware. This will ensure that AI remains a dynamic and transformative force in the years to come, while understanding how quantum technologies will integrate to it with a more strategic outlook. Our aspiration is to achieve a 100-1000x improvement in performance/watt when all the pieces come together. Our focus is on ease of use, affordability, adaptability, and ubiquity at very large scales.

Emerging compute-intensive workloads, especially driven by the disruptive new AI/ML techniques, are increasingly complex workflows with wide-ranging compute and data requirements. From the remarkable rise of large-scale distributed training leading to self-supervised models (also known as foundation models), to complex workflows mixing workloads with different characteristics, such as asynchronous, batch jobs (e.g., data ingestion, pre-processing, and training), and synchronous, interactive computations (e.g., model inference), these workloads commonly span multiple steps in different computing environments. They leverage and benefit from an increasingly wide range of computing resources, from commodity CPUs, to high-end GPUs, to specialized AI accelerators, and upcoming quantum devices.

Cloud computing technology has transformed how heterogenous computing resources can be used through multiple layers of abstraction, from programming models to platform and infrastructure. Cross-layer orchestration technology, driven by these abstractions, enables elasticity, fault-tolerance, and flexibility. Historically, however, the majority of distributed, large-scale applications and commonly used libraries have been written with programming models developed for a fixed, homogeneous architecture in a contained environment (e.g., MPI). While this fixed topology is efficient, it prevents the new class of emerging AI/ML applications to benefit from cloud-native computing's key underlying characteristics, such as elasticity, fault-tolerance, cost and resource efficiency, and portability.

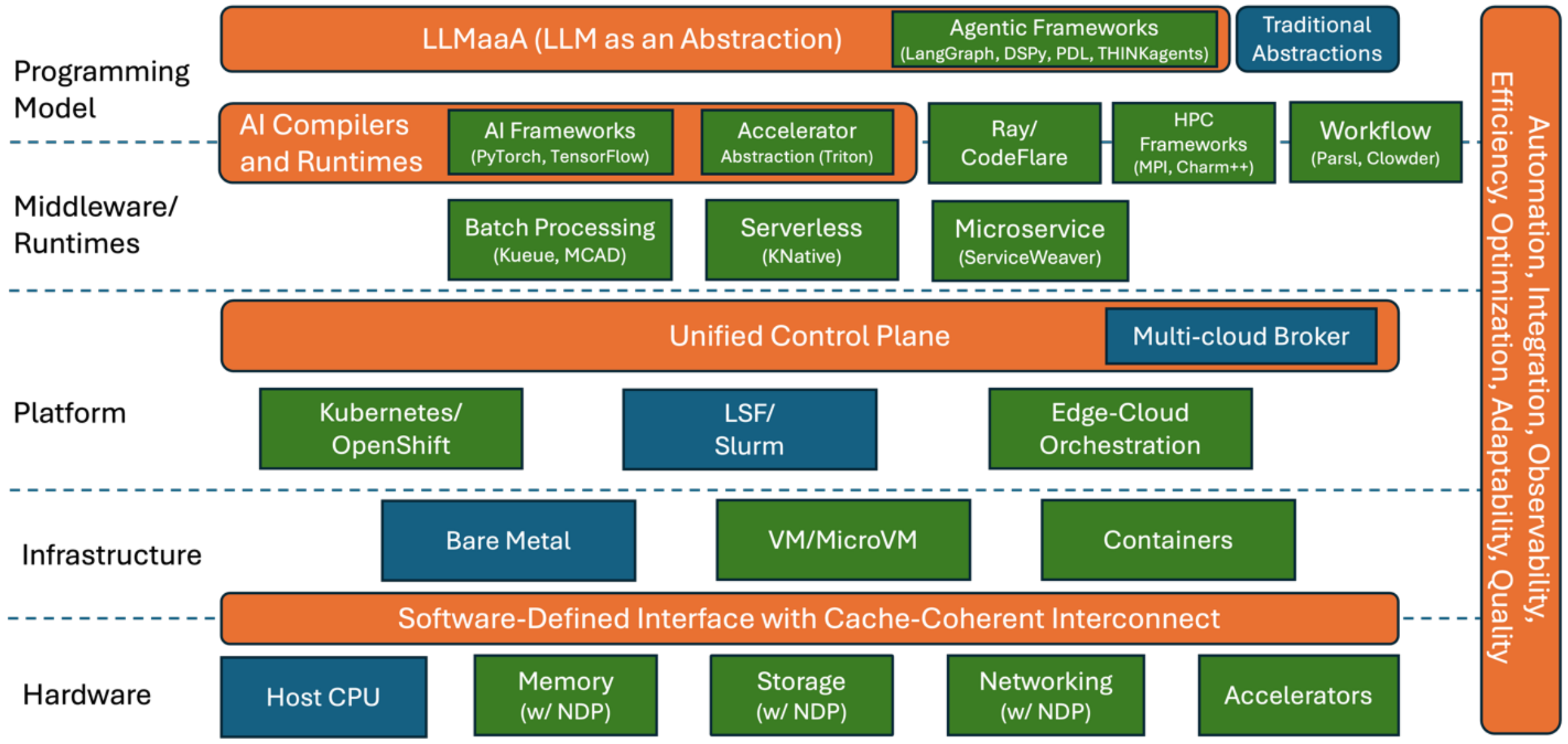

*Figure 6: Transforming the Hybrid Cloud for the future. Orange bars represent new features; green boxes represent enhanced components to work with the newly added features; blue boxes are existing components.*

Meanwhile, with the rapid growth of emerging AI/ML applications, existing cloud-native solutions are facing significant challenges to their relevance for these applications, especially in terms of consumability, affordability, adaptability, and complexity, as outlined in Section 3. Looking ahead, we envision the development of a next-generation hybrid cloud system designed to tackle these multi-dimensional challenges posed by the evolving workflows in AI/ML and AI-centric scientific computing domains. This future system must be capable of addressing the growing application demands while providing greater flexibility, robustness, efficiency, affordability, and sustainability.

As illustrated in Figure 6, this new vision spans across multiple layers of the system stack. First of all, we propose a groundbreaking computing paradigm that fundamentally transforms how humans interact with complex computer systems. At its core is the concept of *LLM as an Abstraction* (LLMaaA), which leverages advanced language models to create a natural, intuitive interface between humans and machines. This vision centers on an enhanced cloud-native system that integrates heterogeneous computing resources under the governance of LLMaaA. This new cloud system will incorporate AI-driven middleware/runtime for sophisticated workflow orchestration, provide dynamic resource allocation and robust error handling to ensure continuous operation, offer a unified interface and abstraction to diverse computing environments, from edge devices to powerful servers and accelerators, and enable seamless communication and coordination between different system layers, from hardware to application.

The key to realizing this future vision lies in the development of intelligent AI agents that leverage large language models, including a newly proposed framework called THINKagents. These agents will plan and execute complex workflows, debug issues across the system stack, generate comprehensive reports, observe system behavior and performance, control resource allocation and task scheduling, and communicate with each other and with human users.

To fully harness the power of LLMaaA, we must reimagine and re-engineer existing abstractions and tools. This involves implementing intelligent, vertical cross-layer automation flows to optimize resource utilization and workflow execution, developing mechanisms across different system components and layers for integration, efficiency, and adaptability, as well as creating comprehensive monitoring and analytics capabilities that span the entire system stack, thus providing accurate assessment and timely delivery of desired quality of results. Our aim is to ensure efficient scheduling, adaptive workload management and orchestration, flexible IaaS (Infrastructure as a Service) and PaaS (Platform as a Service), cross-layer robustness and security, and guaranteed SLOs (Service Level Objectives).

Horizontally, a unified control plane (Figure 6) is essential for managing heterogeneous computing resources across different environments, such as on-prem data center, private cloud, public cloud, and edge. As resource managers like Kubernetes and LSF/Slurm become specialized for distinct workloads, there is a need for AI-powered middleware/runtime to automatically orchestrate resources across these systems while working with different platforms seamlessly and efficiently, enabled by this unified control plane. A multi-cloud broker can decompose computing jobs and map them to the most appropriate resource managers across the edge-cloud system.

At the infrastructure and hardware level, emerging cache-coherent interconnects like Compute Express Link (CXL) and Ultra Accelerator Link (UAL) offer revolutionary potential for improving data transfer and enabling cooperative heterogeneous computing. These technologies provide cache-coherent host-to-device and device-to-host memory access, simplifying data movement with load/store semantics rather than complex DMA transfers. Coupled with software-defined interface between the infrastructure and hardware layers, future systems can integrate memory, storage, and network devices with near-data processing (NDP) capabilities, allowing fine-grained cross-device cooperation. This approach promises to dramatically reduce ML training and inference costs by optimizing resource use across devices based on specific computational requirements of individual ML model layers, offering unprecedented efficiency and flexibility with better affordability.

Overall, we desire to offer a more cohesive, forward-looking approach to address the challenges and opportunities in hybrid cloud environments for emerging AI/ML workloads and enhance and reinvent the latest hardware and software technologies, while ensuring the systems are scalable, highly performant, energy-efficient, and robust enough to handle increasingly complex workloads distributed across the edge and cloud systems. With this broad vision, we have identified the following important future research directions to drive the next generation of innovation and creativity.

*THINKagents: Advancing Collaboration and Intelligence in Agentic AI Systems.* A major research focus is improving AI agent collaboration through transactive memory, enhancing specialization and group intelligence. We propose THINKagents – a new agentic AI research framework. Leveraging this framework, future research should explore how AI agents collaborate like human collective intelligence, enabling them to avoid collective mistakes and achieve higher levels of intelligence. By leveraging memory systems, specialized tools, and planning mechanisms, future agentic AI systems should achieve better task decomposition, chaining, and self-improvement, offering new research pathways in cognitive and AI systems design.

*LLM-as-an-Abstraction (LLMaaA).* Our newly proposed system interface, called LLMaaA, is a novel paradigm for engaging with cloud computing and services in the future, with a natural language-based interface for building, deploying, and managing complex applications. The system uses a Master Agent – an LLM-based orchestrator that intelligently coordinates LLM and non-LLM agents (e.g., simulation agents, solvers) to perform tasks efficiently. This architecture ensures scalability, security, and continuous evolution by tracking agent performance. Future research should focus on enhancing agent collaboration, improving the adaptive programming model, and advancing secure, scalable cloud systems that flexibly integrate LLM and specialized agents for evolving real-world applications.

*AI Compilers and Runtimes.* Scaling foundation models to handle larger context lengths is crucial for advancing AI applications in areas like NLP, climate prediction, and geospatial data. Achieving this requires innovations in neural architectures, AI frameworks, and compiler optimizations. Additionally, sparse machine-learning models play a pivotal role in predictive analytics for fields like drug and material discovery and quantum chemistry, but are challenging to scale due to irregular data. Future research should focus on compiler, framework-level optimizations, common accelerator abstractions and higher-level kernels (e.g., Triton) as well as hardware innovations with a focus on cross-stack co-design to unlock synergies that will enable efficient scaling and optimization of emerging AI models, thus maximizing their impact on real-world applications.

*Adaptive Middleware and Runtime in Hybrid Cloud Systems.* Future research should focus on developing adaptive and intelligent middleware and runtime solutions to optimize the interplay between compute and communication across distributed AI workloads. The goal is to make future hybrid cloud dynamically adjust to real-time workload demands, evolving system topologies, and edge-cloud coordination, through AI-driven workload management prioritizing efficiency, scalability, and fault tolerance. One important direction is to design a unified control plane for AI-powered management of the resources in heterogeneous and dynamic cloud environments.

*Cross-Layer Automation and Integration.* To optimize cloud infrastructure for complex computations, cross-layer automation, integration, and observability are essential. Automation across layers of the cloud stack enables efficient resource allocation, scheduling, and SLO optimization, ensuring minimal delays and maximum availability. Cross-layer observability provides crucial performance monitoring, helping identify bottlenecks and automate diagnostics. Future research should focus on developing automated frameworks for seamless orchestration and monitoring across cloud layers, which will improve efficiency, flexibility, robustness, adaptability, and scalability in next-generation cloud systems.

*Unified, Programmable, and AI-Optimized Networking for Distributed AI Workloads.* Future research should focus on designing unified, reconfigurable, and AI-optimized networking infrastructures that facilitate smooth and efficient data transfer across hybrid cloud environments, eliminating inefficiencies in inter-node and intra-node communications. One important direction is to develop AI-driven network orchestration schemes that intelligently allocate bandwidth and resources, minimizing latency and cost while maximizing throughput for AI frameworks like PyTorch and TensorFlow, and ensuring secure data flow across diverse networking protocols (e.g., RoCE v2, InfiniBand).

*Data Management and Storage Efficiency for Large Models.* Future large-scale AI training will rely on techniques to automatically place and migrate data in real-time, ensuring efficient storage use while reducing bottlenecks – not just within a single cloud, but also across edge devices, private clouds, and public clouds. Additionally, new AI-enhanced security frameworks should ensure data integrity and encryption during data transfers across these environments. Future research should focus on innovating data management systems that intelligently distribute and manage a large amount of data (e.g., those used by foundation models and LLMs) across AI-Accelerator/CPU memory, SSDs, and node-local storage in hybrid cloud environments securely and efficiently.

*Advancing AI Systems through Co-Design and Emerging Hardware Innovations.* In order to dramatically enhance AI system performance and energy efficiency, leveraging system co-design and emerging hardware technologies such as Compute Express Link (CXL), Advanced Matrix Extensions (AMX), and GPUDirect will be essential. These technologies will streamline data transfer, optimize memory access, and facilitate cooperative heterogeneous computing. Future research should focus on novel data compression, memory management, and orchestration techniques for efficient AI/ML training and inference with lower cost. Software-defined interface and cache-coherent interconnects need to be co-designed to improve fine-grained computational efficiency across AI workloads.

*End-to-end Edge-Cloud Transformation and Optimization.* The future of AI system design lies in flexible SLOs, optimizing the balance between edge and cloud computation, and fine-grained, composable acceleration. Emerging applications like extended reality and robotics require a better understanding of trade-offs between latency, accuracy, and power. The design of fine-grained accelerators will boost energy efficiency without the recurring cost of monolithic designs. Research should focus on unified programming models, specialized data communication methods, co-design of neural architectures and accelerators, and integrated offline and online optimization techniques. End-to-end system prototypes and benchmarking are essential to validate these ideas, driving the future of AI systems with advanced flexibility, functionality, performance, and energy efficiency.

*Robustness, Security, and System Health Monitoring for AI in Hybrid Cloud.* Future cloud systems should include advanced fault detection, intrusion detection and containment, and self-healing algorithms and mechanisms to ensure their long-term health, resiliency, and dependability. Future research should focus on developing AI-driven robustness and security solutions for AI models running in hybrid cloud environments. Additionally, robust security protocols must be integrated into the AI orchestration layer, leveraging AI-enhanced intrusion detection, encryption, and access control techniques to prevent data breaches and system attacks. These mechanisms should be designed to allow systems to adapt to evolving security threats and ensure the integrity of distributed AI workloads.

*Energy-Efficient AI Workloads and Sustainability in Hybrid Cloud Systems.* Next-generation novel approaches will allow AI frameworks (e.g., Triton, PyTorch, Ray/CodeFlare) to make energy-aware decisions during model training and inferencing, contributing to carbon efficiency while maintaining high performance. Future research should focus on developing AI-driven energy management systems that dynamically optimize energy consumption across hybrid multi-cloud systems powered by diversified energy sources, including renewable sources, while working with various AI frameworks. This includes creating adaptive techniques to balance energy and performance, efficient compute/memory/storage management, and smart workload distribution between edge devices and multi-cloud resources.

*Adaptive and Reconfigurable Cloud Infrastructures for AI Workloads.* One exciting future direction is to enhance reconfigurability and adaptability in hybrid cloud system through technologies such as programmable SmartNiCs and in-network switches, reconfigurable hardware and interconnects, software-defined programmable interface, and workload-adaptive control strategies. Coordinated reconfiguration and specialization tailored towards specific workload characteristics will enable clouds to achieve significant performance gains (e.g., up to 100x). These innovations will allow smart and flexible adaptation to various workloads, from lightweight services to large-scale AI tasks, positioning future clouds as highly efficient, affordable, dynamic platforms.

The following sections elaborate on these exciting research directions in more detail. Section 4.2 introduces the key concepts of our proposed agentic system, called THINKagents. Section 4.3 introduces LLM as an Abstraction (LLMaaA). Section 4.4 discusses various AI model optimization techniques. In Section 4.5, we present the newly envisioned programming model, middleware, and platform, and then Section 4.6 covers the underlying infrastructure and hardware. Section 4.7 delivers our vision for an end-to-end edge-cloud transformative approach. Section 4.8 delves into critical system-level optimization tasks, including robustness, dependability, and security (Section 4.8.1), as well as energy optimization and sustainability (Section 4.8.2). Lastly, Section 4.9 explores opportunities for application-adaptive system architecture designs aiming for high efficiency and performance for the hybrid cloud system.

## 4.2 THINKagents: A Research Framework for Agentic Systems

### 4.2.1 Introduction

Agentic AI is a class of AI systems that are designed to act autonomously – as agents – to perform general-purpose work like completing a task end-to-end using planning and software tool-calling skills. Building on base AI technologies such as LLMs or reinforcement learning, these larger systems are constructed so they can make decisions, interact with their environments without constant human vigilance, and even interact in a multi-agent manner to learn from each other, cooperate, perhaps even developing their own language. As an example, AI agents have been created to model each role on a software development team and collected into a functioning multi-agent structure. These agents then collaborate using standardized operating procedures and a shared memory to complete complex tasks, to produce software engineering artifacts such as project requirements, design documents, and functional code. As another example, AI agents can be used to run large-scale simulations in a game-theoretic formulation to address notoriously difficult policy design questions such as design of dynamic tax codes, regulations to protect information environments, or optimize supply chains to ensure reliability and resilience. Going forward, one might imagine automating cybersecurity, where agents would scan for vulnerabilities, patch them, and further develop/test possible solutions in controlled environments. There is strong interest going forward in automated design of agentic systems (ADAS) where one combines building blocks to automate design.

Broadly, agentic workflows empower generative AI models to tackle more complex, real-world problems by providing them with a structured approach, increased autonomy, and the ability to learn and adapt. As AI continues to advance, agentic workflows will likely play a crucial role in unlocking the full potential of generative AI technologies. Indeed, as compared to current foundation generative AI that often requires step-by-step human guidance, AI-enabled agents may take a prompt, break the goal down into subtasks, take action, check work, and adapt the approach as needed. Recent research on software engineering tasks, such as resolving Github issues [8-9], has shown state-of-the-art retrieval augmented generation (RAG) algorithms [17] with modern foundation models, e.g. GPT-4, perform much worse than agentic workflows with the same underlying LLM.

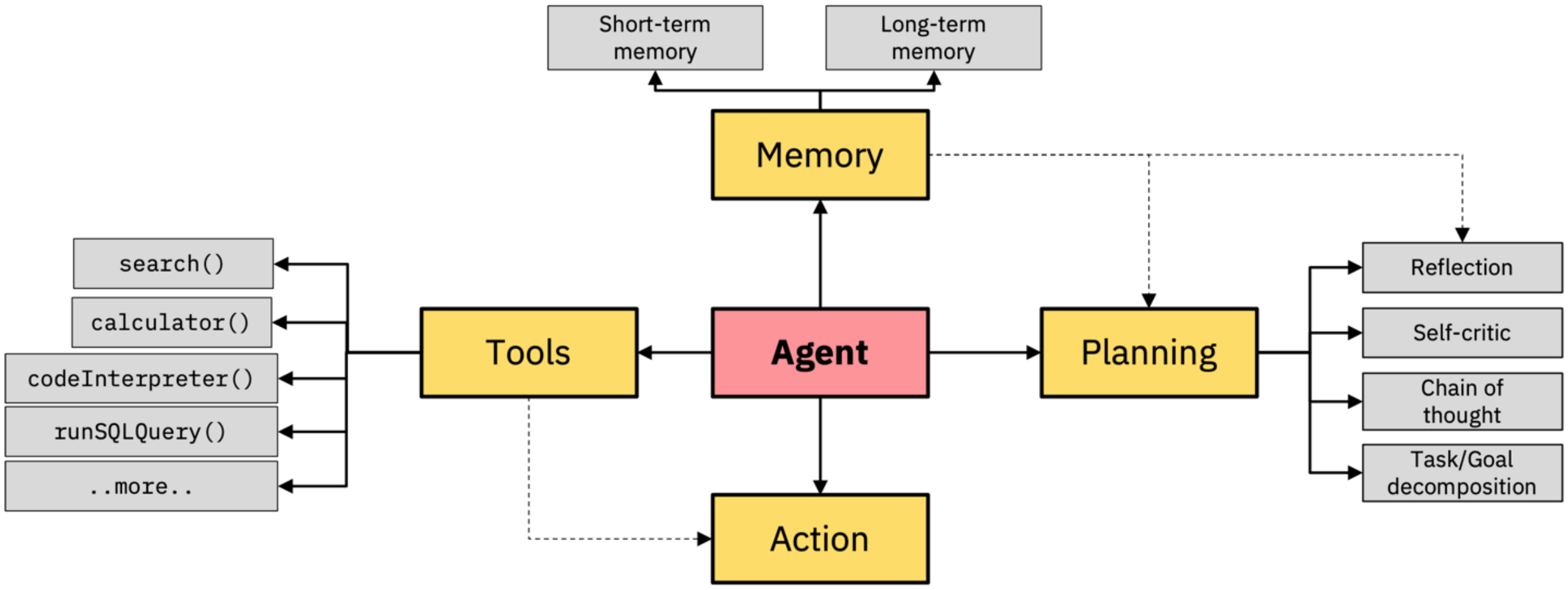

*Figure 7: THINKagents, an agentic AI research framework that mirrors the structure of an agentic AI system [This figure was reproduced from a blog post by Lilian Weng: https://lilianweng.github.io/posts/2023-06-23-agent/].*

To advance agentic AI workflows for tackling real world problems, there is a need for agentic AI research infrastructure, as shown in Figure 7. Such an infrastructure would enable addressing numerous research questions: an agentic AI research framework would have different components, corresponding to research questions of interest, and how these components are connected in the agentic infrastructure would itself provide insights. In a way inspired by cognitive science, some of the key components of the THINKagents framework is to have short-term and long-term memory as well as the use of a variety of tools that are engineered for specific purposes. To bring these capabilities together, there are forms of planning that enable decomposition and well-planned chaining of tasks, together with reflection and self-criticism so as to correct and improve.

#### 4.2.2 Broad Research Questions

An agentic AI system must be highly reliable to successfully achieve complex goals and take actions in complex environments. Consider a travel-planning agent that must coordinate calendars, find destinations, check the weather, book flights and hotels, and plan activities. Even a small mistake such as overlooking time zone changes when booking connecting flights could render the whole trip untenable, and there may be 40 or 50 steps over which even small individual error probabilities might compound. Now imagine the same for programming tasks that may require thousands of lines of code, perhaps in several modules: even perfection in almost all individual steps might yield faulty outputs. Current agentic AI systems are good only at simple tasks such as those that may take people a few minutes, but struggle significantly for complex tasks that may take a person a few hours. To move beyond narrow and specialized tasks such as software development with testable solutions or math problems with provable answers, high reliability is needed.

How might one make agentic systems perform better planning and reasoning? Further, how can one develop agentic foundation models that are capable of making decisions and interacting with the world to solve long-horizon tasks through extensive interactions? Agentic foundation models have achieved great success in recent years and are widely recognized as a promising way forward, by enabling the foundation models to frequently interact with the world, improve with outside feedback, keep growing with self-reflection, and eventually address tasks that require many steps to accomplish. While there are already successful examples such as OpenAI's o1 model, the mechanism for how to build such models is still largely unknown to the foundation model community.

One may further wonder how one can bridge the gap between state-of-the-art and open-source models for agentic tasks. In most agentic tasks, e.g. Github issue resolution [8] or answering financial questions, we have seen frontier models, e.g., GPT-4, performing significantly better than open source LLMs, e.g., Llama 3.1 405B. As such, it is important to determine whether there are effective training mechanisms for open-source models for agentic tasks, so as to determine whether one should investigate better training objectives or better neural architectures. Moreover, one might wonder whether one can generate feedback-based training/instruction-tuning data (feedback can come from evaluation or interpretation of the agents, and can be adjusted to their abilities) that improves training. Indeed, it is of interest to determine what extent such feedback would improve the performance of open-source models compared to frontier models. All such studies can be carried out in the THINKagents framework.

Another application of agentic AI systems is in the area of regulatory compliance to empower organizations to navigate regulatory complexities and mitigate risks more effectively. Compliance agents not only enhance operational efficiency and accuracy but also expose agentic collaborative capabilities to foster the leverage of other agents' skills, for instance to outsource non-compliance resolution to SRE (Site Reliability Engineering) agents for incident remediation or to GRC (Governance, Risk and Compliance) agents for exception handling and audit. The focus on agentic AI for regulatory compliance is motivated by the extremely high pace of regulatory compliance programs development. Our approach to compliance Agentic AI brings together the latest technology on compliance as code, that enables programmatic expression of how regulatory controls are satisfied, with LLM-assisted code generation. Our aim is to empower a compliance team in accelerating the adoption and operation of new regulatory programs by automating the generation of code for the evidence collection and for its posture validation against the requirements, based on the compliance rules and RAG technology templates. We aim to demonstrate the end-to-end ecosystem from Compliance as Code to automatic generation of Policy as Code and its real-time deployment and execution to generate compliance posture, handle non-compliance resolution, and support adaptable audit reporting.

In addition to acting in real world and software development settings, agentic AI systems can be used within complex world simulations to design policies to meet a variety of objectives. One example is wargaming and multilateral decision-making in international relations, where agents can be role-prompted to act like certain countries or bureaucracies, and various scenarios played out. This can be done purely with LLM-based agents, but another possible approach is to combine LLM techniques with reinforcement learning techniques within agents: such an approach would allow the combination of qualitative and quantitative role prompting, based on notions of value-driven rationality together with language-driven idiosyncrasies. One might develop multilateral regulatory regimes by considering not just agents for each country, but also a special governance agent that sets the rules of play in a two-level Stackelberg game setting, so that the policies of the governance agent emerge as a dynamic regulatory regime. The same basic approach only considering reinforcement learning agents, has been used to design taxation policies that balance productivity and equity in a manner that far exceeds static analytical solutions. Going forward, an especially compelling setting where agentic AI built into agent-based models of the world may be impactful is in policies for securing supply chains, e.g. for food or semiconductors. Again, agentic AI systems may find equilibrium problem solving policies that yield efficient and resilient supply chains. There are numerous related embodied agentic AI problems that can be considered.

### 4.2.3 Initial Results from IBM-Illinois

One promising direction that has achieved initial success in existing works is to build agentic foundation models with Monte Carlo Tree Search (MCTS). More specifically, one may implement a tree-based search mechanism that allows the foundation model as an agent to explore different branches of its decision, select the most promising actions by comparison, and correcting its perception with feedback from the outside world. Such a mechanism has been applied on several prior works, such as RAP [31], LATS [37], and LLM-MCTS [32], and has shown success in improving the model's reasoning, acting, and planning ability.

Figure 8 demonstrates the performance of LATS. We believe such a method, when combined with existing LLM techniques such as reinforcement learning with human feedback (RLHF), can be quite promising for building stronger agentic foundation models. To thoroughly evaluate such a proposed solution, one may test methods on a variety of benchmarks such as NaturalPlan [33] and VisualAgentBench [34], and build novel environments if necessary.

Code-empowered LLMs enable better reasoning and can perform more complex tasks [10-12]. They take the task instruction as a prompt in natural language and synthesize an executable program as a response/action. The performance of code-empowered LLMs and agents relies on the quality of the generated code. However, several studies demonstrate that LLM-generated code can often be semantically incorrect, especially when the instruction is long and complex (the LLM may generate a syntactically correct code that does not pass the tests) [13-14]. Iterative code improvement with textual feedback from test execution or user feedback is a technique to alleviate the issue. However, this feedback is returned to the model "after" the execution of the entire code, making it hard for the model to pinpoint and resolve the root cause of semantic correctness violation. More importantly, in all existing Code LLMs, the rationale for a response/action is explained in natural language, making it hard to validate the reasoning or pinpoint hallucinations. Given that such feedback is specifically important for Software Engineering (SWE) agents [25], one direction is to enable LLMs to better understand and reason about code execution [15-16]. Such models understand control and data flow and hence, perform programming tasks that depend on them better. As a first step in this direction, one may investigate techniques to analyze code execution reasoning abilities of the models, and based on the insight from that analysis, we aim to instruction-tune models that can better simulate code execution and use the results promptly in solving programming tasks. These techniques are especially useful when agents are used to perform the tasks proposed in the LLMaaA section below (Section 4.3).

Several current applications of agentic AI are in AI4Code. SWE tasks are inherently connected, and performing them separately can in fact make individual tasks harder. For example, to fix bugs, one needs to first show the existence of the bug in the program, and then pinpoint/localize the statements culprit of manifesting the bug. Without proper bug localization, fixing the bug is hard, if not impossible [19-22], specifically for LLMs that usually do not consider the intra- and inter-procedural dependencies in programs [18]. The synergy between agents in the agentic SWE framework can help overcome this limitation. We first may identify the tasks that can benefit from this unity in agentic frameworks, e.g., the oracle problem in test generation [23], bug localization and repair [24-25], and code translation [13], and then explore how to unify agents so that their collaborations overcome the limitations of performing SWE tasks separately. These research tasks are also essential for enabling the LLMaaA vision.

While LLMs, and specifically agents, have shown emerging abilities in programming tasks, evaluating their abilities is still in its infancy. Except SWE-Bench [8], which is a collection of real-world projects and is used for program repair task, researchers still use HumanEval and similar datasets for every other code-related tasks. Such datasets do not reflect complexities of the real-world, and using them for evaluation can mislead the state-of-the-art and state-of-the-practice concerning true abilities of the models [26]. As a result, we explore automated techniques for construing benchmarks to better evaluate Code LLMs, in general, and agentic programming frameworks, specifically. Our initial results shown in Figure 9 demonstrate the value of agentic approaches.

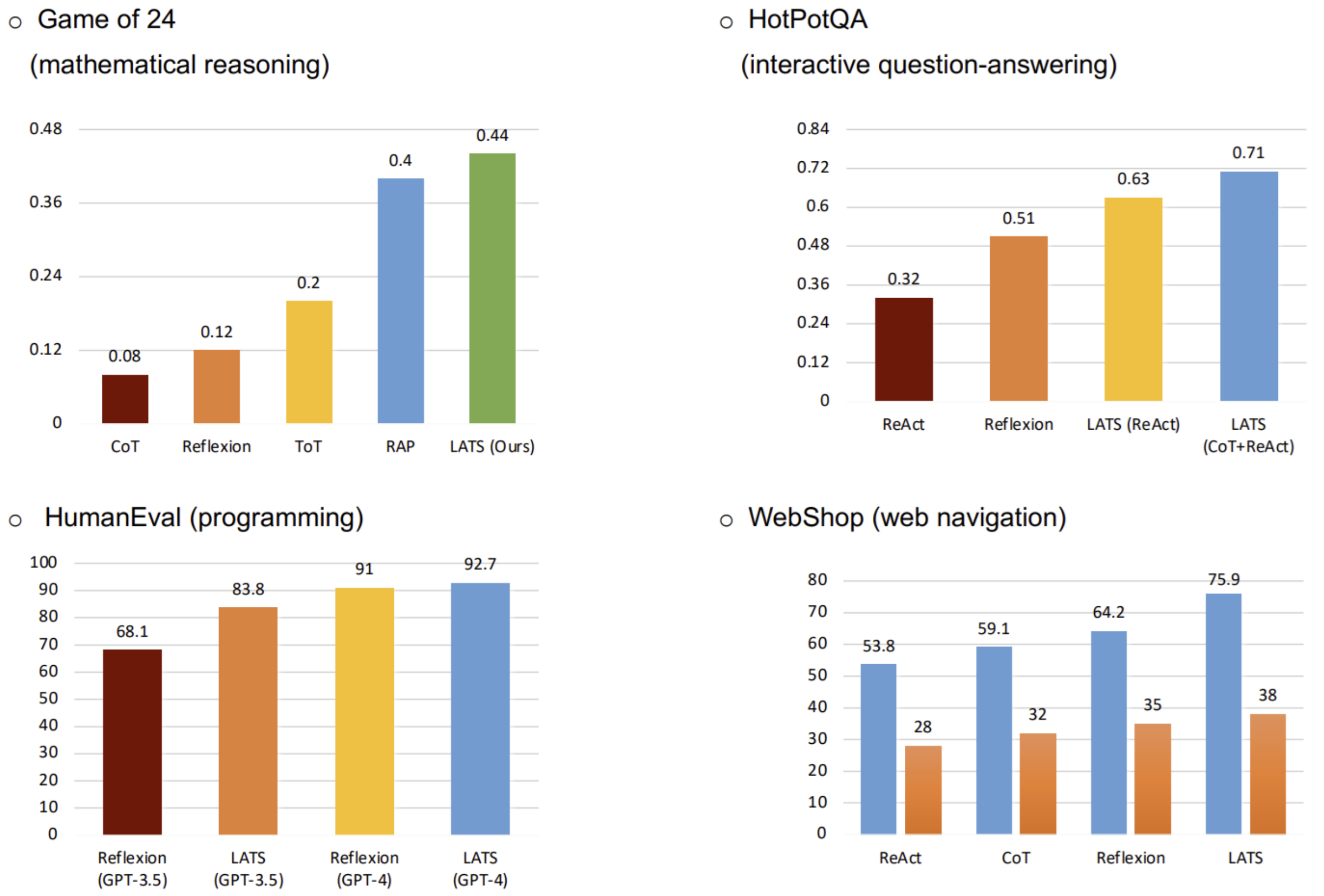

*Figure 8: Language agent tree search (LATS) improves agentic performance over simpler prompt programming techniques, on a variety of tasks including mathematical reasoning, programming, web navigation, and interactive question-answering.*

### 4.2.4 Long-term Direction and Vision

Following the common task framework, there are several existing datasets that may be used to evaluate the performance of agentic AI systems, especially in the area of software engineering. We may consider using existing benchmarks, e.g., HumanEval [28] and its extensions such as MultiPL-E [27] and HumanEvalPack [36], SWE-Bench [8], ClassEval [18], CodeNet [29], and R2E [30]. Going forward, however, it is very important to construct new benchmark datasets in a variety of application areas. These will provide methods of evaluation. As an example, there is a need for open benchmark tasks and datasets in the area of IT automation. This benchmark would be designed to evaluate how well analytical solutions can handle real-world incident management challenges, particularly by resolving real faults that may manifest in IT environments. More broadly, to drive progress, it is important to release open-source code and synthetic data on platforms such as Github, Zenodo and HuggingFace to help the advancement of the academic community on agentic foundation model research. Indeed, researchers at IBM and Illinois have a track record of open-sourcing their research artifacts and datasets, and obtaining official ACM artifact evaluation badges [35].

Moreover, it is of interest to extend systems-theoretic concepts from information theory, control theory, game theory, and statistical learning theory to understanding the fundamental limits of problem solving that any system could do, so as to provide absolute scales against which to compare. We believe existing challenge tasks, further challenge tasks that may be developed, and corresponding fundamental limits will push the field forward and lead to new ideas in how to orchestrate agentic AI systems to accomplish complex tasks.  Beyond software development and IT operations, another key direction that will drive progress is embodied agentic AI, where notably safety is of utmost importance not just in supply chain optimization but in all kinds of settings where there are interactions with the physical world.  Physical embodiment also introduces novel information sources and novel constraints that my help ground agentic systems and advance numerous dimensions, whether tool use, memory use, or planning.

Of particular importance is improving the ability of AI agents to delegate to other AI agents that may be more skilled at a certain task than themselves. This requires endowing AI agents with transactive memory capability, which seems to thus far be absent.  If this capability can be developed, it will lead to gains from even more hyperspecialization of AI agents and tools in context, just like efficiencies in the labor market. As there are more agents involved in working together, it will be important to understand what kinds of communication and collaboration patterns yield collective intelligence rather than collective stupidity.  In studies of human collective intelligence, patterns such as turn-taking and human abilities in theory of mind and emotional intelligence are key for collective intelligence, often even more than the individual intelligence of the group members.  It is important to understand the analogs in agentic AI systems, so these dimensions can be enhanced.

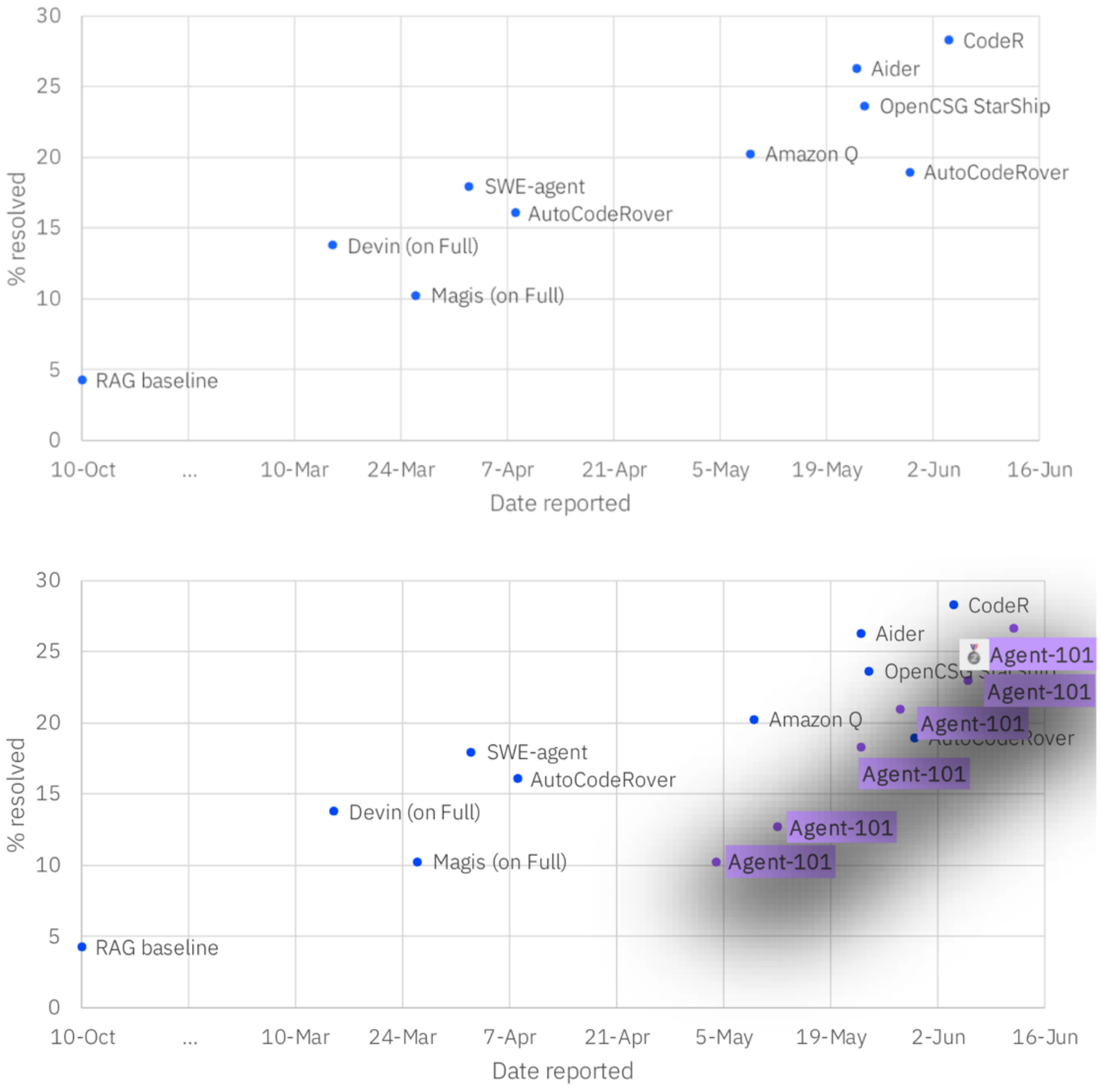

*Figure 9: Agentic techniques demonstrate superior performance (and are improving over time) for real-world software engineering (SWE) tasks.*

### 4.3 LLM as an Abstraction (LLMaaA)

#### 4.3.1 Introduction and Challenges

The recent emergence of Large Language Models (LLMs) has revolutionized AI-driven language processing, transforming how we interact with technology and information. Over the past few years, trends such as the scaling of foundation models, the rise of LLM as a Service (LLMaaS), and the fine-tuning of models for specific tasks have made LLMs more powerful, accessible, and widely applicable in fields such as code synthesis [38], design verification [39], hardware design [40], social networks [41], cloud system operations [42], and scientific workflows [43]. With LLMs playing the pivotal role, these developments are driving innovations across diverse domains in both academia and industry.

This rapidly growing field still faces significant challenges in automation, flexibility, reliability, reconfigurability, and evolvability. Specifically, existing LLMs often struggle to adapt to task-specific requirements without human intervention, necessitating prompt engineering to elicit precise and high-quality responses. Moreover, they typically require finetuning on domain-specific datasets to acquire expert-level knowledge, which can be time consuming and labor intensive. Furthermore, LLMs are prone to hallucinations, generating factually inaccurate outputs that compromise their reliability. Their lack of reconfigurability makes it difficult to adapt them to different tasks without extensive engineering efforts. Finally, they often fail to demonstrate evolvability, as they do not automatically learn from new data in an on-line fashion and may struggle to keep pace with rapidly evolving environments, requirements, and developments. Some of these issues have been discussed in the previous section.

#### 4.3.2 Long-term Vision

To address these challenges, we envision a future where LLMs not only become more capable and reliable themselves but also become building blocks for establishing a brand-new abstraction equipped with user-friendly application interfaces. This paradigm shift will enable a new class of applications that can adapt to changing computational demands in real-time, efficiently utilize diverse computing resources, and provide intuitive, natural language interfaces for complex tasks. By abstracting away the complexities of traditional programming and system management, LLMaaA will democratize access to advanced computing capabilities, fostering innovation across industries and disciplines. In addition, such an abstraction can also create a common framework that allows applications to be deployed across different cloud platforms. Our vision is characterized by the following powerful and novel features:

- LLM as an abstraction (LLMaaA) is a new concept that leverages advanced cloud application platforms that host LLM and non-LLM agents, provides user-friendly, natural language-based, secure, scalable, and evolvable services for building, deploying, and managing complex user applications. LLMaaA offers intuitive interfaces for interacting with users (a comprehensive term that can include developers too) and contains an LLM-based Master Agent, which works and coordinates with other LLM-based worker agents and specialized non-LLM agents (e.g., simulation agent, ILP solver agent, etc.). The Master Agent will intelligently select the right agent(s) to perform the desirable user tasks while tracking each agent's service quality and updating the agents as needed. Security is ensured with measures such as data encryption, access controls, and regular vulnerability scanning. LLMaaA can enable the continuous evolution of both the agents and the programming model used to integrate different agents within the cloud platform. Please refer to Figure 10 for the overall design of LLMaaA. More details are introduced below.
- Interface Agents are human-friendly and can carry out iterative conversations with users. Each Interface Agent offers automated prompt engineering that either helps a user to clearly define problem statements, goals, and objectives for various tasks or directly generates high-quality prompts that can be verified by the user. Input and output formats are standardized according to specific task API requirements. Interface Agents also engage with the Master Agent: pass user requests to the Master Agent and channel intelligent suggestions back to the user.
- LLM-based Master Agent communicates with the Interface Agents with a language that can precisely capture the user tasks. It interacts with various surrounding agents, selects the right agents (customizing them, when necessary, e.g., through fine-tuning), and carries out inter-agent coordination in order to complete specific user tasks aiming for multiple targets, including correctness, efficiency, robustness, and security. The Master Agent will monitor the service quality of individual agents and the overall application, offering optimization suggestions to improve result quality, increase problem-solving capability, and reduce solution generation latency. The Master Agent can also help refine problem statements and assist in constructing and testing task flows composed of various agents while working with multiple users simultaneously (Figure 10 illustrates task flows for different users).

- A heterogenous agent pool enables the capability of combining multiple LLM and non-LLM agents to tackle complex problems or tasks. Notably, non-LLM agents can be leveraged to perform high-precision simulations, execute complicated mathematical calculations, and run reinforcement learning (RL) solvers (e.g., AlphaZero for board games, and GNN+RL for graph combinatorial optimizations), just to name a few. The Master Agent, working with the agent pool, will coordinate the interactions of the selected agents from the pool, verify responses from these agents, and ensure the accuracy and reliability of the entire task flow. One important novel feature is the plug-and-play nature of these agents which is enabled by the agents conforming to a standardized communication interface defined by LLMaaA. This feature enables seamless integration of suitable agents into the task flow as well as the removal of individual agents on demand. By harnessing the strengths of different agents, we create a more adaptive, reliable, robust, and flexible application ecosystem.
- Our vision is to build and create all these aforementioned agentic features in the THINKagents framework (Section 4.2) so THINKagents becomes the enabler of LLMaaA, and LLMaaA becomes a highly efficient new user interface for future cloud applications.

LLMaaA can be realized as pods running on large-scale Kubernetes (K8s) CPU/GPU clusters: each LLM/non-LLM agent instance is packed inside a pod with CPU, RAM, GPU, and storage (LLMaaA can accommodate dedicated AI chips as well, such as IBM AIU). These agents talk to each other through inter-pod APIs. While each pod processes only one task request (from a user or from another agent), multiple pods (each with specific user customization settings) can process many requests in parallel. A nice feature of this design strategy is that each agent type can be handled by a K8s deployment where multiple replicated pods will auto-scale based on number of incoming requests.

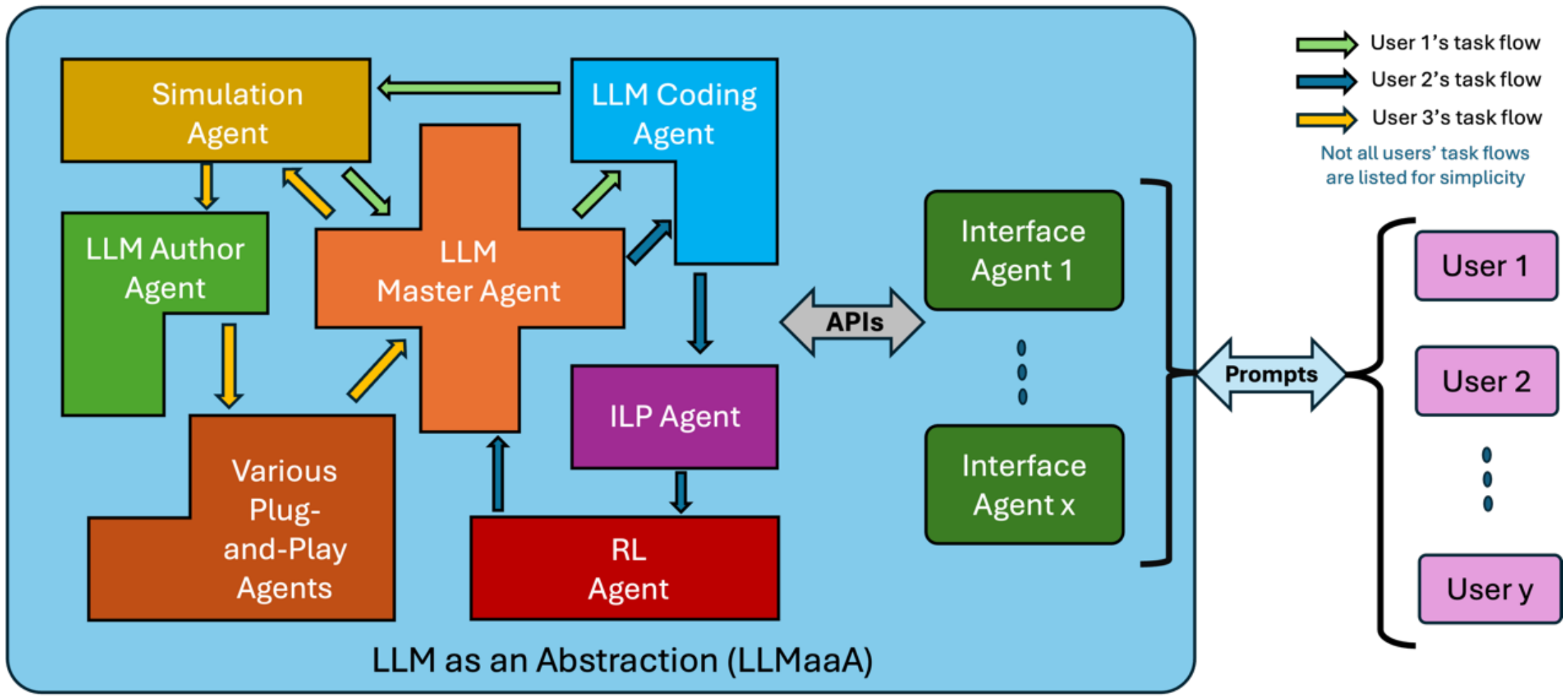

*Figure 10: The Concept Design of LLM as a New Abstraction.*

With LLMaaA, a new paradigm emerges that significantly enhances problem-solving capabilities, streamlines operational and design processes, improves application efficiency, and boosts development productivity. For instance, in October 2029, a UIUC graduate student majoring in astrophysics needs to run a black hole simulation on NCSA's Delta system. The student has the necessary simulation scripts on a Windows laptop but has no experience with Slurm, K8s, or Linux bash scripting. Instead of spending hours or days learning Slurm and navigating Delta's official user guide for job script formatting, the student simply requests Delta's LLMaaA to convert the Windows .bat simulation scripts into the appropriate .Slurm job submission format, customized according to Delta's guidelines. The problem is solved in <10 minutes.

### 4.4 AI Model Optimization across System Stack

#### 4.4.1 Introduction

This section focuses on techniques aimed at enhancing the efficiency of AI models themselves, thereby improving their performance, scalability, adaptability, and reducing their operating costs. Deep neural networks have become an important workload that many systems, programming language, compiler and hardware architecture researchers are optimizing for. IBM has been pioneering work on foundation models for different domains achieving significant productivity boosts. We foresee novel neural architecture designs and training methodologies to evolve as foundation models grow in popularity and artificial intelligence (AI) will be pivotal in enabling new frontiers in future scientific discovery. We identify three main research directions that are paramount to advancing the body of knowledge in this area.

First, among different neural network architectures, transformer models and its variants have become exceedingly important due to the popularity of large language models such as GPT-4, LLaMA, Claude etc. Therefore, optimizing these architectures for both accuracy and runtime performance have become an important topic. Second, we notice that sparsity is becoming ever more important to both encode information (e.g. knowledge graphs) as well as to scale existing machine learning models (e.g. sparse attention). However, getting good performance across multiple sparse inputs has remained challenging and we believe innovative optimization techniques are needed to bridge this gap. Third, we notice an explosion of novel hardware architectures proposed to accelerate deep neural network models. Examples include Google's Tensor Processing Units, Amazon's Inferentia, IBM's AIU Spyre and numerous other academic proposals. Providing compilation and programming support for such architectures are paramount to get wide scale adoption.

#### 4.4.2 Research Challenges and Opportunities

##### *4.4.2.1 Scaling and optimizing large language models*

Researching models that can handle long context lengths, particularly beyond 1 million tokens, presents significant computational and algorithmic challenges. Traditional transformers struggle with scaling due to the quadratic complexity of the self-attention mechanism, which limits their ability to manage such extensive sequences efficiently. Addressing this requires novel architectures or optimization strategies that reduce the computational overhead while maintaining or improving the model's performance on long-context tasks. We expect two categories of innovations to address the above challenges.

Novel model architectures. There has been work on improved model architectures such as SSMs [44], memory-augmented neural networks, linear attention mechanisms that reduce the computational cost of LLMs, while preserving the accuracy. Selective state space models such as Mamba [45] have demonstrated promising results in sequence modeling with linear attention complexity, significantly reducing the memory footprint compared to Transformers. These architectures eliminate the quadratic complexity of self-attention and thereby allowing large context length models. However, a key challenge is to preserve accuracy at large context lengths. Mamba can struggle with extremely long context tasks, because it processes sequences in a recurrent manner, leading to difficulties in maintaining information over very long sequences. We believe there is more innovation to be done here, specially on hybrid architectures that combine traditional transformers with their approximate versions. Combining both model architectures, such as a mixture of Transformer layers and Mamba layers, can help mitigate some of these limitations by leveraging the strengths of both models. This hybrid approach can improve performance in tasks that require both long-term context and efficient sequential processing. The goal of the hybrid model architecture is to obtain the best of both worlds of different architectures while maintaining the training stability and convergence speed. Therefore, it requires careful design to ensure that the components work seamlessly together, which involves significant engineering and experimentation efforts.

Systems-level scaling. While hybrid architectures can optimize memory usage, they may introduce additional computational overhead because of the complexity of integrating multiple components. Novel hybrid parallelism strategies are needed to accommodate different components in the hybrid model. Long context evaluation itself remains a big challenge both due to limited task coverage (e.g., primarily retrieval capability) and high inference computational cost. Therefore, it is also important to come up with advanced distribution techniques to accelerate the long context inference and evaluation.

#### *4.4.2.2 Scaling and optimizing sparse machine learning models*

Optimizations and programming support for sparse-attention models. There is relatively little work on optimizing or supporting programming interfaces for sparse attention mechanisms [46] and they are left underexplored in the compiler community. We believe there is room to innovate powerful optimization techniques for these models. OpenAI's Triton implementation of Sparse attention is an example of what we can achieve if we hard-code the optimizations. However, such approaches are brittle and do not generalize to newer attention patterns (Figure 11). Therefore, it is important to build user abstractions, compiler representations and optimizations that can cover a wide range of approximate attention patterns.

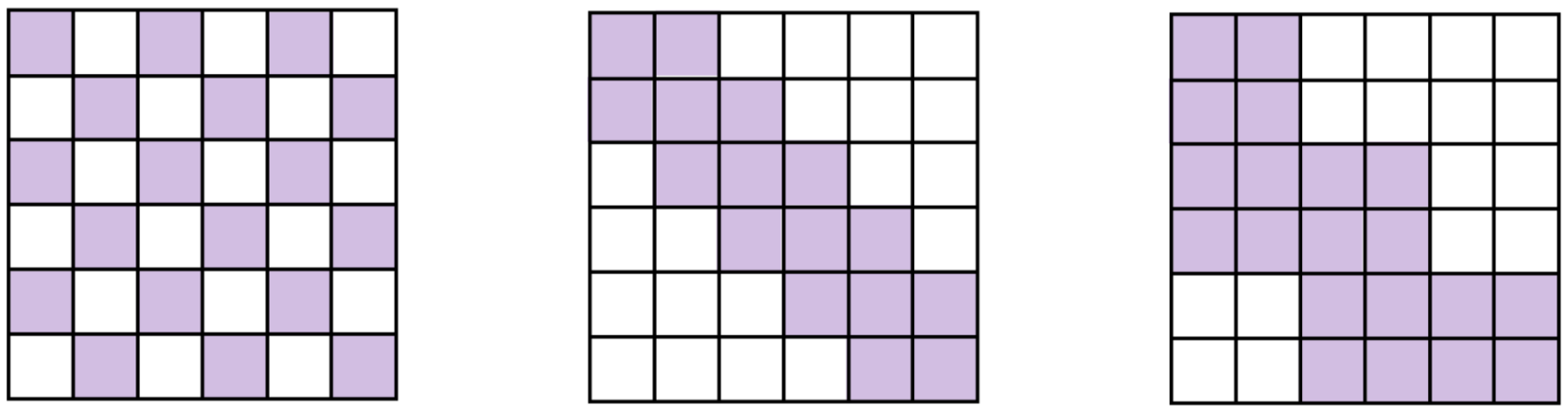

*Figure 11: Different sparse attention patterns used in linear attention mechanisms.*

Optimizations and programming support for graph neural networks. Graph neural networks (GNN) have become popular in fields such as drug discovery, financial services, and knowledge mining. Multiple variants of GNNs exist for each task ranging from static, temporal and streaming GNN variants. Getting performance across diverse graph inputs, embedding sizes as well as different GNN variants have remained challenging. We believe innovation in programming models, adaptive optimization techniques as well as novel distributed learning mechanisms are needed to enable better programmability and large- scale training of graph machine learning models.

#### *4.4.2.3 Programming and compiling to accelerator platforms*

Most ML models are compiled for hardware accelerators such as GPUs, TPUs, and custom ASICs. However, optimizing workloads across diverse hardware architectures introduces challenges in ensuring performance portability and maintaining efficiency. While frameworks like Triton and Pallas are capable of providing hardware agnostic high level languages, they are still premature. Triton targets GP-GPU architecture and provides a high level Pythonic interface for users to write kernels in. It works well for NVIDIA and AMD GPUs. Whereas, the design of Triton does not allow it to target emerging AI accelerators such as TPUs and IBM AIU Spyre. These chips leverage data flow architecture, disaggregating compute and data flows. Such elements are also showing up in newer GPUs such as NVIDIA H100, bringing the primitives needed for cross hardware support closer. To address these complexities, we envision three different areas of innovation.

Agile and retargetable compiler construction for accelerators. Companies spend many man hours perfecting compilers for their own accelerators. However, such manual compiler construction techniques are not agile or evolvable enough to keep up with the explosion of hardware platforms that are coming up. Recent work on automated compiler construction techniques using either formal methods or machine learning have shown promise on building compiler components that are easily retargetable across many platforms. Most of this work is done for commodity hardware and relatively little work has been done for tensor accelerators and we believe there are unique opportunities and challenges in replicating such for accelerators. For example, accelerators don't have predefined hardware software interfaces. On the other hand, accelerators don't have complex control logic making them easier for performance modeling.

Programming interfaces for new hardware features. Frameworks such as Triton and Pallas allow programmers to use advanced hardware features that you find in accelerators. However, it is unclear how to build these frameworks to support newer hardware features without significantly changing their programming models and user interfaces. For example, TMA was introduced by H100 GPUs and Triton has only enabled one of the features and lacked multicast features. A key challenge is designing frameworks and programming interfaces that are evolvable and portable.

Unified programming model for heterogeneous hardware. Currently, we have multiple frameworks for programming accelerators of different types. Pallas targets data flow accelerators, while Triton targets many core GPUs. As a user it is easier to program at an abstraction that can cover many of these accelerators. However, without exposing low-level details it is unclear how to provide a sufficiently high-level programming abstraction for users to directly program these accelerators for ML workloads. We expect multiple innovations on this front on programming languages for heterogeneous sets of accelerators.

### 4.4.3 Representative Contributions

UIUC and IBM have been at the forefront of addressing the aforementioned research challenges. Many works have focused on scaling the performance of LLMs. DeepSpeed-Ulysses [47] introduces novel distributed attention that enables highly efficient and scalable LLM training with sequences over a million tokens, 4x larger than existing systems. DeepSpeed-MoE [48] enables end-to-end 4D parallelism for novel sparse MoE model architectures, which open opportunities for training and deploying high-quality models with fewer resources and lower cost. ZeRO-Offload [49] enabled training massive LLMs with heterogeneous memory (Figure 12) and has been widely adopted in industry (e.g., NVIDIA, Microsoft, HP, Meta PyTorch).

UIUC PIs have proposed multiple approaches to optimize and bring better programmability to sparse machine learning models. TGLite [50] is a temporal GNN programming framework that introduces novel constructs to both program and optimize temporal GNN models. SPLAT [46] is a GPU code generation algorithm that produces high performance sparse attention kernels for a variety of sparse patterns.

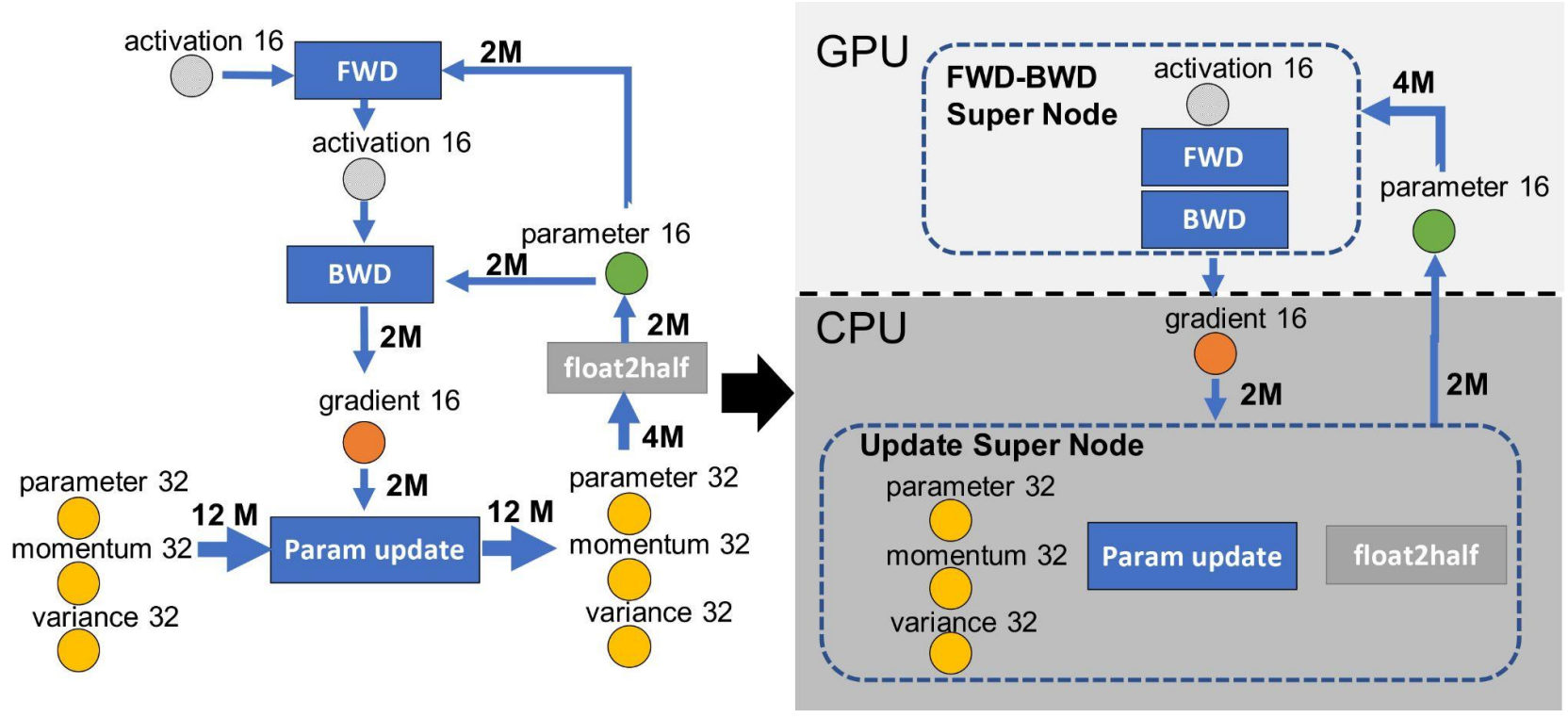

*Figure 12: The workflow of model training on heterogeneous memory using ZeRO-Offload.*

UIUC is at the forefront of creating automated compiler construction algorithms. Vegen [51] was the first vectorizer generator that automatically generates vectorizers given instruction set architecture semantics. Hydride [52] (Figure 13) is the first compilation framework that automatically generates a compiler IR. These works mainly focus on the commodity hardware platforms, and we believe that innovations are needed to bring the same benefits to accelerator platforms.

### 4.4.4 Optimization for Foundation Models

Given the importance and significance of foundation models, this subsection discusses various optimization techniques for foundation models. We focus on the coordination and adaptation between the cloud and edge to achieve satisfactory results targeting different applications.

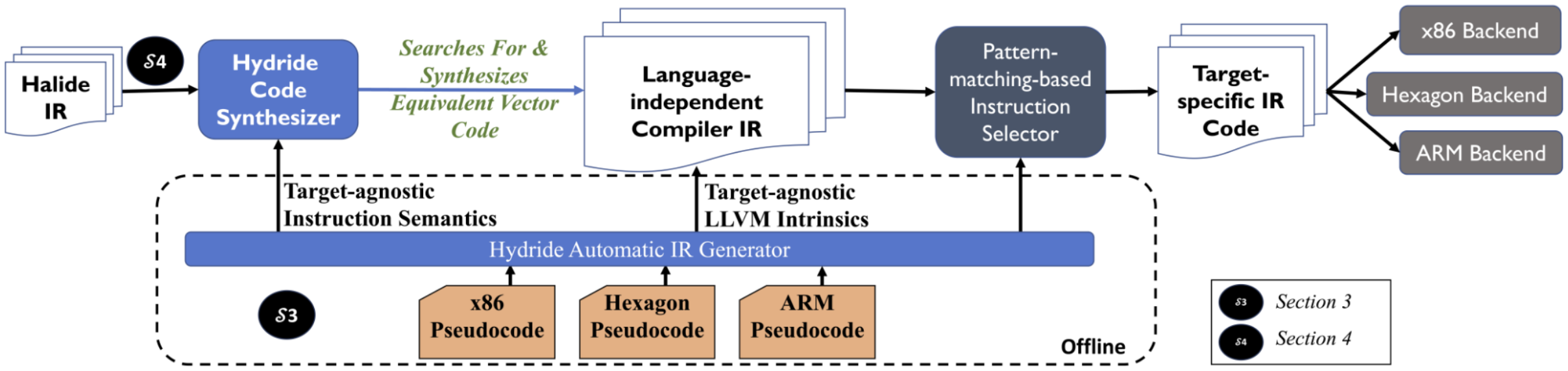

*Figure 13: Hydride workflow of generating compiler IR and code generators automatically.*

When foundation models are developed with hybrid cloud techniques, a typical case is that we pre-train the foundation model on cloud computation nodes, and then deploy it on edge devices, as illustrated in Figure 14. However, a challenge usually arises from the discrepancy between training on the cloud and inference on the edge (more details in Section 4.7): the real-world application on the edge side may face data distributions, target tasks, computation resources, or other environments and conditions that are drastically different from the training time. For example, the foundation model may require large memory and efficient GPU computation support, but such computation resources cannot be equipped on the edge device. In such cases, if the foundation model is directly applied, it would not be able to achieve satisfactory results.

Therefore, for foundation model inference on the hybrid cloud, it becomes essential to develop a comprehensive set of algorithms to seamlessly adapt the foundation model for diverse inference settings, including varying data, task, and computation capacities. Moreover, the constraints of training data and labels need to be considered carefully – oftentimes, we do not have abundant labeled data for the inference setting. We may only observe raw, unlabeled data in the inference setting, but annotating these new data samples can be both time-consuming and resource-demanding. As such, when adapting the foundation model, we cannot expect a massive, curated, and fully labeled dataset as in the pre-training setting. Instead, we may only rely on new data in the target setting with limited or even no human annotations.

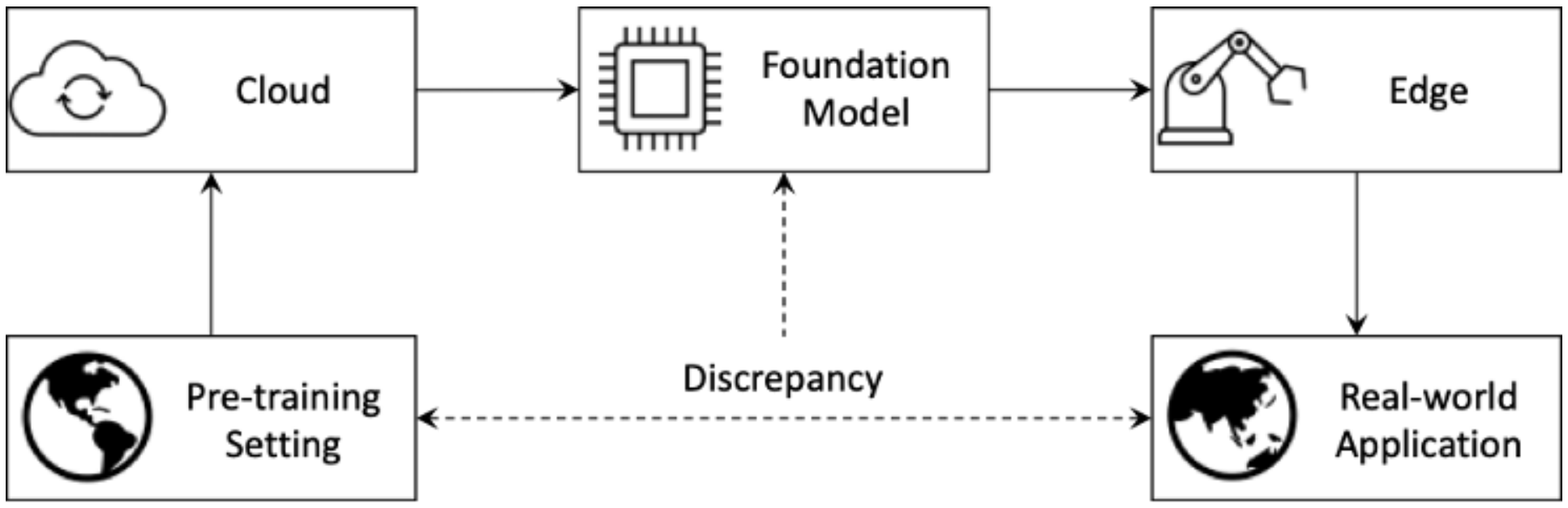

*Figure 14: Typical training-inference paradigm of foundation models on hybrid clouds.*

To address such challenges in deploying foundation models on the hybrid cloud, we have proposed a variety of algorithms to adapt foundation models for different inference settings with minimal human supervision. In particular, we investigate the computation constraints on the edge, and accelerate foundation models for better accuracy-latency trade-off; we adapt foundation models to bridge the data domain gap between training and inference, and even exploit foundation models for novel tasks that the models have never seen during training; we also explore how to develop foundation models with domain-specific knowledge and data. Specifically, we are pursuing the following novel studies on foundation model inference.

1) Accelerating Foundation Model Inference: The computation platform may be less powerful on the edge side, as compared with the abundant cloud computation resources. Therefore, foundation models need to be accelerated for deployment on the edge. We can develop new smaller models based on the learned knowledge of larger foundation models, and aim to preserve the accuracy, robustness, and generalizability of the powerful large models. Existing techniques including knowledge distillation, model compression, and quantization have been studied to create a small model that can function like a well-trained large model, but with significantly reduced computation demands.

2) Balancing Accuracy and Latency of Foundation Models in Real-Time Applications: In real-time applications, such as autonomous vehicles and immersive computing, the demands on AI can be multi-faceted. While high-accuracy models are always being pursued, the temporal demands of real-time applications focus on the latency of foundation models. A strategic computation allocation is critical to strike a balance between accuracy and latency. Furthermore, the models must be capable of forecasting future scenarios and compensate for any discrepancies emerging due to state changes during the inference time.

3) Adapting Foundation Models to New Data Domain with Limited Supervision: Another challenge in real-world edge-side applications is the data domain gap between training and inference. The real world often exhibits an environment where the data follow a new distribution, different from the training time. For instance, the training data may only instruct the foundation model to predict in normal weather conditions, but when the weather turns foggy or rainy, the change in visual observations may lead to impaired performance. Furthermore, the lack of labeled training data in the new data domain adds to the challenge, requiring the model to adapt dynamically without human supervision.

4) Exploiting Full Potential of Foundation Models Beyond Fine-Tuning: While foundation models are typically pre-trained for certain tasks such as text generation or visual recognition, they have implicitly obtained capabilities beyond the pre-training tasks. For example, a language-based foundation model may be tuned for conversations and act as a chat bot. The traditional approach of transfer learning exploits such capabilities through fine-tuning, i.e., further training the foundation model on a new task or dataset. In contrast to this transfer learning approach, our novel exploration reveals potentials of foundation models without further fine-tuning.

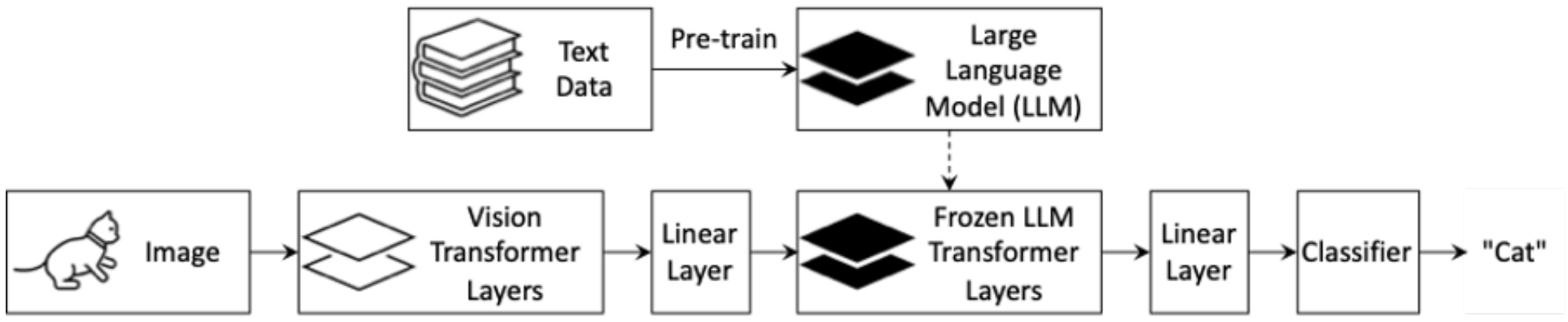

*Figure 15: Frozen large language model (LLM) layers in a visual classification model.*

Figure 15 shows one example where we surprisingly discover that frozen LLM layers can effectively function as visual encoding layers [53]. We insert LLM layers into a vision classification model, keep its parameters frozen, and only learn the other components in the classification model. Even though pre-trained for a different data modality and task, the LLM layer can improve the visual recognition model's performance. In another study, we build a Monte Carlo search tree that enables AI agents to reason, act, and plan in a unified framework [37]. Although the LLM is pre-trained for text generation, we can utilize its capabilities and knowledge to guide an AI agent. Our exploration demonstrates future directions in exploiting foundation models beyond the traditional pre-training and fine-tuning paradigm.

5) Self-Directed Enhancement of Foundation Models for Novel Downstream Tasks: In inference time, the foundation model may face a previously unseen target task. For example, the visual foundation model may be pre-trained for image classification, but the edge application requires object detection, i.e., recognizing and localizing all objects in a complex scene. Such task transition prohibits direct deployment of the foundation model, since the model has not been trained for the novel task. Typically, architectural changes, additional data labels, and further fine-tuning are necessary when the target task differs.

To address this challenge of novel tasks, we propose a fully self-directed or self-supervised multi-stage learning paradigm [55] to enhance the foundation model, as shown in Figure 16. In the first stage, we discover potential objects in complex scenes by clustering visual features extracted by the pre-trained model into semantically coherent regions, and then build structured representation of objects and their constituent parts in the complex scene. In the following second stage, we learn the enhanced perception model for the object detection task based on the discovered objects. We adopt self-training strategies to refine the model itself and correct previous errors, improving the overall detection accuracy.

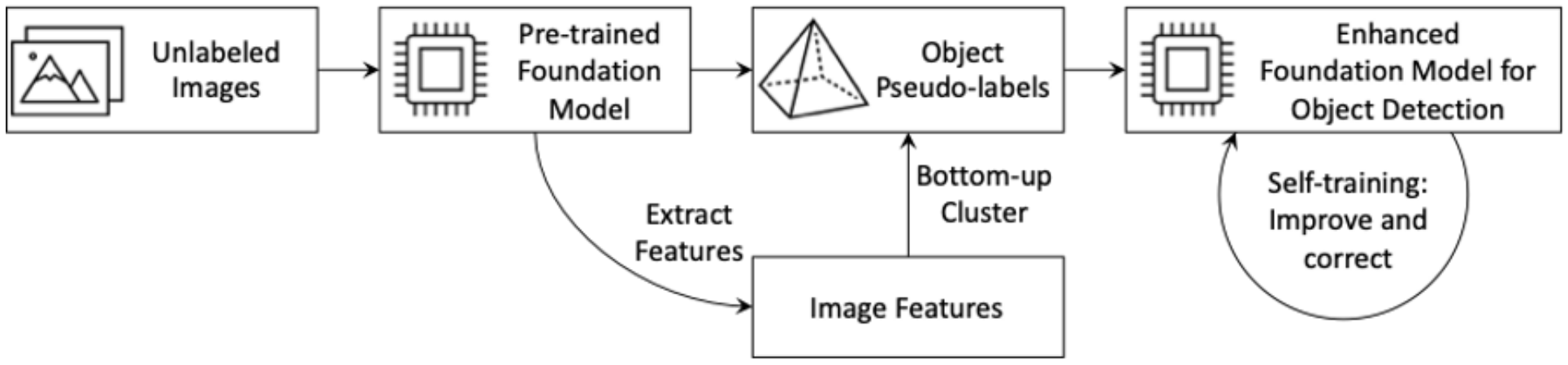

*Figure 16: Self-supervised learning approach for enhancing visual foundation models for structured perception tasks of object detection.*

6) Developing Foundation Models with Domain-Specific Knowledge: Although usually pre-trained on general data, foundation models have potentials in specific domains, such as answering scientific questions. However, compared with typical AI tasks (e.g., vision and language in general scenarios), certain domains may only contain limited data samples (e.g., materials science). With data scarcity in consideration, domain-specific knowledge becomes the key to learn foundation models for domain tasks. By injecting domain knowledge into the learning process, models can be augmented with the necessary information and produce more accurate, reliable predictions.

Additionally, in certain domain tasks, understanding the factors that decide the model's predictions is critical and provides scientific insights. Integrating domain knowledge can also enhance model interpretability, allowing us to understand and explain the model predictions.

Our recent research, as exemplified by [56], aims to create foundation models in the hybrid cloud for general-purpose accelerated discovery of inorganic crystalline materials, thus overcoming the traditionally laborious, empirical materials discovery process. In particular, we exploit the domain knowledge imparted by materials physics (together with the interconnections between materials properties) and the pre-encoded knowledge in LLMs to build foundation models from heterogeneous materials data for materials property prediction, design, and synthesis. Section 5.1 is dedicated to exploring the application of AI techniques in material discovery.

In another line of work, we aim to develop largely automated program generation tools that greatly reduce the effort required for program development and testing by automatically generating application code with a correctness guarantee with respect to an automatically constructed formal specification. We combine formal typing guarantees from type-based program synthesis with program generation from LLMs, which would greatly simplify automated code generation, produce well-documented and interpretable generated code, and also give end-users increased confidence in the correctness of their code.

### 4.4.5 Long-term Directions and Vision

Foundation models, including large language models, have shown immense promise in enabling the next generation of AI applications. The transformer architectures that power these models are the cornerstone of multiple foundational models including those invented by IBM [57]. It is our belief that scaling these models to large context lengths will be key to unlocking newer capabilities. This will enable novel applications and new scientific discoveries in the fields of natural language processing, weather and climate predictions, code generation, geospatial data management etc. We believe novel neural architecture designs, systems designs, compiler optimizations are key to achieving this goal as mentioned in detail under the research challenges and opportunities subsection. We also believe that cross-layer innovations and co-design would enable extra synergies to collectively reach these objectives.

Sparse machine learning models efficiently use relationships that exist in data to enable predictive analytics on graphs and other sparse structures. These have already enabled groundbreaking applications in drug discovery, quantum chemistry and we believe they will continue to enable future predictive applications that revolve around sparsely structured data. However, sparse machine learning workloads have notoriously been difficult to scale due to its irregular nature. We believe novel optimizations both at the compiler and distributed framework level are needed to scale these models and enable the next generation AI applications that revolve around sparse data.

Efficient use of hardware accelerators will become key to unlocking generalizable acceleration for all these machine learning models. NVidia's GPU platforms have grown exponentially in popularity and mostly power the machine learning models of the current era. However, emerging classes of new tensor accelerators have shown promise including IBM's AIU Spyre. We believe that the same programmability and compiler support that is available for more established hardware platforms such as GPUs and CPUs should become available to make emerging accelerators more mainstream. To achieve that, innovations on programming languages, models as well as novel compiler construction methodologies will be extremely important.

## 4.5 Programming Model, Middleware, and Platform

### 4.5.1 Introduction

Modern computing applications span a diverse array of computing paradigms, platforms, and infrastructure requirements. Beyond classical simulations typically performed on dedicated HPC systems, there is now extensive use of AI/ML and big data analytics, which predominantly occur on cloud systems. Each application may demand specific platform support, with many requiring massively parallel distributed clusters or specialized hardware accelerators like GPUs, TPUs, and FPGAs to efficiently handle specific tasks. Consequently, these rapidly evolving applications necessitate a critical reassessment of how to develop and operate computer systems to meet a myriad of distinct requirements while delivering flexibility, performance, and security. This underscores the urgent need to innovate in platforms, middleware, and programming models to support the next generation of computing applications.

Our long-term vision is to transform the hybrid cloud system that supports advanced programming models and incorporates intelligent middleware for workflow orchestration, as shown in Figure 6. This platform will be designed to enable emerging applications that span diverse architectures while delivering high performance. It will seamlessly integrate heterogeneous computing resources, offering elasticity, fault tolerance, and hardware abstraction, thereby enabling the efficient execution of intricate workflows across multiple computing environments. Implementing this vision requires developing a new programming paradigm encompassing a broad spectrum of parallel applications, not limited to traditional HPC or cloud-native applications. We aim to create a model that offers high-level abstractions for parallel and distributed computing while ensuring performance and ease of use.

### 4.5.2 Research Challenges, Opportunities and Future Directions

At the platform level, a comprehensive suite of services and tools is essential to provide the infrastructure, software, and resources for developing, deploying, managing, and scaling applications and services over hybrid cloud systems. The platform-level software must support the collection of physical and virtual resources used to deliver efficient computing environments. This presents a number of opportunities to innovate at the platform level, including developing a unifying cloud-native runtime system, developing hardware abstraction layers, and providing a path for integration of emerging technologies.

- Cloud-Native Runtime System that extends beyond traditional models like MPI. This runtime needs to support dynamic resource allocation, enable fault-tolerant execution, and manage communication across heterogeneous resources. Incorporating advanced scheduling algorithms and communication protocols to optimize performance is critical to enable such a runtime system.

- Hardware Abstraction Layer is needed to address the challenge of hardware heterogeneity. This layer needs to abstract the complexities of various accelerators and specialized devices as well as CPUs. This layer will present a unified interface to the runtime system, allowing applications to leverage GPUs, AI accelerators, DPUs, and quantum devices without requiring device-specific code. Partitioning program execution appropriately among different computing units and continuously evolving sets of accelerators without having to involve the application programmer in tedious accelerator-specific code optimizations is going to become an increasingly critical challenge. Future platforms must export the benefits of heterogeneous acceleration affordances while retaining portability of application code and minimizing development cost.
- Integration with Emerging Technologies such as quantum computing [58] and specialized accelerators. Examples of such specialized accelerators already include GPUs for high-end analytics processing, near-memory analytics accelerators that can be leveraged, for example, to reduce compression/decompression cost of memory snapshots and speed up VM cold starts, Data Processing Units (DPUs) that may enable secure and efficient data filtering near storage to improve the performance of query processing and reduce data transfer times, and optimized coherent host-NIC interfaces that allow hosts and network interfaces to exploit shared cache hierarchies for significantly accelerated data transfer [59]-[60].

Middleware acts as an intermediary layer between different applications, services, and the underlying platform and infrastructure, facilitates communication, data management, and integration across various distributed cloud environments. User-facing tools are necessary to access the environment, deploy and manage applications, and monitor their states. Some of these need to be cross-layer to enable better visibility and coordination.

- The inherent use of our integrated platform lies in its ability to merge various computing patterns across multiple clusters within a single application distributed in space and time. We envision AI-powered design tools for creating computational workflows, such as a conversational assistant for assembling workflows, as well as new AI-driven solutions for interactive workflow execution automation [61]. Typically, workflow languages are crafted with the presumption that workflows operate within a uniform environment. We foresee the emergence of advanced languages to articulate workflows that require heterogeneous computing resources, thereby enhancing programmability and adaptability.
- Adaptive Execution Engine will pave the way for workflow automation, lifecycle management, and the seamless handling and optimization of data. Within this framework, a new orchestration layer will facilitate workload adaptation and optimization while considering cost, performance, energy, and carbon efficiency. We have already demonstrated an adaptive HPC-cloud bursting system [62]-[63] that seamlessly places workloads across multiple compute systems. Our prior work on workflow-aware scheduling [64] enables tailoring orchestration and optimization processes to align with high-level goals and specific Service Level Objectives (SLOs), accommodating distinct phases of the workflow lifecycle and diverse user needs.
- Unified Control Plane is needed to enable AI-empowered middleware and runtimes to interface with heterogeneous resource managers such as Kubernetes and LSF/Slurm. Given the heterogeneity of the resource management for different workloads (e.g., microservices versus HPC) and infrastructures (e.g., edge versus cloud), we envision that different resource managers will continue to exist and become more specialized for target workloads (e.g., Kubernetes for datacenter workloads and k0s for edge and IoT workloads); meanwhile, there is a strong need to further orchestrate them for heterogeneous computing jobs. A recent trend in edge-cloud continuum workloads is such an example. The core component of the unified control plane is the multi-cloud broker, which enables the decomposition of a given computing job (e.g., based on its specification) and maps each component to the corresponding resource manager(s). We envision that the decomposition and mapping (including resource allocation and configuration) is done automatically in an AI-assisted manner. At a high level, the broker plays a similar role as that proposed in Sky Computing [174], which orchestrates workloads across multiple cloud providers. Differently, we envision a more generic and automated broker that supports more diverse workloads, is more composable, and supports more dynamic resource allocation for higher efficiency and better cost.

On top of this, flexible programming models are required to enable application developers to easily command the underlying hardware to perform complex computational tasks without needing to write excessive amounts of highly specialized and platform-dependent code. Even though our vision calls for LLMaaA, we still need to provide support for many existing programming models that are widely used on today's systems.

- High-Level Parallel Abstractions are needed to simplify the development of parallel applications. Inspired by the ease of use of languages like Python and the efficiency of C++, we investigate developing a programming interface that allows developers to express parallelism and data movement without delving into low-level details. One way to accomplish this is through high-level computational constructs that can be translated to specialized hardware by runtime systems, without revealing the complexities of the underlying hardware.
- Autonomous Runtime Optimization is needed to address the challenge of hardware specialization. As common patterns and libraries arise, such as advanced architectures for distributed training, an autotuning runtime should be employed to encode optimizations for more efficient computing, while keeping the underlying complexity hidden from users. This runtime will leverage machine learning techniques for autotuning and performance optimization.
- Library of Optimized Computation Primitives are commonly used in AI/ML and simulation workloads. These primitives will be optimized for various hardware accelerators and accessible through the high-level programming interface. These libraries can then be utilized with standard APIs, irrespective of the offloading mechanism behind them.

Finally, cross-layer automation, integration, and observability are critical components needed to efficiently execute computations on the envisioned platform. Cross-layer automation enables the coordinated management of resources and workflows across different layers of the cloud stack. By automating the selection and configuration of Infrastructure-as-a-Service (IaaS) and Platform-as-a-Service (PaaS) resources, cloud platforms can better meet application-specific Service Level Agreement (SLA) goals. This includes minimizing processing delays, maximizing availability, and optimizing throughput. Cloud applications often involve multiple heterogeneous processing frameworks and cross-layer automation can help to manage the vast configuration diversity among available cloud resources and big data processing frameworks, which is challenging to handle manually. Cross-layer Integration is important for coordinated scheduling and service delivery. By integrating the application layer with the networking layer, cloud systems can perform coordinated scheduling that optimizes both computational and network resource utilization. Cross-layer resource orchestration also enables seamless service delivery by providing service awareness through session control and integrating it with network control and management. Cross-layer observability is critical for performance monitoring and automated diagnostics. It enables comprehensive monitoring of applications across multiple cloud layers. This helps in identifying performance bottlenecks and optimizing resource usage. Various frameworks can be developed to provide automated, cross-layer instrumentation for diagnosing issues in deployed distributed applications. This capability is crucial for maintaining the health and performance of cloud-based systems. The integration of these components will lead to improved efficiency resulting in more optimal use of the resources, enhanced flexibility and adaptability and simplified management.

## 4.6 Infrastructure and Hardware

### 4.6.1 Introduction

Generative AI, FM, and LLM workloads require significant systems infrastructure to deliver their full potential and democratize access to their capabilities [65]–[67]. At present, these AI models are growing exponentially, requiring AI infrastructure comprising numerous compute/storage boxes like Figure 17.

However, such high-end on-prem clusters are unsustainable for most organizations [68]. As discussed in earlier sections, the future hybrid cloud systems supporting such AI workloads must be affordable to ensure broad accessibility and adoption. Furthermore, a system co-design approach that incorporates end-to-end optimizations based on the unique characteristics of AI workloads will enable dramatic reductions in system cost, improved resource utilization, and development of energy-efficient systems. Some key emerging trends to be considered in this co-design approach include AI models with unlimited context windows, models that incorporate the mixture of experts paradigm, and models developed for multimodal datasets such as text, video, audio, images, and time series data [69]–[73].

As an example of such co-design, LLM inference needs multiple expensive GPUs because a single GPU with limited memory capacity cannot store hundreds of gigabytes of parameters and intermediate values generated during serving. As a cost-efficient alternative to using multiple expensive GPUs for LLM inference, the latest framework first stores (or offloads) all parameters and intermediate values to large CPU memory, and then transfers a subset of them that a single GPU can store and compute with at a time [74]. However, such CPU-GPU transfer over slow PCIe creates a bottleneck for low-latency and high-throughput LLM inference [75]. To ease the bottleneck, the framework also offloads certain (empirically determined) computation to the CPU, which reduces the amount of CPU-GPU transfer but achieves limited success due to the CPU's insufficient throughput.

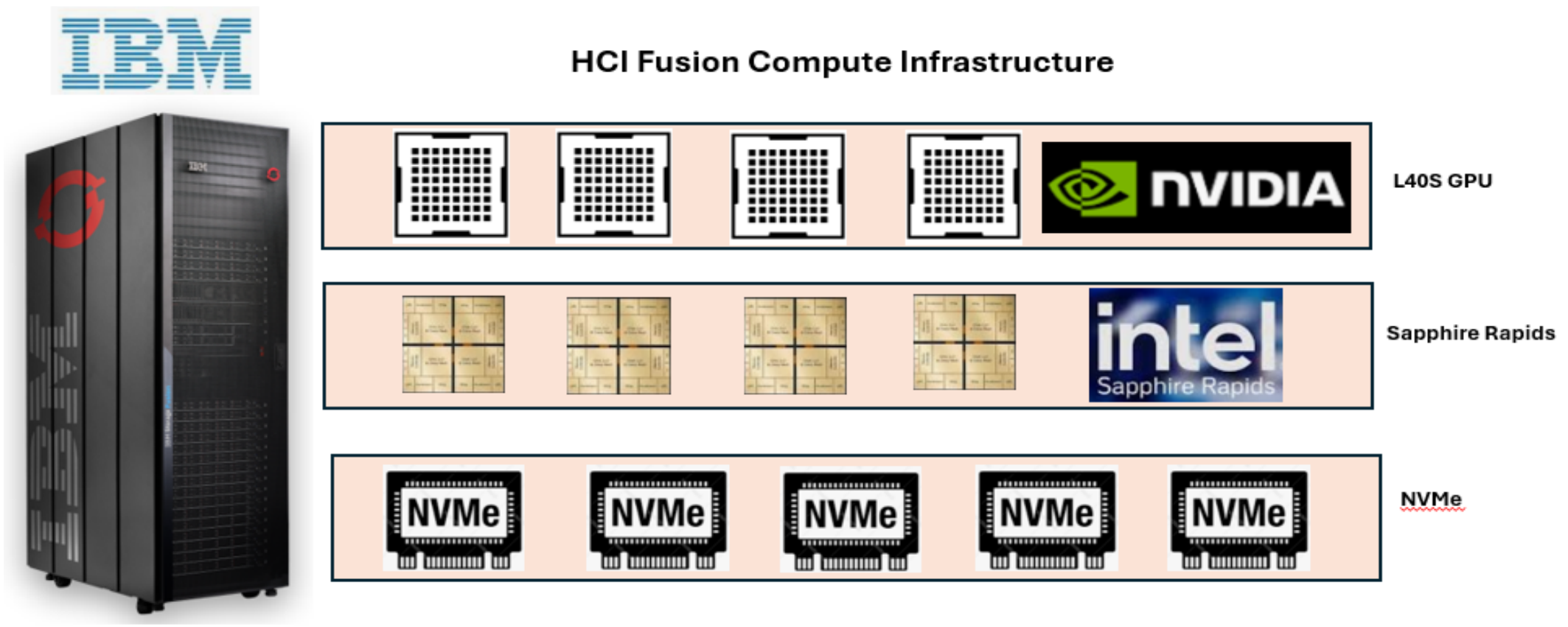

*Figure 17: HCI Fusion Compute Infrastructure.*

A primary requirement for this hardware/software co-design is to dramatically improve cost-performance of AI workloads. As models grow larger and customers increasingly use multiple models, compute capacity is expanding by 2-3 times per generation [68]. However, memory capacity and memory bandwidth are not increasing at the same rate [76]. Consequently, multiple accelerators must work together to train and infer on these models. Tackling these challenges, emerging technologies like UALink, UltraEthernet, SmartNICs, acceleration on CPUs using technologies like AMX, emerging cache-coherent interconnects like CXL can be exploited to alleviate this memory wall problem and improve utilization of accelerator system [77]–[80].

### 4.6.2 Recent Advances and Developments

LLM inference with Intel AMX and CXL. The latest Intel CPU has integrated Advanced Matrix Extensions (AMX) for acceleration of AI/ML applications [79]. Our evaluation of AMX can offer significant matrix-multiplication throughput, even comparable to that of some GPUs [75]. This allows us to offload more computation to the CPU (i.e., less CPU-GPU transfer) than past CPUs for LLM inference. Besides, the industry has introduced Compute Express Link (CXL) and supported by the latest Intel and AMD CPUs. CXL is built on the PCIe physical layer and allows Hyperscalers to recycle DDR4 DRAM for the latest CPUs supporting only DDR5 DRAM and expand memory capacity and bandwidth inexpensively [81]. Such an inexpensive memory capacity expansion capability becomes very useful for running large LLMs with large batch sizes which often require more than terabytes of memory capacity.

Unified GPU memory and storage architecture. A promising and practical approach to expand the limited GPU memory is using flash memory, which provides larger memory capacity at a low cost. With this approach, a few architecture solutions have been developed in both academia and industry [82]–[84]. Unfortunately, the limited bandwidth of flash chips is still the performance bottleneck, in comparison with the high-bandwidth memory in GPUs. Although we can scale up the SSD bandwidth by using multiple SSDs or flash chips, the aggregated bandwidth is still limited by the PCIe interface. Even though we can employ faster interconnects such as NVLink, the data transfer bandwidth is still much lower than the GPU on-board memory bandwidth. To tolerate slow flash accesses, developers have to carefully manage the data across the heterogeneous memories to explore the data locality [85]–[88]. This inevitably complicates the GPU memory management and hurts the development productivity.

We present a unified GPU memory and storage architecture named G10 driven by the fact that the tensor behaviors of deep learning workloads are highly predictable. G10 integrates the host memory, GPU memory, and flash memory into a unified memory space, to scale the GPU memory capacity while enabling transparent data migrations. Based on this unified GPU memory and storage architecture, G10 utilizes compiler techniques to characterize the tensor behaviors in deep learning workloads. Therefore, it can schedule data migrations in advance by considering the available bandwidth of flash memory and host memory. The cooperative mechanism between deep learning compilers and the unified memory architecture enables G10 to hide data transfer overheads in a transparent manner.

Accelerators for LLMs. Currently, most LLM workloads run on GPU platforms. However, it is likely that inference accelerators will take over as the preferred platforms, following a model like Google's TPUs [89]-[90]. How to integrate such accelerators with the CPUs is a matter of current research. Ideally, accelerators would be integrated as separate chiplets into a package with the CPUs. One possible model is Intel's integrated accelerators in the Sapphire Rapids platform, which target datacenter functions [91]. In such a model, the accelerators are not simply PCIe devices programmed via memory-mapped I/O operations. Instead, their ISA has instructions for dispatching work and for signaling; the accelerators operate with virtual addresses, exploiting the IOMMU for address translation; finally, the accelerators support virtualization to make them usable in a cloud environment.

An important consideration in a CPU with integrated accelerators is how to organize the communication between the CPUs and the accelerators [92]. Currently, the designs involve a CPU core or a centralized hardware manager orchestrating the communication between the CPUs and the accelerators. Other proposals allow direct accelerator-to-accelerator coordination, either in hardware or software. This appears a reasonable approach going forward.

### 4.6.3 Research Challenges, Opportunities and Contributions

Research challenges and opportunities lie across the entire stack from cost-efficient LLM training to high performance LLM inference. Two key technologies are changing at the same time:

1. AI model architecture, and thus their resource utilization characteristics are changing rapidly as they try to address needs in multi-model domains, support large context windows, and support emerging agentic use cases.
2. Systems technology such as accelerator, accelerator interconnects, networking, storage, protocols, software stack are changing at the same time. So, a primary challenge for this cluster is to pick workload characteristics that are invariant and demonstrate prototype technology that addresses these workloads.

Our representative contributions provide novel initial solutions to address these challenges.

LIA: a full-system CPU-GPU-CXL cooperative computing framework for cost-efficient LLM inference. Tackling the cost and performance issues of the past solutions for LLM inference, we propose LIA, an AMX-aware framework for CPU-GPU cooperative LLM acceleration. Specifically, it systematically determines what computation of a given model to offload to the CPU for lower latency and higher throughput of LLM inference. It also extends Intel Extension for PyTorch (IPEX), currently implemented for CPU- or GPU-only LLM acceleration, to seamlessly facilitate CPU-GPU cooperative LLM acceleration. LIA is also implemented to effectively use CXL memory alongside DDR memory. LIA offloading parameters to CXL memory not only provides higher throughput than LIA using only DDR memory but also uncompromised latency for LLM inference. We demonstrate that LIA offers up to 12.4× lower latency and up to 6.0× higher throughput than the latest framework for latency- and throughput-driven inference of OPT-175B, respectively.

G10: Breaking the GPU Memory Wall with Tensor Migrations via GPU Direct Storage. We present a unified GPU memory and storage architecture named G10 that enables smart tensor migrations for scaling the GPU memory transparently using flash memory, while tolerating the performance overheads of slow flash accesses. G10 [93] consists of three major components: (1) a tensor vitality analyzer for extracting the semantic knowledge of tensors in a deep learning model, (2) a tensor migration scheduler for planning the tensor migrations in advance, and (3) a unified memory system for simplifying the GPU memory management and enabling transparent tensor migrations. The tensor vitality analyzer works with deep learning frameworks like PyTorch to track all the tensors in a DNN model. It leverages the execution graph generated by the compiler to learn the size and lifetime of each tensor as well as its dependency on other tensors. Based on the extracted semantic knowledge of tensors, the tensor migration scheduler of G10 will plan the tensor migrations in advance before executing the model training process. To maximize the benefits of tensor migrations, G10 prefers to migrate large tensors that will be inactive for a long time to the flash memory. Therefore, the precious GPU memory can be best utilized for active tensors. For the inactive tensors whose inactive time is short, G10 will make the best effort to keep them in the GPU memory to avoid unnecessary tensor migrations. G10 also plans intelligent data prefetching in advance with its tensor migration scheduler. To facilitate the tensor migration, G10 integrates the GPU memory, host memory, and flash memory as a unified memory space by extending the Unified Virtual Memory (UVM) of GPUs. G10 extends the page table of UVM by storing flash page addresses in its leaf-level page table entries. The unified page table can point to an address in either host memory, GPU memory, or flash memory. The unified memory system will conduct the transparent address translation at runtime. This significantly simplifies the GPU memory management and the compiler optimizations.

EcoFaaS: An energy-efficient framework for serving LLM inference requests. While most works focus on LLM performance, LLM energy- and power-efficiency are as important as performance. Inference requests execute in opaque virtualized sandboxes, and are co-located in a highly-dynamic manner with many other invocations of diverse properties. These features are a radical shift from more monolithic application environments and require a new approach to manage energy and power. EcoFaaS [94] takes a user-provided end-to-end Service Level Objective (SLO) and tries to execute each request in the most energy and power-efficient manner without violating the SLO. Based on the computed deadlines of the different LLM invocations, EcoFaaS sets the optimal per-invocation core frequency using a prediction algorithm. The algorithm performs a fine-grained analysis of the execution time of each invocation, while taking into account the specific invocation inputs. To avoid the overhead of continuously changing the frequency of cores, EcoFaaS splits the cores in a server into multiple Core Pools, where all the cores in a pool run at the same frequency and are controlled by a single scheduler. EcoFaaS dynamically changes the sizes and frequencies of the pools based on the current system state. Our experiments show that EcoFaaS can reduce the total energy consumption of requests substantially while simultaneously keeping the tail latency of the requests within the SLO limits.

### 4.6.4 Long-term Directions and Vision

The future vision is to develop system co-design ideas, test and validate them to dramatically improve cost-performance, utilization, and energy efficiency of AI systems. Specifically, it will be important to exploit new emerging hardware technologies such as CXL, AMX, GPUDirect, and various accelerators integrated with CPUs and NICs. Specifically, the latest Intel CPUs include Data Streaming Accelerator (DSA), Dynamic Loader Balancer (DLB), In-Memory Analytics Accelerator (IAA), and Advanced Matrix Extensions (AMX). They can greatly improve the performance of data movement between memory regions or devices, network packet distribution among CPU cores, data compression and decompression, and matrix operations. We have already demonstrated the benefit of using AMX in our collaborative work. In the future, the data compression and decompression capabilities can be used to reduce the system memory requirement for running LLMs without hurting the latency or throughput of LLM services. In addition, we also have shown the potential of GPUDirect capability to provide an illusion of having large memory capacity for the GPUs which often suffer from the limited memory capacity. Our proposal provides a novel data transfer and orchestration algorithm that can greatly reduce the need for using more GPUs to train LLMs.

To facilitate not only more efficient data transfer among these compute, network, storage, and memory devices but also easier and broader cooperative heterogeneous computing among the compute devices and network/storage/memory devices potentially with near-data processing (NDP) capabilities, we can exploit the emerging cache-coherent interconnect (CCI), such as Compute Express Link (CXL) and Ultra Accelerator Link (UAL). For instance, CXL can provide cache-coherent (CC) host-to-device (H2D) and device-to-host (D2H) memory accesses. This can simplify the programming effort for the data transfer among these devices as these devices can transfer data with simply load/store semantics, instead of cumbersome and inefficient DMA requiring manual cache coherence management under various constraints. Such H2D and D2H memory access capabilities give us unprecedented opportunities in revolutionizing the interface not only between the host and traditional storage and networking devices, but also among the compute, memory, storage, and network devices. Specifically, we can envision a system with memory, storage, and network devices with NDP capabilities and connect and use them through a CCI-driven software-defined interface (see Figure 6). Such a software-defined interface will not only improve the data transfer efficiency among compute, memory, storage, and network devices, but also enable efficient fine-grained cooperative heterogeneous computing among these devices, which will dramatically reduce the cost of AI/ML training and inference with combinations of the most cost-effective devices for specific compute requirements of individual layers of target AI/ML models.

## 4.7 Enabling the Edge Transformation: An End-to-end Edge-Cloud Approach

### 4.7.1 Introduction

The rapid advances in foundation models, generative AI, and LLMs are fundamentally changing how we interact with computing at the edge; e.g., with immersive computing or extended reality (XR), robotics, autonomous vehicles, drones, etc. These new modalities for computing have the potential to transform most human activities. However, these new edge devices and applications are increasing in complexity, with stringent but heterogeneous and multidimensional performance requirements (e.g., heterogeneous compute, real-time latency, data bandwidth, power, and multidimensional quality of experience or QoE). Achieving their transformative impact will require a far greater integration and co-design of the edge with the cloud than we have today.

For example, immersive computing or XR, including virtual, augmented, and mixed reality (AR/VR/MR), has the potential to change medicine, education and training, industrial maintenance, telepresence, entertainment, etc. in a fundamental way, from single user (a virtual personal assistant) to large collaborative scenarios (a mixed reality hybrid conference with thousands of attendees). However, this is still far from reality. As a powerful example of the advances made and the road still to be covered, we see that foundation models are enabling truly photorealistic reconstruction, rendering, and user speech directed editing of mixed reality environments [95], but today this is far from real time and from the power budget of a comfortable headset. For example, ideal headset power budget is in the range of 100s of milliwatts while today’s state-of-the-art expend 10s of Watts and are still not able to run the state-of-the-art neural models for rendering and reconstruction. Offloading computation and remote collaborative XR services and workflows have been difficult because of the stringent latency and bandwidth requirements. Further, generative AI is fundamentally changing the traditional vision and graphics XR pipelines, making today’s GPUs less than ideal for these requirements.

Our goal is to develop the interfaces and underlying software and hardware implementations that will foster the needed co-design between the edge and cloud to enable the edge transformation.

### 4.7.2 Research Challenges and Opportunities

A key source of opportunity (and challenge) arises from the heterogeneity and multimodality within even a single edge application (e.g., a rich XR experience includes speech, graphics, audio, haptics, perception, etc.) with heterogeneous, stringent, and multidimensional performance requirements for each, and the ability to trade off these different requirements based on the limits of human perception and the current resource availability. Thus, SLOs for these applications are a (currently ill specified) Pareto frontier of heterogeneous requirements. Exploiting this flexibility in a coordinated way throughout the hardware and software layers of the edge and cloud will be critical.

The above observations open up a number of research challenges and opportunities as follows. How to specify flexible SLOs and how to distribute computation between edge and cloud so these SLOs are satisfied within the constraints of the edge? How to design accelerators for the diversity and evolution of emerging workloads, in an integrated edge-cloud ecosystem? How to design a uniform communication interface that serves the needs of diverse accelerators and cores in a system-on-chip (SoC), system-in-package (SiP) or chiplets, and further systems of systems, while also exploiting energy efficiencies from specialization. How to co-design neural architectures, accelerators, and inferencing workflows; integrate online and offline model optimization techniques; and optimize online training coupled with inference for our integrated edge-cloud view with flexible SLOs. Exploiting the flexible SLOs and flexible systems described above will require an innovative compiler and scheduling infrastructure to orchestrate the computation and data movement among the available edge-cloud resources. Finally, such an integrated edge-cloud system view motivates novel prototyping infrastructure to evaluate the ideas.

### 4.7.3 Representative Contributions

UIUC faculty and IBM researchers have made many contributions already towards the vision described here, laying a strong foundation for future work. For example, the ApproxHPVM, ApproxTuner, and ApproxCaliper line of work [96]-[98] has explored quality-latency-power tradeoffs and developed compiler and scheduling infrastructure to properly leverage these tradeoffs for heterogeneous edge computing. Catan presents a scheduling framework to ensure all tasks in robotics and XR applications are provided adequate resources to achieve their latency goals [16]. Other work discusses and exploits energy, latency, and quality tradeoffs in scene reconstruction and eye tracked foveated rendering [99]-[101]. The ILLIXR work [102] has laid the foundation for distributed edge-cloud compute distribution. The Mozart project [103] explores composable disaggregated accelerators for an SoC, leveraging the Spandex coherence line of work [104]-[105] for coherence specialization with a familiar coherent address space in heterogeneous systems. Several projects within IBM and UIUC have prototyped complex end-to-end systems, e.g., the EPOCHS project [106], HPVM [107], ILLIXR [102], etc.

### 4.7.4 Long-term Directions and Vision

Our long-term vision is to develop a co-designed edge-cloud architecture where applications are oblivious to which component is being run where, and the hardware and software infrastructure is designed from the ground up to have an integrated view of the edge and the cloud. This will enable complex emerging edge applications, with stringent compute, latency, and bandwidth requirements to effectively exploit all resources, enabling a transformation of how we interact with computing at the edge. Specific research directions follow below.

Specifying flexible SLOs: Emerging edge applications such as XR and robotics have complex workflows. Although there are objective metrics used for assessing the goodness of these systems, they often don't correlate to the subjective end-user experience as determined from user studies. Furthermore, converting the end-user quality of experience metrics to clear SLO requirements is by itself difficult; e.g., end-to-end latency from head motion to display in XR can be compensated by prediction and warping techniques in XR; accuracy in image analysis in a robotics pipeline can be compromised in favor of latency if it can later be compensated by the motion control algorithm. Understanding the dependences between different application subsystems and pipelines and determining how to apportion latency, power, etc. across these different components is a challenging problem. Finally, as evident from the above examples, latency, accuracy, power, and quality offer tradeoffs that can be made based on current resources. Understanding, quantifying, and specifying end-to-end user-level goodness metrics, their relationship to system level and component level requirements, and the tradeoff space is an important research question for emerging edge applications. Effectively conveying these requirements and the associated tradeoff space to the cloud, in a way that ensures appropriate services are provided and the requirements are met, poses an additional challenge.

Distributing computation between edge and cloud: Fundamentally, there is a tension between high fidelity/accuracy compute on the cloud vs. low latency on the edge. How to distribute the computation across the edge and cloud and how to compensate for latencies incurred? For example, recent techniques consider using expensive generative models in the cloud to provide high quality frames to the client at a low frame rate, with cheaper models on the client filling the remaining frames for an effective high frame rate. Similarly, some modalities have higher latency thresholds than others, but considered as interdependent pipelines, it is unclear where to place what.

Disaggregated and composable fine-grained acceleration: With increasing diversity in the workloads, rapid changes in models, and diverse sources of performance bottlenecks from infrastructure functionality to AI models [108]-[109], achieving energy efficiency through highly specialized monolithic accelerators is difficult to sustain – each new algorithmic advance entails a large new investment in specialization. Further, various specialized components in current systems are already known to be under-utilized, wasting area and exacerbating leakage. We therefore advocate finer-grained accelerators for common primitives, and an architecture that enables them to be transparently composed to give the performance benefit of monolithic acceleration without their recurring design cost. Such an architecture must enable accelerators as first-class virtualizable citizens that can directly communicate and synchronize with each other, without the need of CPU orchestration. Furthermore, the architecture must provide a uniform programming model that is transparent to whether a computation is being done on a monolithic accelerator, a CPU, or some subset of software-composed fine-grained accelerators. This concept of composable disaggregated acceleration applies to SoC's, emerging chiplet based ecosystems (systems in package), and other hierarchies of systems of systems, connected with appropriate interconnects. Our recent work on Mozart [103] for an SoC provides a foundation, but much work is required for modern foundation models, new chiplet/SiP, CXL, NDP types of technologies, and emerging applications.

Data communication: Similar to the philosophy of disaggregated and composable accelerators for compute specialization above, we also advocate a similar vision for data communication. A coherence shared-memory or a global address space has provided a portable popular programming model; however, more specialized communication models such as point-to-point messages, bulk transfer DMAs, and private memory structures such as scratchpads provide for higher efficiency through specialization. Can we have a uniform programing interface providing the ease of coherent shared memory but with the specializations as above? Our work on Spandex [104]-[105] and its use in Mozart provides a foundation but much needs to be done to adapt these ideas for modern AI models and their use for emerging applications.

Co-Design of neural architecture, accelerators, and inferencing workflows. A key to efficiency will be co-design methodologies where neural network architectures and accelerators are jointly optimized for specialized AI inferencing tasks in the cloud and on the edge. Fine-tuning AI inferencing workflows with tightly coupled software-hardware optimizations can enable significant performance and energy efficiency gains across both edge and cloud environments.

Integrated offline and online model optimization techniques. Creating a hybrid optimization framework that integrates offline techniques (e.g., model pruning, quantization, graph fusion) with online techniques (e.g., multi-user/multi-request management, adaptive batching) has the potential to expose new opportunities for performance improvement. The goal is to establish seamless model management that adapts to real-time user requests while simultaneously maintaining model efficiency and accuracy across edge and cloud infrastructures. This system should include AI-driven mechanisms that adaptively choose the optimal optimization path, depending on the hardware, workload, and model type.

Compiling and scheduling with flexible SLOs on flexible systems. We have advocated for a vision where there is flexibility in the end-application requirement specification, model architecture, hardware architecture, and variation in workload. Underlying such a system is a compiler and scheduling runtime that can find the correct mapping of an application request to a model architecture and hardware.

Call for end-to-end system prototypes and benchmarking: Many of the above problems are inter-related. We believe that systems research that seeks to break entrenched barriers and interfaces must go hand in hand with end-to-end system prototypes and benchmarking.

The above outlines a rich, long-term research agenda to enable transformative impact on the edge. It requires collaboration and co-design between all layers of the system stack across the edge and cloud, with a final outcome of enabling new classes of applications that transform most human activities.

## 4.8 System-level Optimizations

### 4.8.1 Robustness, Dependability, and Security

Robustness, dependability, and security are essential pillars of large-scale computing systems, and have been the grand challenges for many decades due to the ever-increasing complexity, dynamics, and heterogeneity of cloud systems. Emerging computing paradigms like microservices and serverless computing, deployment patterns like hybrid cloud and edge-cloud continuum, and drastically increased scale and heterogeneity of AI-oriented infrastructures, expand existing challenges and introduce new challenges. Traditional techniques, which focus on individual programs/systems, are no longer effective.

Another important trend is to develop novel infrastructure for trusted and confidential computing. As many applications targeting AI workloads, scientific discovery, and classical supercomputing applications are being moved to cloud environments, the end user relegates many security and trust related functions to the cloud provider, and hence has to rely on that the cloud provider does not intentionally or unintentionally violate that trust. However, such blind trusts are no longer viable given that a whole class of applications and use cases emerges that requires a higher degree of verifiable trust, confidentiality and integrity, largely driven by regulatory or business requirements to keep data confidential. Overall, we identify the following challenges:

- Correctness and fault-tolerance of scalable infrastructures is difficult. Modern cloud infrastructures like Kubernetes are composed of loosely coupled microservices (called controllers), each independently reconciling the system to its desired state. However, without strong consistency for scalability, implementing correct, fault-tolerant controllers is challenging [110]-[111]. Our work [112]-[114] discovered hundreds of critical bugs in existing controllers.
- Software dependencies and cross-system interactions are of immense complexity. Cloud platforms are orchestrations of independent, interacting systems. The reliability of their interactions is critical to whole-system dependability, yet are hard to validate. Traditional testing and verification techniques can hardly reason about interdependent systems collectively. Cross-system interaction failures become dominating failure patterns today [115].
- New fault domains are exposed to cloud-based applications. Although the cloud-based, serverless programming model significantly simplifies application development, it makes application reliability much more challenging due to more complex fault domains [116]. Seatbelts and airbags are needed by emerging applications.
- Automated system operations become single points of failures. Management of large-scale production systems are automated by programs (e.g., Ansible playbooks and Kubernetes operators) and increasingly with AI-based policies. Their reliability is concerning – software bugs and misconfigurations can easily lead to disastrous consequences [113,114,117,118].
- AI/ML robustness and interpretability are critically important when used for critical operations or system components. With AI being increasingly used to make runtime policies (AI4Systems) and to automate operations (AIOps), their robustness becomes critical.
- With the complexity of the system and their interactions, security and confidentiality continue to be grand challenges. Bandages like bug fixes may not yield a trustworthy system.
- Traditional reliability and security problems are still unsolved and become more so under AI-generated code and configurations. Software correctness like safety and liveness are still highly desired building blocks for modern systems and infrastructures [119]. IBM and Illinois have worked together for system-level security with AI accelerators [125]-[126]. However, security challenges still remain when various workloads of diversified types and sizes need to share the multi-cloud system efficiently in a dynamic fashion. Sophisticated sharing scenarios, such as a large foundation model shared across different users with different down-stream tasks, make the situation even more complicated.
- Multi-cloud systems pose new challenges. While attestation and integrity measurements only establish trust at load in time, multi-cloud systems pose new challenges to provide continuous attestation capabilities and handling unreliable connections to remote systems. Attestation of dynamic systems usually leads to a large number of false positive alerts. Practicality of whole-stack integrity attestation at scale depends on our ability to drastically reduce false positives while maintaining high detection rate of true security threats.

Our long-term vision is to build truly reliable and secure cloud systems and infrastructures that can meet the robustness, dependability, and security requirements of next-generation cloud computing paradigms. We construct the following research directions:

- Formally verified, provably correct cloud infrastructures. Formal verification of software and hardware systems was considered moonshots, but has become increasingly practical not only due to the maturity of the toolchains but also thanks to modular architectures that enable us to verify components individually and compositionally. In the Anvil project [121], we show that it is possible to develop formally verified Kubernetes controllers with feature and performance parity to existing controllers. We believe that compositional verification techniques can be developed to verify the interactions and dependencies of interacting controllers. Certainly, it is impractical to replace all existing systems with a formally verified form at once. We envision using model checking and principled testing to continue hardening existing components as what we have been doing in projects [112]-[114][122]. We believe that it is possible to reason about compositions of formally verified components and model-checked/tested ones.
- Cloud-native applications made reliable. We have shown in the Rainmaker project [116] that modern cloud-based applications are not ready for cloud-native environments and programming models due to new fault domains, error-propagation patterns, and state management. Moreover, existing cloud service APIs make application reliability harder, e.g., the lack of idempotency leads to inevitable semantic violations. Rainmaker only shows a specific type of cloud-based applications (applications that use cloud services) and modern cloud-native applications are much more diverse with a broader scope. Unfortunately, the system principles of reliable cloud-native applications are unclear and require research. We envision research efforts that holistically understand the reliability challenges of cloud-native applications and automated tooling to check the reliability of such applications.
- Safe and reliable automated and AI-based operations. The vision of self-driving infrastructures relies on safe and reliable operation programs. In the Acto and Sieve projects [112]-[114], we show that today's operation program, as exemplified by Kubernetes operators, are significantly lacking in reliability and fault tolerance. Similar results are reported by other studies [117] as well as studies on Ansible Playbook [118]. We believe that the problem will become even more emergent given the trend of AIOps and AI-based codegen. We envision principled techniques to validate (if not verify) the safety and reliability of operator programs including those with AI-based policies. Besides ongoing efforts on testing and model checking like Acto and Sieve, we envision research on AI/ML that is specialized for safe and reliable AIOps with robust and interpretable models.
- Agents and LLMs for incident management. With the advance of AI technologies such as LLMs, agent-based incident management is a promising direction. We envision to use LLM/agent-based techniques throughout the cycle of incident management, such as fault detection, root-cause analysis, and failure mitigation. Innovations are needed to effectively leverage the power of LLMs (Section 4.2). The key challenges include effective prompt engineering to provide useful input data such as logs, traces, and metrics and to constrain the search space as well as to validate the correctness and safety of AI-generated code and policies.
- Security from the ground up. We will take security as a first principle throughout system design and implementation, instead of fixing bugs and vulnerabilities with bandages. We plan to continue investigating safe, expressive extensibility of existing systems and infrastructure such as kernel extensions and controller extensions. We also plan to investigate confidential computing technologies to harden the interactions of different systems across software-hardware boundaries for different tenants and protection domains. We anticipate that new hardware features will provide stronger isolation through hardware reconfiguration and partitioning and plan to attest the entire software stack on top of secured hardware in a continuous fashion to ascertain that each software component is indeed the one intended to run and that it has been verified through the software supply chain. This also requires secure interactions between hardware and software in the cloud system.

- Confidential computing with dynamic workload distribution. Distributed applications require orchestration of multiple security capabilities in compute, networking, storage, key management, and access control. Therefore, confidentiality must be considered as a workload abstraction/orchestration artifact to be built on top of our transformed hybrid cloud system. Finally, we need to develop new security and confidential computing solutions to work with workloads of diversified types and sizes that share the multi-cloud system in a dynamic fashion, aiming for guaranteed security and SLOs.

Figure 18 illustrates our vision. We take an incremental approach towards a verified cloud platform starting from existing ones (Kubernetes). We will verify and validate individual components of Kubernetes. For controllers that follow the state-reconciliation principle, we will write formally verified ones to replace the existing ones using the technique described in [120,121,123]. Certain controllers like the scheduler may need more advanced approaches for verification and we will continue validating the existing ones using approaches similar to [112]-[114]. For components like etcd, we will use model checking to verify them (see [122]). We will then investigate compositional verification to verify multiple controllers and the entire platform collectively. The managed applications will be done in a similar fashion and the key is to model and specify the application-platform interface. A key challenge is to accommodate code and policies generated by LLMs (e.g., using LLMs to generate simple operators). We will test and model check the code and investigate ways to generate formally verified code with proofs. Finally, we will build hardware-software solutions to support continuous run-time integrity monitoring and attestation (see [124]) to ensure trustworthy software and environment.

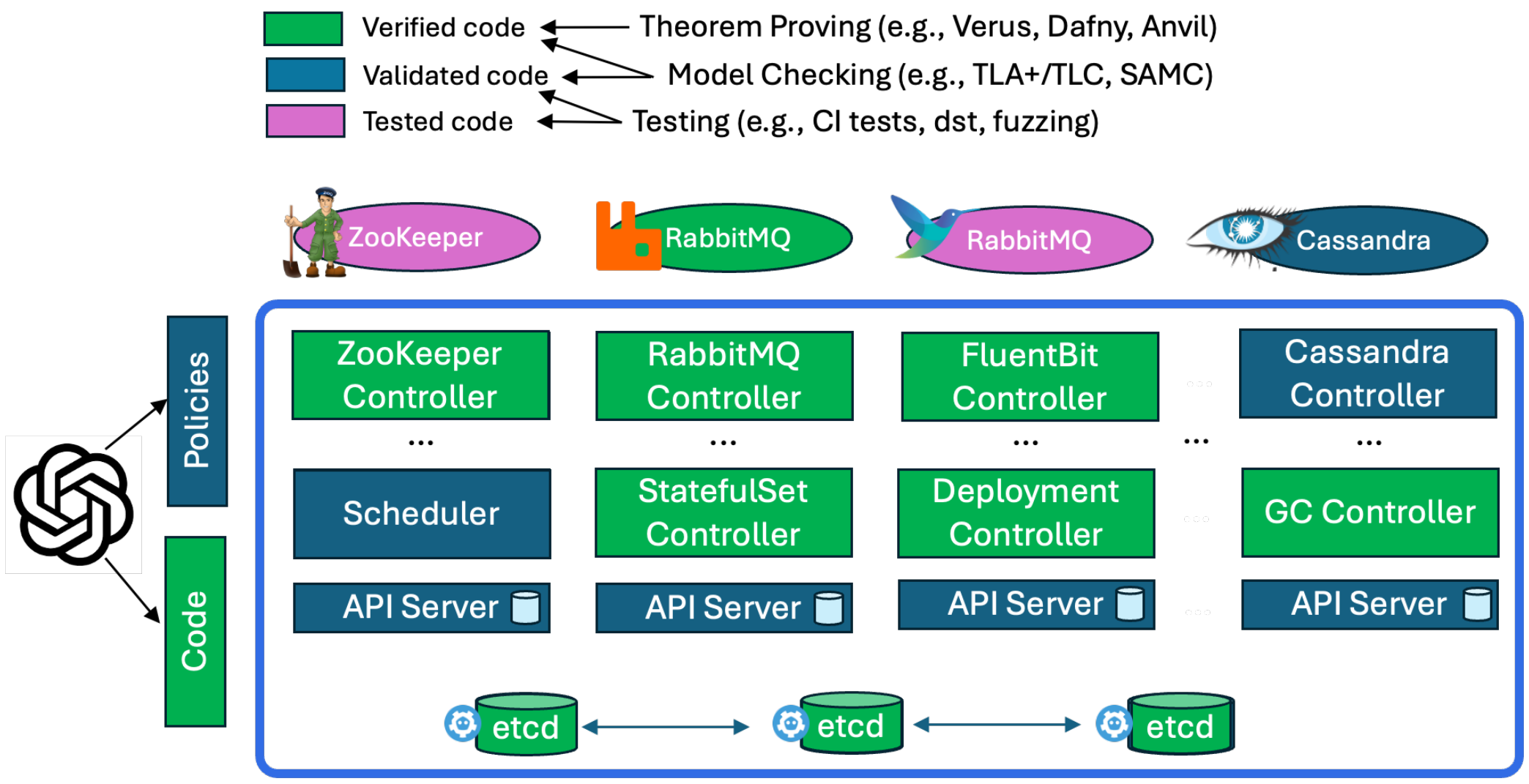

*Figure 18: A roadmap towards a truly reliable and secure cloud infrastructure.*

### 4.8.2 Energy Optimization and Sustainability

Datacenters today consume more than 2% of the total U.S. power and emit even more carbon than the aviation industry. Hence, it is of crucial importance that datacenters, consisting of hybrid multi-cloud infrastructures, look for energy-optimized and sustainable solutions. The grand challenge is to develop holistic, integrated, and sustainable solutions under highly dynamic and constantly changing AI workloads' demands, running on large scale heterogenous hybrid multi-cloud systems, and powered by a mixture of renewable and non-renewable energy sources. We are exploring multiple solutions ranging from energy optimization and sustainability solutions of individual servers, single and multi-cloud systems within datacenters to renewable-energy-based data centers.

*Research Challenges:* To solve the grand challenge, we need to dissect the challenge into sub-challenges, using the 'divide-and-conquer' approach ranging from enabling highly efficient energy optimization on individual servers of the hybrid cloud, energy optimization on multi-cloud systems and energy-carbon optimization among datacenters.

The individual serving nodes in hybrid cloud are facing a significant challenge of heterogeneous hardware including multiple CPUs, GPUs, and complex memory structures that run AI workloads such as Gen-AI, and federated learning and consume large amounts of energy during their training phases. Individual servers' AI model training involves intensive cooperation across multiple hardware processing units, i.e., CPU, GPU, and memory controllers, hence we need optimized local training control for energy-efficient learning workloads.

The hybrid multi-cloud systems are facing a significant challenge in terms of holistic resource management that needs to integrate workload scheduling, placement, server power state control, datacenter cooling system controls, and renewable versus non-renewable managements to optimize various multi-cloud systems adaptively at different scales due to a huge solution space.

The challenge of using renewable energy for datacenters is their variability across time and space. For example, solar power production varies across time and geographical locations. To address this issue, the community proposed to co-locate data centers with renewable farms and powering datacenters with renewable energy. However, this approach also has its own challenges such as achieving stable energy production of renewable energy farms for co-located datacenters and efficient placement and migration of computational workloads.

Failure mitigation and service recovery protocols in cloud systems are developed to handle various hardware and software failures (e.g., network link failures and power outages). However, classic system resilience introduces disruptions to power optimization by incurring additional energy consumption (due to redundancy, migration, and checkpointing). In addition, classic system resilience does not consider the impact of errors of ML inference, out-of-distribution situations, and data/model uncertainties in the ML inference engines that are increasingly integrated with today's cloud systems. Co-designing power and resilience management is required to provide fast failure recovery and differential treatment to critical/non-critical services to minimize disruptions while optimizing carbon footprint.

*Representative Contributions:* We have multiple representative contributions within the Hybrid Cloud Thrust of the IIDAI institute.

Bayesian Optimized Local Training Pace Control for Energy-Efficient Federated Learning: To overcome the challenge of individual node's heterogeneous processing units (GPU, CPU, memory, and their expensive energy usage during AI models training phases, we investigated an energy-efficient training pace control framework, called BoFL, of federated learning (FL) workloads over multi-axes of DVFS (Dynamic Voltage and Frequency Scaling) configurations [130] and with energy-efficient FL straggler handling (FedCore) [131]. The training speed and energy efficiency can be drastically affected by different operational frequencies of CPU, GPU, and memory controller. BoFL develops three techniques to reach proper DVFS configurations: (1) BoFL operates in an explore-then exploit manner, i.e., in limited rounds of FL tasks, BoFL first explores the DVFS configuration space with a few trials, and then exploits the remaining rounds with the best configurations observed; (2) BOFL strategically explores the large configuration space with multi-objective Bayesian optimization (MBO) framework which searches for a set of Pareto tradeoffs in the energy-latency performance space efficiently in just a few steps; (3) BoFL uses a safe exploration algorithm to make sure every training deadline is being met. Experiments on multiple real-world hybrid cloud edge nodes with different FL tasks show that BoFL reduces energy consumption of model training by 26% and achieves near-optimal energy efficiency.

Deep Graph Reinforcement Learning-based Holistic Multi-Cloud Resource Management: To overcome the challenge of providing a holistic multi-cloud resource management, we investigated a hierarchical resource management framework [127]. This framework utilizes a two-level graph combinatorial optimization. The top level involves job-to-datacenter graph mapping and datacenter cooling management. The bottom level includes pod-to-server graph mapping, server, or rack power state controls and renewable versus non-renewable energy usages. Graph neural network (GNN) and reinforcement learning (RL) solvers are employed at both levels to collaboratively optimize the overall multi-cloud system. Furthermore, state-of-the-art (SOTA) RL-safe techniques are applied to establish safety decision boundaries towards minimizing risk of suboptimal decisions and maximizing overall reliability. The GNN-RL framework has been validated through simulations and real-world K8 experiments using AI workloads such as LLM and diffusion models. The preliminary results on single cloud without cooling system control show savings by 2.13 times when compared to previous solutions on 100-1000 server systems.

Towards Building Scalable Modular Data Centers with Renewable Energy: To overcome the challenge of renewable energy-datacenters co-locations, we defined renewable energy-based modular data centers (MDCs) where MDCs can elevate the heavy use of batteries and power transmission lines and potentially achieve zero carbon emissions [128]. The MDCs deploy the SkyBox novel solution, a framework that uses a holistic and learning-based approach to enable efficient use of renewable energy at scale. SkyBox develops three important techniques: (1) SkyBox quantifies the coefficient of variation of the power production of renewable energy farms based on historical traces. This allows the system to identify sites with relatively stable energy supply and capacity; (2) SkyBox includes a dynamic subgraph identification algorithm that groups individual renewable energy sites into subgraphs with complementary power production patterns within the same subgraph. Each subgraph then represents a stable aggregated power production as a whole and allows datacenter operators to decide the sites where MDCs will be deployed; (3) SkyBox includes a Mixed-Integer-Programming model for enabling optimized placement and migration of workload, encapsulated within virtual machines (VM) [129]. The evaluation results of SkyBox show that the carbon footprint of MDCs can go down by 46% with low VM migration frequency in comparison to conventional datacenter deployment approaches.

*Long-Term Directions and Vision:* Hybrid multi-cloud systems must be considered within a holistic framework if efficient energy optimization, sustainability, and robustness should be achieved. The grand challenge still stands since

1) We need to consider energy and carbon emission optimization within (a) the individual server nodes with their heterogeneity of hardware and system platforms supporting diverse workloads, (b) clusters of heterogeneous nodes forming datacenters, (c) multiple datacenters geographically distributed.
2) We need to consider very different and fast changing AI models that are used (a) by diverse cloud applications ranging from federated learning, CNNs, Deep Learning to transformers, attention models and others, and (b) by underlying systems themselves ranging from Graph Neural Networks (GNN), Reinforcement Learning (RL) and others.
3) We need to consider diverse energy sources ranging from renewable sources to non-renewable energy sources which bring a major challenge of variability in energy supply and need for efficient migration of workloads.
4) We need to co-design power and resilience management to provide fast failure recovery and differential treatment to critical/non-critical services to minimize disruptions while optimizing carbon footprint.

Hence the long-term direction and vision should be to investigate:

- Cooperative energy-optimization approaches between techniques: (a) AI for Systems: AI techniques used for hybrid cloud systems that aim to optimize energy usage for AI workloads (e.g., GNN-RL within cloud resource management) and (b) Systems for AI: system techniques used for AI workloads in cloud applications and their energy optimization (e.g., DVFS for FL training workloads). At this point the community optimizes mostly the AI workloads in cloud applications and develops systems for AI but discards the usage of energy in AI algorithms used within cloud systems.

- Energy optimization and sustainability when AI workloads such as LMM bring exponential growth of data and model sizes and infrastructures: We will need energy-aware AI model compression, distillation, and other reduction techniques.
- Fine-grained energy/ carbon measurement tools: Kepler provides energy usage measurements at coarse level of cloud systems, but as some of the studies show, finer-grained energy optimizations of hardware-software configurations could yield further energy savings. Carbon tools are almost non-existent and available only at very coarse levels of the datacenters [132]-[133]. We need new energy/carbon tools.
- Sustainable modular hardware and software: (a) high churn in the supply chain increases carbon footprint of devices causes throw-away mentality and repair and replacement of computing components becomes difficult; (b) mismatches between software component upgrades and between software and hardware lifecycles cause that devices are often retired or put offline because software support has ended even though the hardware has significant usable lifespan left [134]. We need modular and energy efficient hardware/software architectures and tools to assist with aging components in an energy-efficient manner.
- Split reward models and failure recovery acceleration for SLO-aware energy optimization: Split reward functions allow the ML models to learn differential policies under various failure recovery procedures and for applications with diverse levels of criticality. We need to coordinate power management and resilience management to minimize disruption to energy optimization. This requires multidisciplinary work that brings together power systems and cloud systems engineering to achieve significant progress towards dependable green computing.
- Access to cloud testbeds with renewable energy and access to corresponding datasets/traces: We need to get access to testbeds and corresponding datasets/traces of AI workloads, system resource usage, carbon usage from cloud providers to develop realistic solutions for energy optimization and sustainability.

## 4.9 Application-Adaptive Cloud System for Dynamic AI Workloads

### 4.9.1 Introduction and Challenges

Inspired by the application-specific design paradigm such as ASIC (Application-Specific IC) or FPGA (Field-Programmable Gate Array), one bold strategy is to build an application-adaptive hybrid cloud system. Just like how an ASIC or an FPGA design can beat the general-purpose CPU in terms of performance and energy efficiency by a large margin (sometimes by 100-1000x), we envision that an application-adaptive cloud system would demonstrate a large advantage over the conventional general-purpose system as well. However, such a system should still be flexible enough to serve different types of workloads, being highly adaptive to their different characteristics, priorities, and demands. The key idea to achieve both application-specificity and flexibility is through reconfiguration and programmability at different levels of the new cloud system (Figure 6).

The current cloud system has some limited ability for reconfiguration and programmability. We propose new reconfiguration and programmability capabilities at different levels of the cloud system and significantly improve future cloud's capabilities to reconfigure, reprogram, and adapt itself to serve dynamically changing workloads. Such design innovations would range from a single node, to a single rack, to the entire infrastructure, including interconnections of different levels, and all the way to cloud platform, middleware and programming models. The goal is to provide specialized solutions adaptive to the specific needs of a workload, delivering high computation and data management efficiency, high resource utilization, shortened latency, and higher level of SLO guarantees, leading to reduced cost and improved affordability.

The critical challenges to achieve this vision will include overcoming the limitations of current reconfiguration and programmability capabilities in cloud systems. Achieving the desired balance between application specificity and flexibility will require innovations across multiple layers of the cloud stack. At the hardware level, ensuring the seamless integration and co-design of technologies like CXL, programmable SmartNICs, and reconfigurable accelerators will be key. At the higher cloud system levels, managing dynamically changing workloads while optimizing for energy efficiency, resource utilization, and performance will require advanced orchestration and scheduling algorithms. Furthermore, enabling cross-layer programmability and reconfiguration across compute/storage/networking devices and platform/middleware frameworks will be essential to creating a system that can adapt in real-time to workload requirements. Finally, security, robustness, and maintaining SLOs across diverse applications, ranging from lightweight services to large-scale AI tasks, will present additional challenges that need to be addressed for sustainable, scalable, and adaptable cloud systems.

### 4.9.2 Existing Solutions and Contributions

In previous sections, we have already demonstrated advantages through specialization and adaptivity, such as those described in LLMaaA (Section 4.3), foundation model optimization (Section 4.4.4), and edge transformation (Section 4.7). We will introduce application-adaptive solutions and contributions focusing on other system layers in this subsection.

At the programming model/platform levels, we have already demonstrated an adaptive HPC-cloud bursting system [62]-[63] that seamlessly places workloads across multiple compute systems. In [62], our system integrates automated data management with learning-based scheduling at the function level, using a dynamic label-based design. It automatically prefetches data files based on demand and detects data movement and execution patterns for future scheduling decisions. In [63], our framework demonstrates advantages in several aspects: users can provide their own cloud resource; the framework provides the Python-level abstraction that does not require users to interact with job submission systems, and allows a single Python-based parallel workload to be run concurrently across an HPC cluster and a cloud system. In [64], we develop task splitting rule to set the level of parallelism dynamically depending on task duration, cluster capacity and carried load at the time of task arrival. Our experiments demonstrate a better balance of weighted workflow completion time and resource utilization compared to the existing heuristics.

At the middleware/runtime level, multiplexing of compute resources across microservices is still challenging in production because contention for shared resources can cause latency spikes that violate the SLOs of user requests. We offer a new solution, FIRM, an intelligent fine-grained resource management framework for predictable sharing of resources across microservices to drive up overall utilization [144]. FIRM leverages online telemetry data and machine-learning methods to adaptively (a) detect/localize microservices that cause SLO violations, (b) identify low-level resources in contention, and (c) take actions to mitigate SLO violations via dynamic reprovisioning. Experiments across four microservice benchmarks demonstrate that FIRM reduces SLO violations by up to 16x while reducing the overall requested CPU limit by up to 62%. Moreover, FIRM improves performance predictability by reducing tail latencies by up to 11x.

To delve deeper into the cloud infrastructure, reconfiguration and adaptivity can take place at the top-of-rack (ToR) Switch [135]-[136]. Meanwhile, as the backbone technology, software-defined networking (SDN) allows network operators to configure and manage network resources through programmable switches. The programmable switch usually has reconfigurable hardware such as a programmable ASIC that supports domain-specific languages like P4. Similar to SDN, software-defined flash (SDF) enables upper-level software to manage the low-level flash chips for improved performance and resource utilization, and a programmable SSD can be virtualized into two types of vSSDs: hardware-isolated vSSDs, and software-isolated vSSDs. Both SDN and SDF share a similar architecture – the control plane is responsible for managing the programmable devices, and the data plane is responsible for processing I/O requests. As an initial study, we envision to integrate and co-design both SDN and SDF and redefine their functions in a coordinated fashion to improve the efficiency of the entire rack-scale storage system. Our initial results demonstrate a reduction of the tail latency of I/O requests by up to 5.8× over state-of-the-art rack-scale storage systems [135].

RDMA and SmartNIC highlight the reconfiguration potential of inter-host connections in the cloud. While RDMA emphasizes low-latency memory access protocols, SmartNIC [137] concentrates on near-NIC computations. The former prioritizes data transfer performance, while the latter offloads computations from the CPU to dedicated hardware near the NIC. These distinctions influence the configurations that the cloud adopts for various workload purposes. In our initial study, we have developed UniNet [60], a SmartNIC-based solution that offloads network functions efficiently and is able to support all the major Container Network Interfaces (CNIs). UniNet boosts CNI throughput by an average improvement of 7.08×, cuts tail latency by 41.6%, and reduces CPU usage by up to 5.6× for RX and 4.02× for TX.

At the accelerator level, GPU-FPGA and GPU-GPU P2P protocols and GPU-SmartNIC direct connections [138] are pivotal. They empower the accelerator communications with reconfigurability, ensuring adaptability especially when faced with intra-host data transmission bottlenecks. Tools like DPDK further reduces latency by bypassing the Linux kernel, and MIG (Multi-Instance GPU) has made GPU virtualization and reconfiguration more achievable with potentially large benefits on performance gains through adaptivity [139]. A recent work [126] introduced reconfigurable TPUs based on switchbox-enabled systolic arrays to support rapid dynamic partitioning and re-partitioning of the TPU to adapt to changing ML workload characteristics, achieving up to 42.1% higher performance for realistic ML inference workloads.

FPGA accelerators can be dynamically reconfigured. Our recent work demonstrated their great adaptivity in a cloud setting through FPGA virtualization [140] and shared virtual memory system [141]. Our work described in [142]-[143] represents an exciting new direction for high-level synthesis (HLS) that raises the design abstraction level even higher than what conventional high-level languages can offer. For the first time, we could directly take large PyTorch models, go through multiple levels of optimizations, and generate high-quality hardware designs to be mapped to FPGA accelerators that even outperform RTL-based designs, thus significantly improving programmability and design quality of FPGAs. Such a new framework represents a transformative and automated compilation methodology for the future to map various AI models quickly and flexibly to different types of accelerators in the cloud.

### 4.9.3 Future Vision and Directions

The future of adaptive cloud systems lies in creating even more adaptive and intelligent platforms capable of dynamically distributing workloads based on demand, resource availability, and real-time performance metrics. Key research areas could include the development of more advanced data management systems that go beyond prefetching, incorporating predictive analytics for optimal workload and data placement, and enhancing the ability of platforms to operate across hybrid infrastructures seamlessly.

Moreover, exploring how such frameworks can adapt to varying task complexities, task arrival times, and hardware heterogeneity could lead to more efficient resource utilization and improved completion times for complex workflows. Advancing these frameworks to support the newly proposed LLMaaA abstraction could change the whole application programming landscape and significantly improve the development productivity. Additionally, further innovations could focus on scaling these platforms for handling highly parallel AI/ML workloads, allowing them to split tasks intelligently between various hybrid-cloud resources based on workload types, priorities, and cost-efficiency.

Future research could also investigate more granular levels of resource management, integrating unified software-hardware and cross-layer control technologies to adapt the allocation of not only CPUs and accelerators but also memory, network, and storage resources in real time. Additionally, designing middleware systems that can balance SLO guarantees for diverse, co-located workloads, including AI and data-intensive tasks, will be crucial for improving utilization without compromising performance. Extending these solutions to handle edge-cloud continuum environments, where resource variability and latency tolerance are even more critical, represents another exciting area of research.

Scaling up SmartNiC, cache-coherent-interconnect, and software-defined-interface solutions to large-scale cloud systems will be an important next step. One specific idea is to extend the SmartNiC's functionality so it behaves as a smart and distributed controller with compute capabilities for connecting and managing different ranges of hardware resources dynamically to meet the changing workload demands. Specifically, one focus area could be on foundation model workloads where a wide range of users may have different requirements in terms of throughput, latency, and security with diverse cost preferences and constraints.

In addition to compute nodes, storage nodes also present numerous reconfiguration opportunities. Innovations like SmartSSD and vSSD introduce programmability and configurability, allowing SSDs to be more byte addressable and with better provisioning capabilities. Moreover, they provide better flexibility and facilitate integration with cross-level acceleration opportunities in the cloud.

In summary, we envision to transform the future hybrid cloud into a dynamic, adaptable, and smart system, being able to coordinate different levels of reconfigurations in a coherent way for achieving up to 100x higher performance gains. This new capability will keep future cloud systems at the cutting edge, allowing them to meet the changing needs of AI-driven workloads. Its flexibility enables the cloud to easily shift from handling lightweight ML models to managing more intensive tasks, such as LLMs.

# 5 Applications Powered by Our Vision

Applications drive requirements for the future hybrid cloud platform for emerging AI workloads. From the need for training and fine-tuning of foundation models with different modalities, to infusing AI in simulations, and combining other HPC computations, these applications shape the capabilities of the cloud system to enable advances in multiple domains. In this section, we specifically target two important AI-driven scientific computing applications: material discovery and climate and sustainability.

## 5.1 Materials Discovery

### 5.1.1 Introduction

Materials discovery is pivotal in a wide range of industries, from renewable energy to electronics and pharmaceuticals. For example, solar cells require materials with high light absorption efficiency, mechanical durability, and low environmental toxicity. These characteristics stem from the material's atomic and molecular structures at both microscopic and macroscopic scales. Traditionally, the design of such materials has been a labor-intensive and time-consuming process, largely driven by trial and error and the intuition of material scientists. The parameter space for designing new materials is vast, with only a fraction of potential configurations yielding desirable properties. Consequently, developing new materials can take 10-20 years and cost between 10 to 100 million US dollars [146]. These conventional methods struggle to cope with the expansive design space, making the process of materials discovery ripe for disruption through the integration of artificial intelligence (AI) and machine learning (ML).

In recent years, AI has emerged as a powerful tool to accelerate various stages of materials discovery, including property prediction, material synthesis, and even the generation of new material structures. For instance, predicting properties such as electrical conductivity is crucial when developing materials for electronic devices like organic light-emitting diodes (OLEDs). These prediction tasks are typically framed as regression problems, where neural networks are employed to predict continuous values of specific material properties. Appropriate representations of material structures, such as molecular graphs, crystal structures, or polymer chains, are critical for these tasks [154]. Molecular graphs, for example, map atoms as nodes and bonds as edges, while crystal structures capture atomic configurations within a lattice. These structural representations, combined with spectroscopic signatures, provide essential input data for predicting the physical and chemical properties of materials.

The emergence of foundation Models (FMs) presents a new frontier in materials science by offering the ability to generalize across multiple tasks using multimodal datasets. Unlike conventional machine learning models that are task-specific and typically trained on unimodal datasets, FMs are pretrained on large, diverse datasets and can be fine-tuned for various downstream applications with minimal additional training [153]. In the domain of materials science, data comes in many forms, including molecular graphs, three-dimensional atomic configurations, and spectroscopic data, among others. By training FMs on this rich array of multimodal data, the models can learn more generalized and transferable feature representations, which can be applied to a wide range of tasks, such as property prediction, inverse material design, and chemical synthesis.

One of the key advantages of FMs is their ability to transfer knowledge across domains. Materials science encompasses a vast array of disciplines, from chemistry to physics, and FMs trained on multimodal data can bridge these domains, creating opportunities for cross-disciplinary innovations. The promise of FMs lies in their ability to unify disparate data types and tasks, thereby accelerating innovation across industries that rely on materials discovery.

### 5.1.2 Recent Advances

Several advancements in AI have contributed to the growing significance of FMs in materials discovery. The development of toolkits such as the Generative Toolkit for Scientific Discovery (GT4SD) [147] and the Open MatSci ML Toolkit [144] have laid the foundation for combining models specialized in single tasks into more comprehensive FMs. Both toolkits provide a modular framework that enables the integration of large materials datasets with state-of-the-art AI models. These frameworks support high-throughput experimentation and the training of models that have demonstrated considerable success in tasks like property prediction. By leveraging diverse datasets and advanced architectures, both toolkits facilitate a wide range of materials science applications, from predicting material properties to classifying materials based on their structural and chemical characteristics.

The introduction of MolFormer [149], a transformer-based molecular embedding model, was inspired by advancements in unsupervised transformer models in natural language processing and for chemical reaction predictions [150]-[152]. This model utilizes rotary positional embeddings and a linear attention mechanism, training on large-scale datasets such as SMILES sequences of 1.1 billion unlabeled molecules from PubChem and ZINC. MolFormer demonstrated higher performance compared to existing baselines, and the attention-based analyses revealed that the transformer architecture was able to learn spatial relationships between atoms within a molecule, providing strong predictive capability for various molecular properties, including quantum-chemical properties.

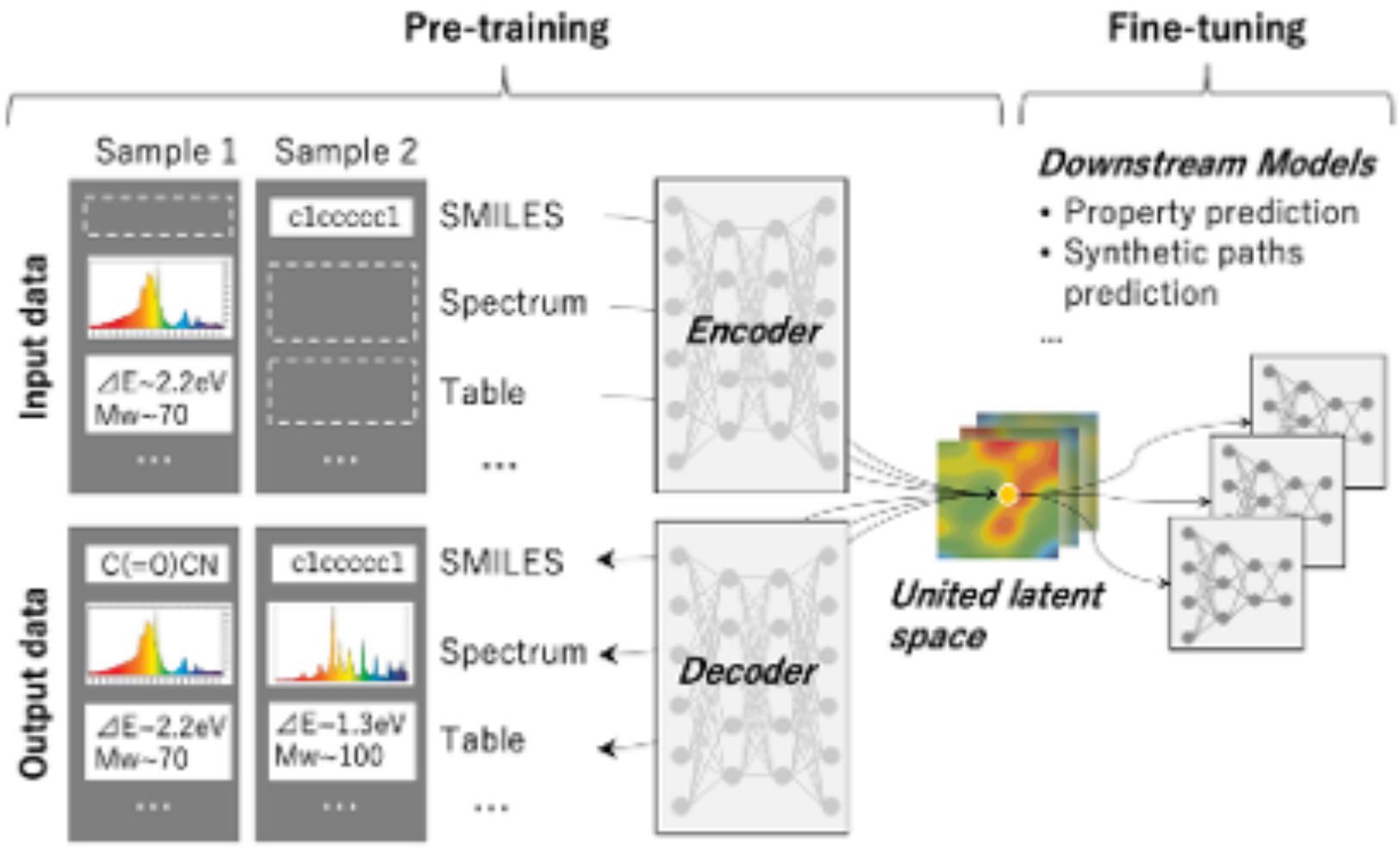

*Figure 19: Prototypical Architecture for a FM for Materials Design (reproduced from Takeda et al., "Foundation Models for Materials Science" AAAI-23).*

Born et al. [154] extend the application of transformers by introducing the Regression Transformer. This transformer relies on hybrid sequence representations combining chemical structure and properties into a unique textual representation, casting regression as a language modeling task. This entangled representation trains models via an alternate scheme to learn the molecular structure and its link to chemical properties concurrently, allowing seamless, controlled generation of materials from small molecules to polymers [155]. Takeda et al. [145] argue that a foundational approach for materials science must incorporate these diverse representations into a unified latent space, allowing for cross-disciplinary insights and overarching modeling across different materials science disciplines, much like how human scientists approach materials research. As shown in Figure 19, the model they propose aims to bridge the gap between various data modalities and materials domains, making it possible to apply the learned knowledge across tasks and contexts.

### 5.1.3 Research Challenges and Opportunities

While FMs offer significant promise, several challenges still hinder their broader adoption in materials discovery. One major challenge is the availability of high-quality and diverse datasets, which are necessary to train these models effectively. The vast design space of materials requires models that can generalize across many types of data, yet most existing datasets are limited to specific material domains or properties [145]. This leads to the challenge of integrating disparate datasets into coherent models that can provide meaningful insights across different types of materials.

Another challenge lies in the complexity of material representations. FMs need to handle various types of material data, including molecular graphs, point clouds, and even multimodal data from different scientific domains. Developing new algorithms that can capture the intricate properties of materials is essential [147]. Despite these challenges, the opportunities for FMs in materials discovery are vast. Future research should focus on expanding the available datasets and improving the architectures used to model complex material properties. Additionally, techniques like active learning and high-throughput simulations may help mitigate the scarcity of data, allowing for more accurate predictions and faster discovery processes [147].

### 5.1.4 Representative Contributions

Researchers at both IBM and UIUC have made significant contributions to the field of materials discovery through their work on foundation models. The Framework for Accelerated Materials Development (FAMD) represents an innovative platform integrating data mining, artificial intelligence, and machine learning to address the challenges of electrochemical materials for energy storage and sustainable separations. The FAMD workbench is designed to gather and process critical unsorted data from the literature and feed it into generative models for materials discovery. One of its promising applications is in the discovery and development of electroactive materials with high longevity and capacity, essential for next-generation batteries and energy-efficient ion separations. This platform also aims to tackle the challenges facing redox-flow batteries, such as finding materials with high solubility, fast electrode kinetics, and increased chemical and electrochemical stability. By employing customized material discovery large language models (MD-LLMs), the FAMD workbench enhances the ability to extract relevant information from scientific literature, supporting the accelerated discovery of materials for energy and environmental sustainability.

A significant contribution within the activities of the institute to advancing materials discovery is the augmentation and extension of IBM's RXN [153] foundation model to cover a broad class of step-growth polymerization reactions. This project leverages the Open Macromolecular Genome (OMG), a comprehensive database developed in a previous funding cycle, to generate a dataset of synthesizable polymerization reactions. These reactions are coupled with high-throughput density functional theory (DFT) characterizations of reaction energetics, providing critical physics-based data that was previously absent from IBM's RXN model, probably one of the very first foundation models in material science specialized on predicting chemical reaction and synthesis. By integrating this dataset, the project aims to enhance RXN's predictive capabilities, allowing it to better handle the complexities of polymer chemistry, including steric interactions and reaction free energy profiles. This work significantly expands the scope of RXN, enabling it to predict general step-growth polymerization reactions and thereby facilitating the design of polymer-based materials for diverse industrial applications. The project is well-aligned with IBM's goal of developing generalizable AI models for closed-loop materials discovery and has the potential to benefit the broader polymer materials community.

Building on the subsequent development of RXN [153], an innovative project proposed the development of foundation models to generate accurate retrosynthetic pathways and experimental procedures, integrating chemical synthesis tools with advanced language models, such as Text+Chem T5 and LLaMA2, to enable the seamless generation of both single- and multi-step retrosynthetic pathways for target molecules. As illustrated in Figure 20, the approach combines chemical structure and transformation knowledge from synthesis tools with the contextual understanding of language models to predict experimental procedures with high accuracy. Reinforcement learning from human feedback (RLHF) is employed to improve the quality of generated retrosynthetic pathways, allowing the model to produce results that are competitive with, or even surpass, those generated by human experts. By incorporating retrosynthetic search algorithms such as Retro* and Monte Carlo Tree Search (MCTS), the model can extend its capabilities to multi-step retrosynthesis, thus broadening its application to complex molecule synthesis.

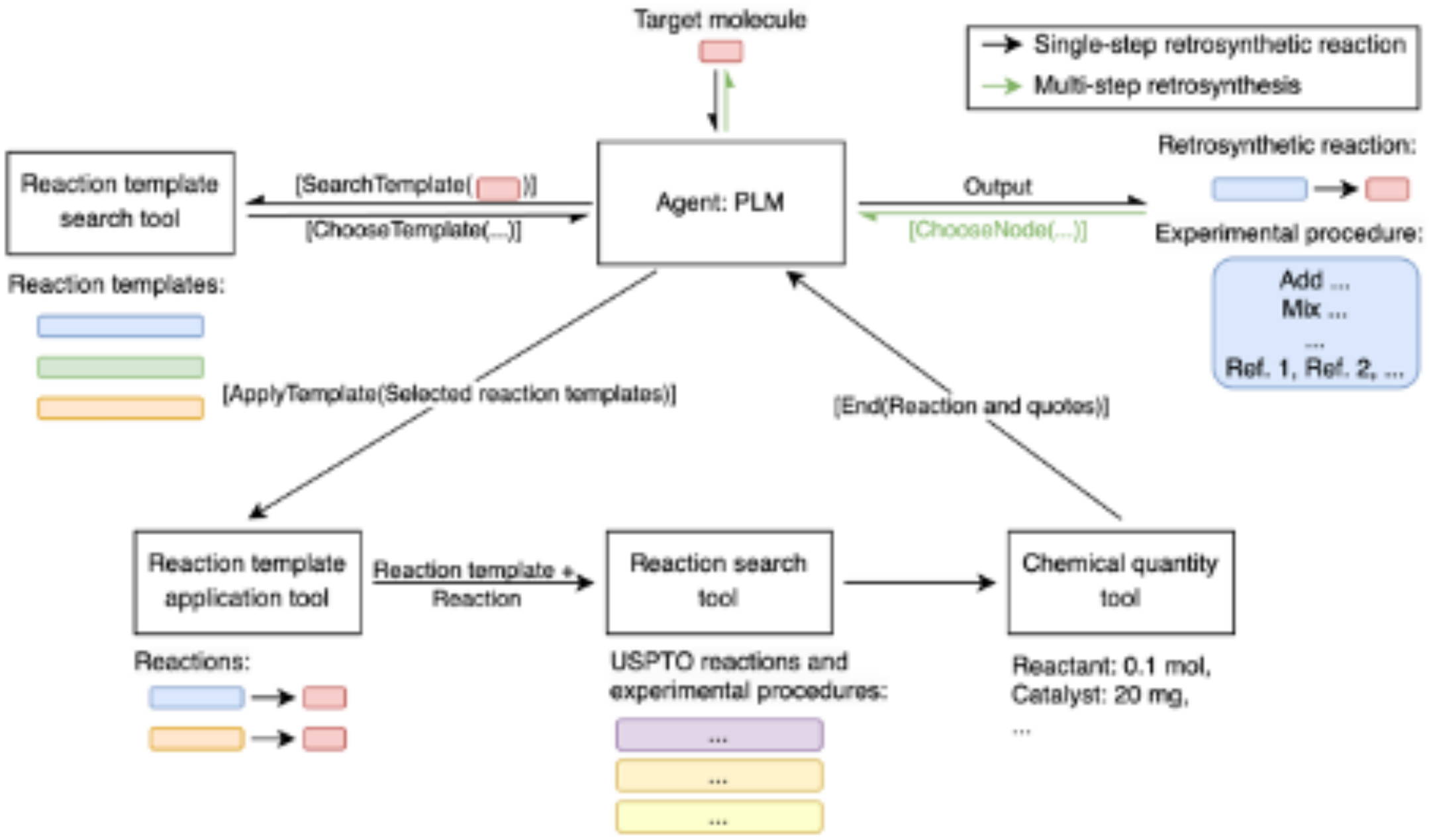

*Figure 20: Overview of foundation models for generating faithful retrosynthetic pathways and experimental procedures.*

### 5.1.5 Long Term Vision

The long-term vision for FMs in materials discovery is to create a unified, overarching framework capable of handling a wide array of material tasks, from property prediction to material generation. This ambitious goal requires the development of FMs that can be trained on vast, multimodal datasets that not only include chemical and physical properties but also structural information, spectroscopic data, and simulations from various domains. By integrating these diverse sources of information through sophisticated fusion methods—ranging from early fusion, where data from different modalities is combined at the input level, to late fusion, where results from individual modalities are merged at the decision-making stage—FMs can unlock a more holistic understanding of materials. This multimodal approach would allow researchers to build models that account for complex interactions between different material properties and design constraints, ultimately enabling the creation of new materials with tailored characteristics such as enhanced electronic conductivity, improved mechanical strength, or superior environmental sustainability.

Additionally, FMs are expected to play a pivotal role in inverse design, a particularly challenging task in materials science where researchers aim to generate novel materials that meet predefined properties. The vast chemical space of possible materials—estimated to include up to 10^60 potential molecular configurations—presents a significant hurdle. However, by leveraging the power of FMs, researchers can explore this space more effectively, narrowing down promising candidates and accelerating the discovery of breakthrough materials.

The development of FMs for materials science is likely to give rise to a range of models designed for different tasks or material domains, such as energy storage materials, polymers, or quantum materials.

These specialized models would coexist, complementing one another and collectively contributing to the accelerated pace of discovery.

Another critical aspect of the long-term vision includes the development of intelligent agents that learn to make use of external computational tools and environments—systems that can handle tasks that FMs alone may not be equipped for. These agents could autonomously recognize when problems require techniques beyond the capabilities of a standalone FM, such as combinatorial optimization, formal verification, or high-fidelity simulations. By integrating these capabilities into the materials discovery workflow, agents could augment the problem-solving capacity of FMs, improving efficiency and accuracy in areas such as the synthesis of novel compounds, the optimization of material properties, and the verification of material behavior in real-world conditions.

Finally, as the scope and complexity of FMs continue to grow, the need for scalable and distributed approaches to model development becomes increasingly apparent. Much like open-source software development, the creation of large-scale FMs will likely benefit from distributed collaborative efforts, where teams across the globe contribute to model development, fine-tuning, and validation. This distributed model of collaboration could accelerate innovation and ensure that models are continuously updated and refined with the latest scientific advancements and real-world data. By fostering an open development ecosystem, the materials science community can ensure that FMs remain relevant, robust, and adaptable to the rapidly evolving demands of materials research.

## 5.2 Climate and Sustainability

### 5.2.1 Introduction

Climate and sustainability represent a very broad topic. As a matter of focus, the Institute has a particular interest in research for climate and sustainability applications which involve geospatial information and data. That is because geospatial is highly relevant for enabling a broad set of climate and sustainability solutions in three key application areas, which are mitigation, adaptation, and measurement & quantifications.

Examples for such geospatial applications in these three areas are nature-based carbon sequestration, flood or wildfire risk and greenhouse gas monitoring for mitigation, adaptation, and measurement & quantification, respectively. To enable these applications, the joint research has been focusing on geospatial foundation models (FMs). Examples of geospatial FMs are by now plentiful. However, IBM Research pioneered jointly with NASA, an initial geospatial FM, which was trained on Harmonized Landsat Sentinel (HLS) satellite data [157]. The work has been extended to include weather data. Different applications and use cases of finetuned geospatial FMs for both satellite and weather data are shown in Figure 21.

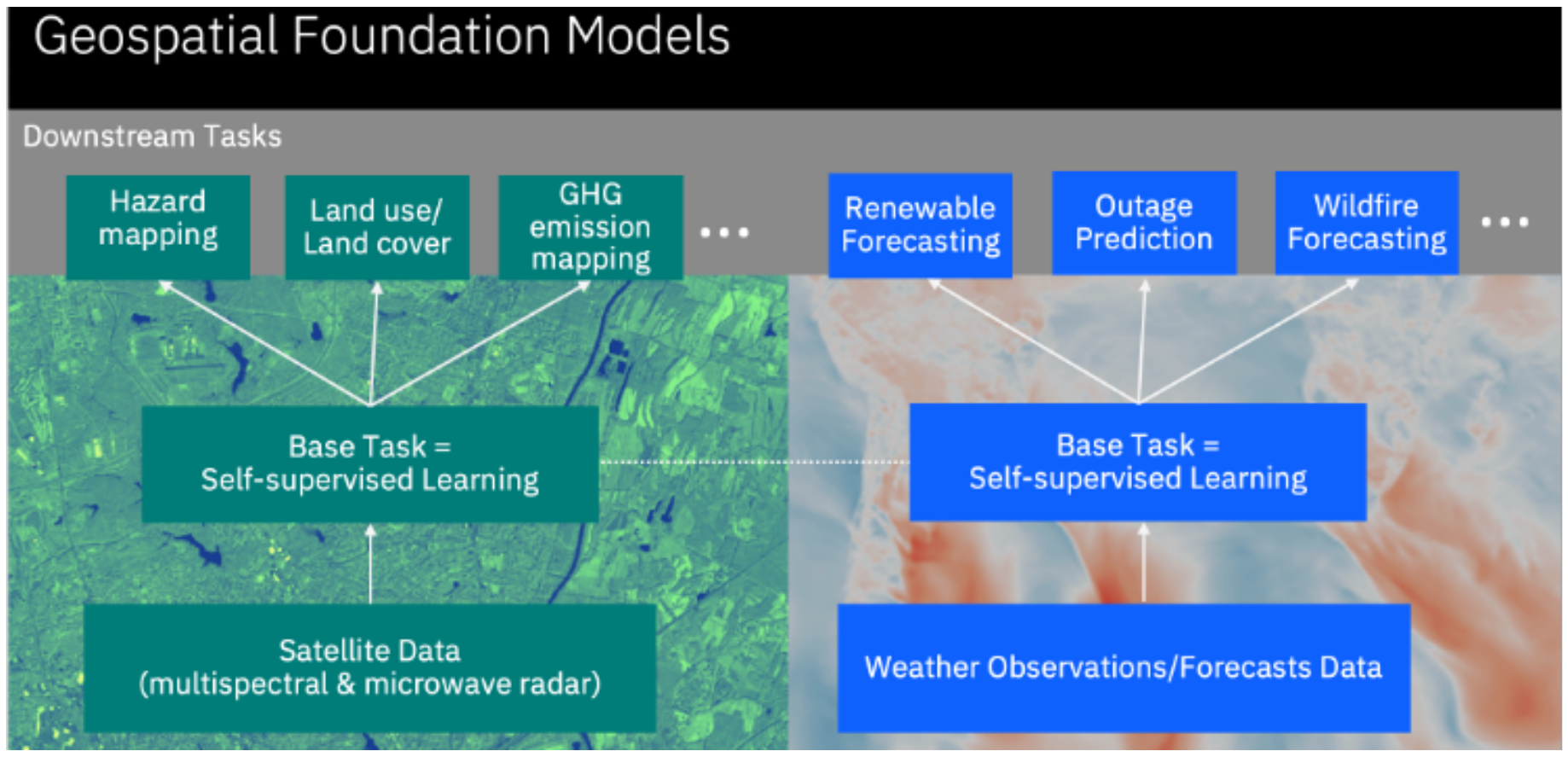

*Figure 21: Use Cases of Geospatial Foundation Models.*

### 5.2.2 Research Challenges and Opportunities

Next, we would like to introduce the identified challenges regarding the design of multi-modal geospatial FMs. Significant efforts have been invested in developing large-scale, data-driven weather models. Current weather FMs, such as GraphCast [158], Prithvi [159], ClimODE [160], SEEDS [161], and FourCastNet [162], are mainly designed to handle single-modal weather data, which leaves room for improvement, as capturing correlations among multi-modal data could further enhance forecasting accuracy. For example, Prithvi focuses on satellite imagery, enabling it to predict masked satellite image regions, but it cannot forecast specific weather conditions in a given location. Similarly, GraphCast, trained on climate reanalysis data, can forecast weather but lacks the capability to extract information from satellite images that might be complementary to the climate reanalysis data. However, integrating multi-modal data is challenging due to the varying spatial and temporal resolutions. GraphCast, for instance, operates on a 0.25° latitude-longitude grid, corresponding to approximately 25×25 kilometers at the equator, whereas the National Water Model provides high-resolution data at grids of 30m/50m/70m. This disparity in resolution presents difficulties in effectively integrating multi-modal data.

### 5.2.3 Representative Contributions

As shown in Figure 22, the initial geospatial FM from IBM Research was pretrained with satellite data using a masked auto-encoder (MAE) [163]. The MAE reconstructs masked images using an asymmetric encoder-decoder architecture with a ViT backbone. Each input image is divided into non-overlapping patches of the same size, and a subset of the patches is randomly masked. The encoder receives only the unmasked patches generating their latent representation. The decoder then receives the latent and masked tokens in order to perform the image reconstruction task. The pre-training task is the reconstruction of masked tokens, for which the loss function is the mean squared error (MSE) between the masked and predicted tokens in the pixel space. To account for the 3-dimensional nature of the input data, 3D positional embedding and the 3D patch embeddings are used.

The initial FM, which was developed by IBM and NASA, was also tuned, to a range of Earth observation tasks that have not been tackled by previous work on FMs involving multi-temporal cloud gap imputation, flood mapping, wildfire scar segmentation, and multi-temporal crop segmentation. Three scenarios were compared, namely 1) fine-tuning the entire model, 2) fine-tuning solely the decoder for the downstream task, and 3) training the model without utilizing the pre-trained weights. The experiments demonstrated that the pre-trained model accelerates the fine-tuning process compared to leveraging randomly initialized weights. In addition, the pre-trained FM compares well against the state-of-the-art on downstream tasks, e.g., outperforming a conditional GAN model [164] in multi-temporal cloud imputation in the structural similarity index. Efficiency of label data was also investigated where the quantity of available labeled data for fine-tuning the model was gradually reduced, which demonstrated that data can be decreased significantly without affecting the model's accuracy.

Based on this initial IBM-NASA work the Institute focusses on three related research questions include

- Novel architectures to address pre-training across all dimensions (i.e., x, y, t, channels).
- Approaches to multi-modal geospatial (vector, raster, text etc.), especially when dealing with large numbers of modalities and channels.
- Composability of geospatial FMs (e.g. how to best compose and combine geospatial FMs such as weather and satellite FMs).

The Institute has researched and developed a framework, which is called Multi-modal Masked AutoEncoder (MM-MAE) which is flexible with the number of input modalities and channels of the geospatial data, as described in Figure 23. This framework entails a variant of a vision transformer (ViT) [165] and a novel model architecture with low-rank spatial-spectral attention blocks which enables effective learning of the relations between spatial and channel information efficiently, as illustrated in Figure 24.

Experimental results demonstrate that the framework surpasses current state-of-the-art multi-modal geospatial FMs, achieving superior performance with less computation and fewer parameters. The flexibility and extensibility of our framework make it a promising solution for future geospatial data analysis tasks that involve a wide range of modalities and dimensions.

To address the resolution mismatch in multi-modal data, we propose to use novel generative models to obtain the unobserved data. However, applying traditional generative models directly can result in suboptimal performance, as these models often assume Independent Identically Distributed (IID) data, overlooking the spatial and temporal dependencies inherent in non-IID geo-spatial data.

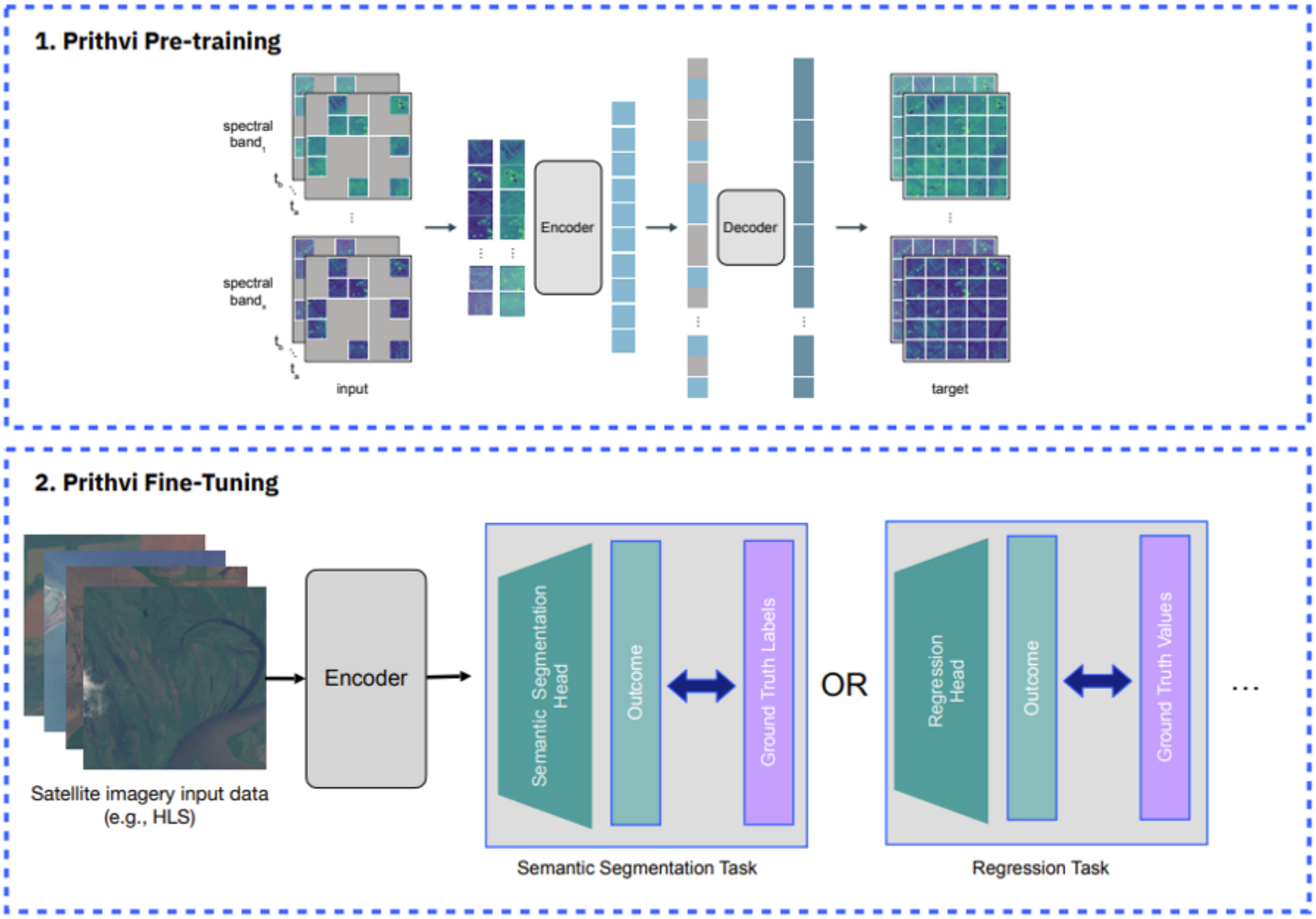

*Figure 22: Masked autoencoder (MAE) for pre-training a geospatial foundation model and finetuning for various downstream tasks.*

To account for these dependencies, we encode the relationships between a geo-location with a low resolution and its spatial and temporal neighbors within the generative model. Unlike traditional generators, which rely on random variables, our proposed generator uses the features of neighboring spatial and temporal locations as well as the representation from another modality to exploit these correlations as shown in Figure 25(a). To ensure the generated features reflect the true distribution, we minimize the difference between the generated and real features based on various discrepancies.

With multi-modal data of the same resolution, the next step involves learning the neural representation of each geo-location. A straightforward method is to use a CLIP-style [166] contrastive loss to maximize the similarity of neural representations across different modalities for a given location. However, this approach overemphasizes the common information shared by different modalities while neglecting their unique contributions to specific weather events. To balance the commonality and uniqueness across modalities, we propose using two projection networks for each modality: one to map the data to a shared representation and the other to map it to a unique representation as shown in Figure 25(b). Contrastive loss [167] is applied only to the shared representations, while an orthogonal constraint is introduced to ensure no redundancy between shared and unique representations.

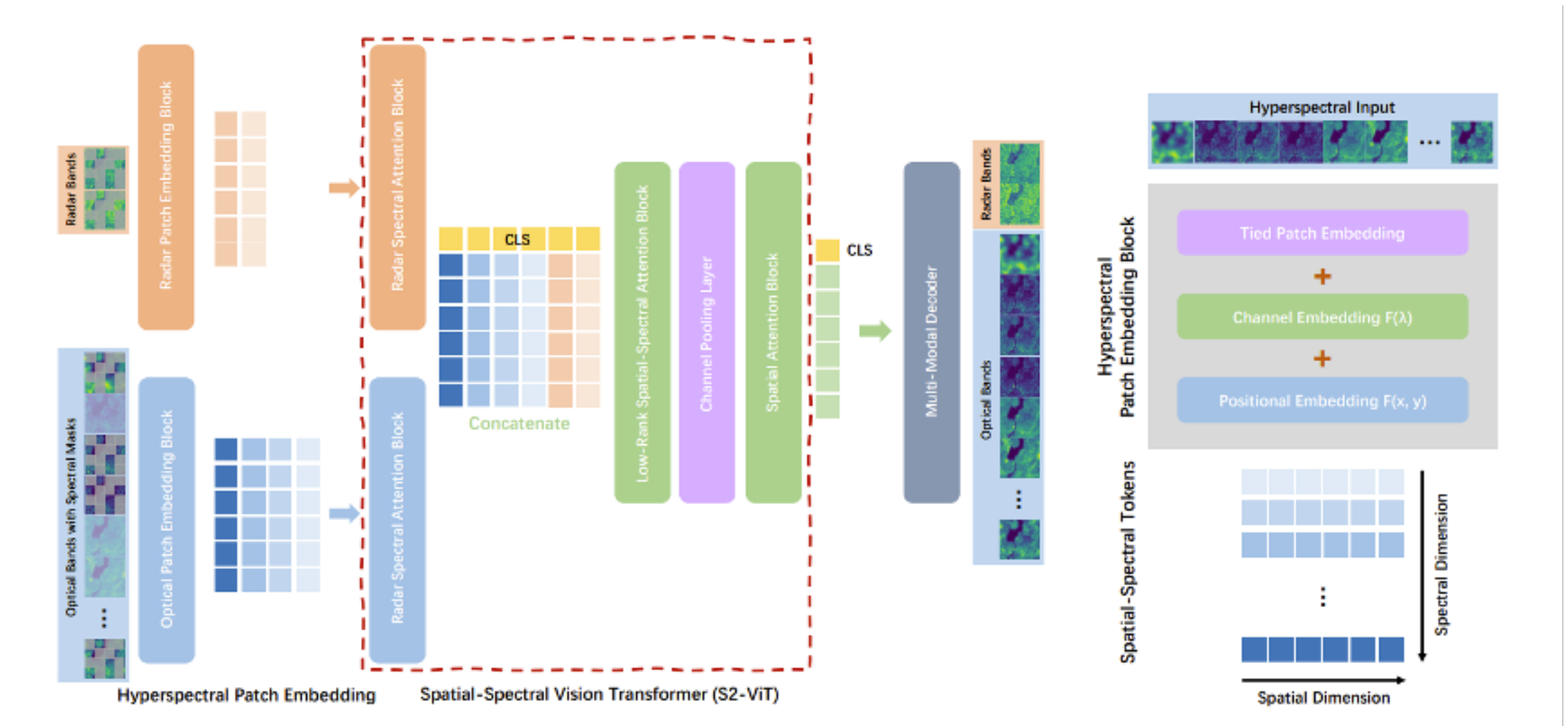

(a) The MM-MAE Framework (b) Hyperspectral Embedding

*Figure 23: (a) Our proposed MM-MAE framework. The hyperspectral patch embedding block addresses the inter-channel relationships. The S2-ViT fuses multi-modal features. The decoder will reconstruct all the channels and modalities. The decoder will be discarded after the SSL training. (b) An illustration of the hyperspectral patch embedding block. The block takes in hyperspectral images as the inputs and generates spatial-spectral tokens.*

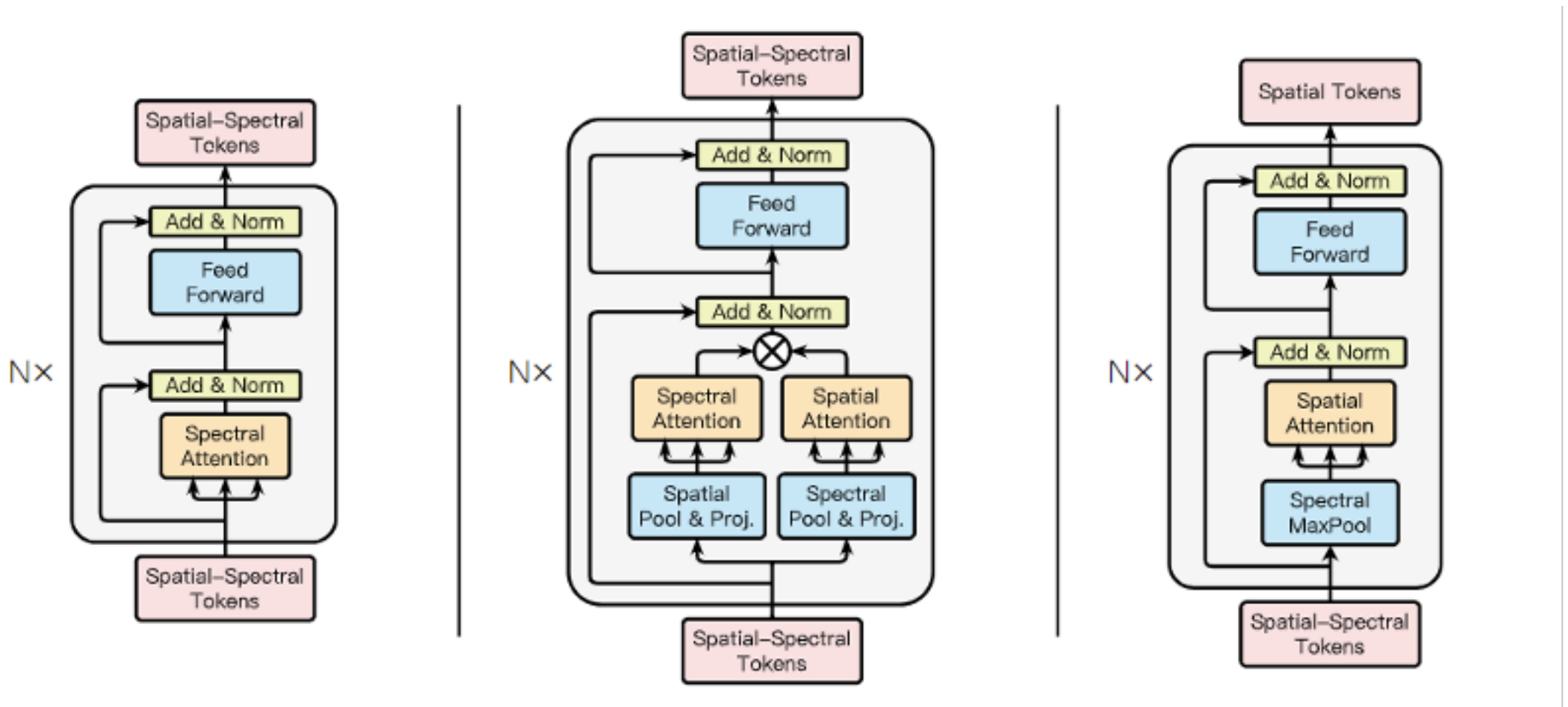

(a) Spectral Attention Blocks (b) Low-Rank Spatial-Spectral Attention Blocks (c) Spatial Attention Blocks

*Figure 24: The three component blocks of our S2-ViT architecture. The blocks are cascaded, with the output of each block serving as the input to the next. The balance between spatial and spectral attention can be adjusted by selecting the number of blocks of each type, allowing for a trade-off between capturing spatial context and spectral dependencies in the input data.*

We conducted experiments to evaluate the performance of our proposed techniques on two tasks: forecasting 2m temperature and predicting masked regions in satellite images. The results in the table below demonstrate a significant reduction in Mean Squared Error with our proposed techniques.

| Tasks | Prithvi + GraphCast | Our Techniques |
|---|---|---|
| Weather Forecasting (2m temperature) | 0.0112 | 0.0067 |
| Masked Region Prediction | 0.4142 | 0.3526 |

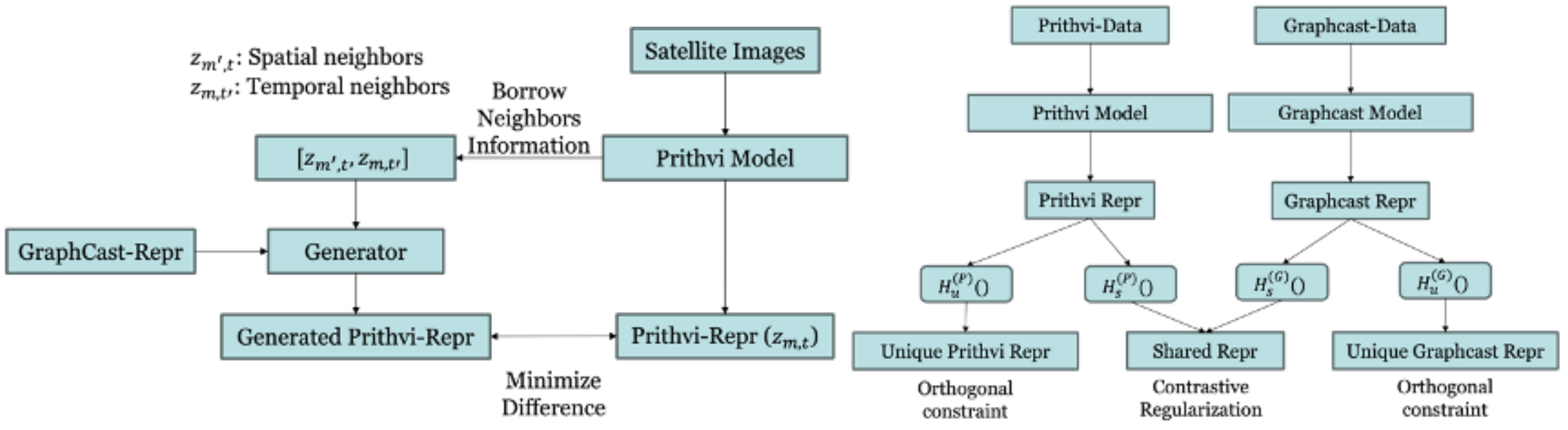

(a) (b)

*Figure 25: (a) Generative Module of Multi-modal Geospatial Foundation Models; (b) Multi-modal Climate Foundation Models.*

### 5.2.4 Long-term Directions and Vision

The long-term directions of the climate and sustainability theme will continue to be pre-dominantly focused on FMs, as follows:

- Continue to explore scalable and efficient architectures and approaches for developing FMs, which can deal with the multi-modal and high dimensional data (as required for climate and sustainability applications); this includes model size, FM composability, loss functions, and efficient strategies for pre-training content generation.
- Explore limits of physics AI emulators (e.g., for weather, power flow, etc.) targeting out-of-sample performances; different forecast horizons; data leakage; and model content.
- Applications of fine-tuned FMs in the area of climate and sustainability with special emphasis on scientific discovery; application areas include climate impacts (physical, economic, etc.), climate mitigation (carbon sequestration, etc.), and greenhouse gas emission quantification.
- Climate and sustainability solutions enabled through quantum computing.

# 6 The Role of Quantum Computing

## 6.1 Why Quantum?

Quantum computation has the potential to solve problems with high algorithmic complexity, which could take millions of years to run on classical computers, helping us explore fundamental concepts of the physical and mathematical sciences. In its essence, a quantum computer is a novel kind of computer, one which harnesses the fundamental quantum-mechanical properties of nature (namely, superposition, interference, and entanglement) and puts these properties to use for the purposes of storing and processing information. As such, quantum computers represent a genuine divergence in the history of computation: prior to their development and deployment, all classical compute – be it CPUs, GPUs, TPUs, or other ASICs – relied on purely-classical means of storing and processing information. Quantum computation is an entirely new paradigm for information processing.

Incredibly, quantum computation offers the possibility of turning heretofore-intractable problems into ones which are much more tractable. Quantum computers are the first credible example of a realistic model of computation that can solve some class of problems exponentially faster than a standard classical device. Quantum computing is at the precipice of enabling breakthrough advancements in various fields: quantum chemical simulation (useful in the pharmaceutical industry), mathematical optimization techniques (useful for a variety of industries), and speedups for search and factoring problems (which could impact current data encryption schemes), to name a few.

## 6.2 IBM and UIUC's Roles in the Field of Quantum Computing and Quantum Technology

Both IBM and the University of Illinois Urbana-Champaign have been deeply involved in the field of quantum computing. In 2018, with an initial $15 million investment, the University started its own quantum technology center, the Illinois Quantum Information Science and Technology Center (IQUIST), to advance the exploration of fundamental science while implementing novel quantum algorithms and state-of-the-art equipment for the fabrication of quantum materials and devices. The Urbana campus is primed to take a leadership role in the coming quantum information revolution as IQUIST develops QIS-focused educational programs for the next-generation quantum workforce.

IBM has a long, well-established history of quantum computing research, starting with Nuclear Magnetic Resonance based systems in the early 2000s, and moving on to superconducting qubits around 2010. In 2016, IBM became the first company in the world to host its quantum computer on the cloud. This helped accelerate the field by allowing anyone around the world – students, practitioners, enthusiasts – to learn and use a real quantum computing system. By open sourcing its software framework, Qiskit, IBM has further opened the doors of quantum technology research for people around the world.

## 6.3 IBM Quantum Development Roadmap and the Era of Utility

Over the years, IBM Quantum has expanded its development roadmap (see Figure 26) to include the software development that must go in conjunction with improvements in hardware. Present-day devices are greatly limited by a variety of errors that deleteriously affect the system.

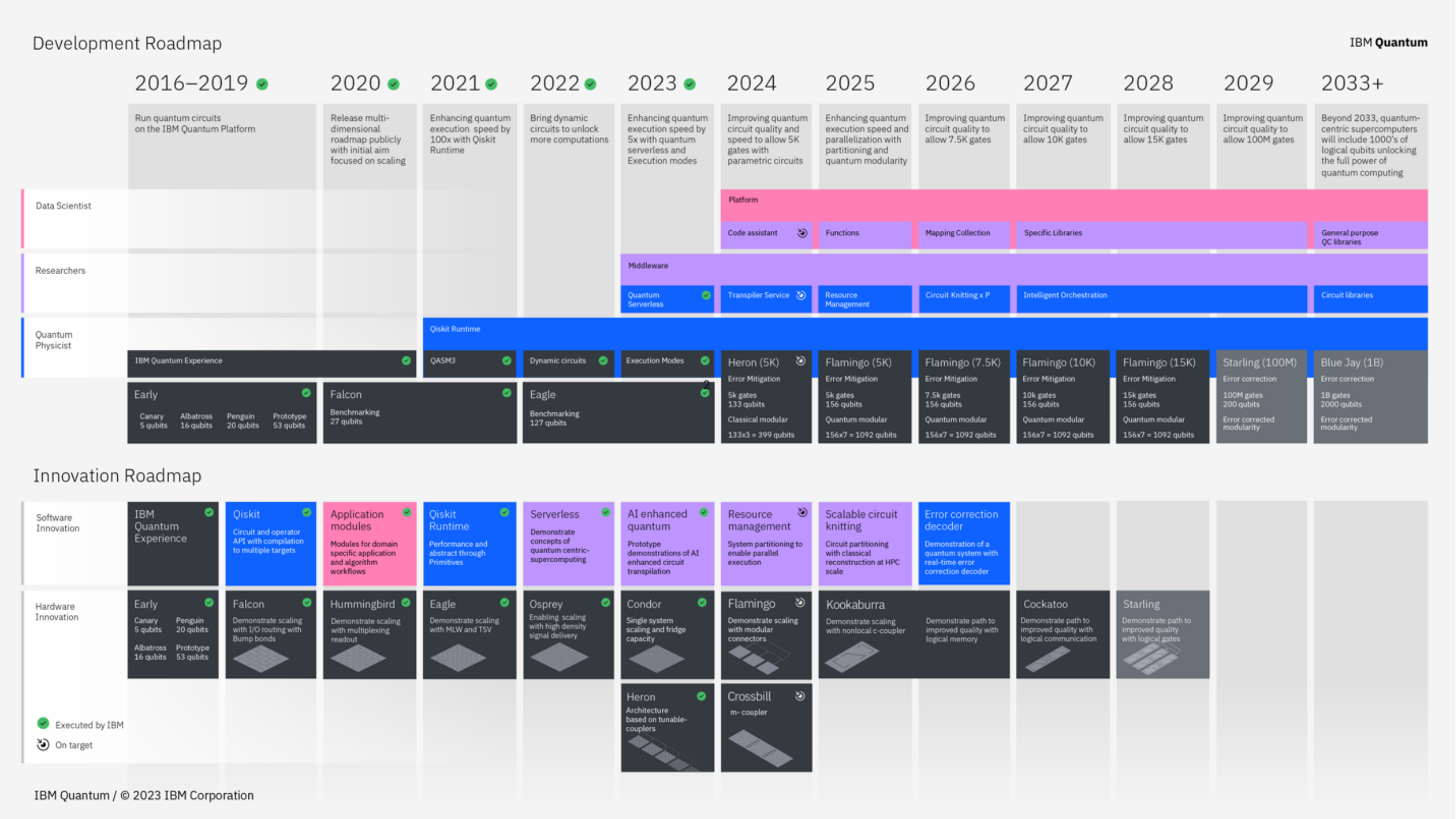

*Figure 26: At IBM Quantum Summit 2023, IBM extended its Quantum Development Roadmap to 2033, and has established an IBM Quantum Innovation Roadmap through 2029. (Credit: IBM).*

In the past decade alone, IBM has greatly improved the state of quantum systems, going from 5-10 qubits to a whopping 1121+ qubits (the Condor processor) not only through cutting-edge research in transmon processor design, but also through software, transpiler optimization, and advancing error mitigation schemes such as Probabilistic Error Cancellation (PEC) and Zero-Noise Extrapolation (ZNE).

These efforts lead to IBM demonstrating the utility of near-term quantum computers through an experiment of unprecedented size and complexity in June of 2023 [168]. This experiment – simulating how a system of interacting spins would evolve in time – validated both that large-scale quantum circuits could be run on its hardware (namely, a 127-qubit IBM Quantum Eagle processor), and that the theory and approach that IBM had developed for doing so were competitive with best-in-class, purely-classical methods known at the time. In particular, by benchmarking the performance of the Eagle processor against a supercomputer hosted by the Lawrence Berkeley National Lab, researchers found extremely competitive performance between both the quantum and classical methods in most experimental regimes. Subsequent work by IBM, the collaborators on the project, and the broader quantum computing community showed that additional purely-classical methods could be benchmarked against IBM's experimental results. These additional results obtained on the quantum system were also competitive with best-in-class classical methods.

In summary, this work opened up the door to the era of quantum utility. Quantum computers have matured to the point of being novel research tools unto their own right, and are capable of producing competitive results on challenging problems. Based on this remarkable accomplishment, IBM has charted a roadmap towards realizing Quantum-Centric Supercomputing. As the first step towards building and deploying a 100,000+ qubit system within the decade, IBM has upgraded its entire fleet to systems of 100+ qubits, namely the Eagle (127) and Heron (133 qubits) processors.

### 6.4 Going Forward: Quantum Supercomputing

Previous quantum systems were monolithic in nature. Scaling these systems required scaling all components simultaneously, which is, long term, not a feasible approach. Quantum-centric supercomputers (QCSCs) represent the next generation of scalable quantum systems, ones which will leverage best-in-class capabilities (both quantum and classical) to perform otherwise-intractable computations. QCSCs utilize a modular architecture to enable scaling. They combine quantum communication and computation to increase the computational capacity of the system and use a hybrid cloud middleware to seamlessly integrate quantum and classical workflows. Developing, prototyping, and fielding such systems represent one of the most important tasks the quantum computing industry must do in the next decade.

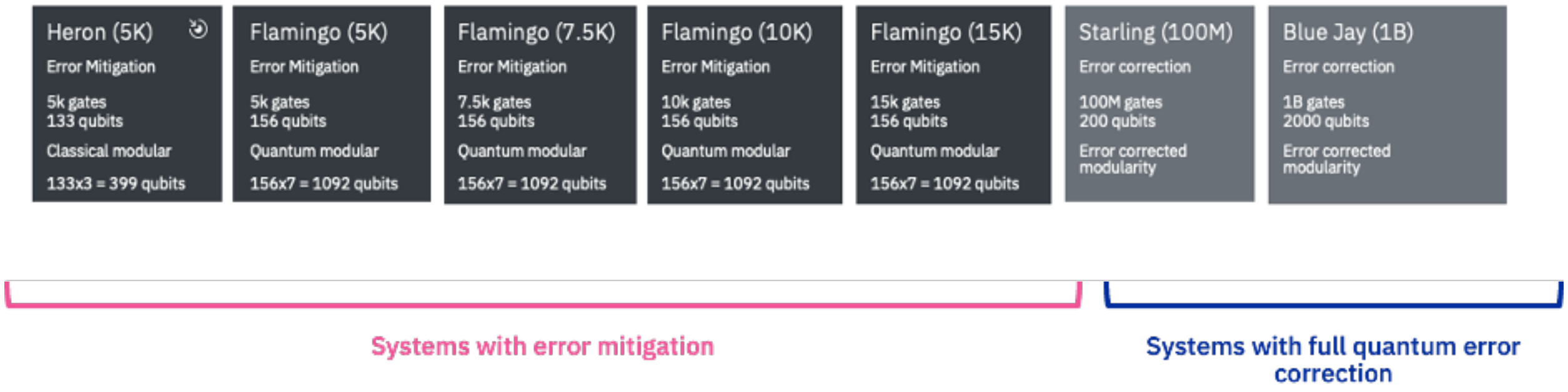

*Figure 27: IBM Quantum systems with error mitigation and error correction capabilities.*

IBM recognizes that the path to developing fault-tolerant quantum computation and quantum-centric supercomputing is not one that it can walk alone. IBM has a multitude of research collaborations with institute partners, including the IIDAI Quantum Thrust. Going forward, the focus of our research will be in the regime of quantum utility.

With advanced quantum processors and quantum algorithms in place, quantum utility is within practical reach and current research challenges center around (i) expanding the realm of application problems that benefit from quantum utility scale systems and (ii) pushing the envelope for error correction as well as error mitigation to reduce noise. The difficulty in the former lies in both identifying and developing quantum algorithms that can be more efficient, in practice, for solving certain problems, than the best classical alternatives. This will require improvements on all aspects of the quantum computation including efficient state preparation, fast algorithms, and intelligent classical post-processing of quantum output (possibly supplemented by using a hybrid cloud system with AI models). Real-world quantum computation is error-prone, statistically noisy, and rate limited in terms of gate speed. To get to the point of a practical advantage, significant experimentation to solve problems on quantum computers, to figure out the details of how to best trade-off between worse computational complexity and limited coherence, will be required. Algorithms such as variational quantum eigensolvers give approximate solutions without the need for deep circuits at the cost of expensive and difficult optimizations; approaches such as qubitization have robust guarantees but require significant improvements to make the circuits practical. Where this space of algorithms competes favorably against the backdrop of strong classical approximations is still very much open. Striking a balance between application needs and algorithms that are beneficial for quantum hardware, involves exploring algorithms such as propagating quantum dynamics, quantum embedding techniques, and variational quantum eigensolvers. Classical simulation workflows heavily rely on linear algebra, such as matrix diagonalization and Fourier transforms. While quantum computers have demonstrated advantages in these areas (e.g. qsvt, qft), the scaffolding using these subroutines need to be quantum optimized to build out full applications. Alternatively, classical simulation workflows need to be structured exclusively or more prominently around efficient quantum algorithms altogether. To best understand the limits of current-day quantum computers, we must focus on applications research on present-day devices that includes running large, high-depth circuits.

Research challenges in point (ii) above are two-fold (see Figure 27): Error correction requires developing and deploying robust, reliable, and high-rate error correcting codes with rapid syndrome measurements and decoding. This requires a multitude of advancements in superconducting couplers, fast decoding schemes, and mid-circuit measurement capabilities, to name a few. Also, decoding in quantum error correction is an exciting usecase in HPC or hybrid cloud in combination with quantum hardware, since efficient decoding is crucial for error correction in the regimes of interest. Error mitigation schemes developed by IBM, such as Zero Noise Extrapolation and Probabilistic Error Cancellation are the best current methods of handling noise on present-day systems without the need for error correction. IBM Quantum is focused on best integrating these methods into Qiskit Primitives, which will allow users and researchers to get the most out of present-day hardware.

Developing these is an ongoing research challenge that attracts the interest of the quantum computing community. The research challenges (i) and (ii) are an immense opportunity for the Quantum Thrust of IIDAI as exploring these leverages respective and mutual expertise and is an ideal path for industry-academia collaboration with great fundamental and applied interest.

### 6.5 Quantum Thrust: The Past Five Years

Since the beginning of the institute in 2021, the quantum thrust has focused on key areas in further advancing and developing quantum computing technology. Previous projects spanned hardware experimentation, applications research, and foundations of quantum information processing.

Major technical and education accomplishments from the technical projects include: A de-mateable cable connection between separated transmon qubits housed in a single dilution refrigerator was demonstrated. A fast (100 ns) and high fidelity (95%) SWAP gate was achieved through the connection. This result is in preparation for publication. Significant progress has been made on making a new qubit platform involving high kinetic-inductance devices. Resonators with inductors greater than 100 nH were characterized, and phase slip rates in 100 nm-wide wires were made.

A paper describing the discovery of a new duality between teleportation and dense coding was published [169], and a paper on exponential separation between quantum statistical query learning and quantum probability approximately correct learning was published [170]. A novel error mitigation technique involving deep learning was developed and used to predict noiseless results for materials simulations. Furthermore, work on quantum computing used for material simulations was published [171], as well as more foundational quantum information theoretic work on multipartite entanglement [172].

### 6.6 Quantum Thrust: Vision and Long-Term Directions

Given the current state of technological and algorithm development, the vision of this thrust going forward centers around the development of scientific use cases as demonstrations for quantum utility. We aim for these to be diverse problems that illustrate to a broad computational community the utility of quantum computers for their work. The vision is to focus on domain science problems, for which a successful use of quantum utility can be generalized to other applications. Naturally, this will start with applications in quantum chemistry, materials science, and physics, but eventually, the impact must be broader across computational fields. This will require implementation of quantum computing aspects into computational workflows as those described in Figure 6, replacing those algorithms that most benefit from quantum computation. This thrust will also work towards building such knowledge and a structured guide of what mathematical operations are best suited for quantum implementations, and under what conditions, such as problem size, error tolerance, etc. Part of this thrust will explore development of new algorithms that can achieve better performance. Numerous adaptive algorithms for measuring observables on a quantum computer might benefit from assisting cloud-computing techniques. Combining cloud-computing techniques with randomized measurement techniques is another interesting avenue [173]. In addition, the development and especially the testing of error mitigation approaches will be critical for this thrust in the future, and will be explored in the context of specific scientific problems and workflows. Finally, this thrust will keep pushing the training of a qualified quantum workforce, starting at the high-school level, but focusing on undergraduate and graduate students, as well as postdocs. The vision is to develop quantum literacy in computational domain scientists, to enable future development of computational software packages with quantum integration.

# 7 Summary

This paper, prepared by the technical leadership of the IBM-Illinois Discovery Accelerator Institute (IIDAI), outlines our vision for the future of hybrid cloud systems in relation to emerging AI workloads, novel AI-driven scientific applications, and quantum computing. We identify key priority focus areas for the technical thrusts of the Institute. Through the launch of a new research program, IIDAI is determined to tackle the challenges facing current hybrid cloud systems and change their trajectory to become more affordable, programmable, adaptive, resilient, and accessible, empowering the next generation of AI-centric applications for novel scientific discoveries.